\documentclass[twocolumn]{aastex63}
\pdfoutput=1

\newcommand\Gaia{{\it Gaia}~}
\newcommand\tess{{\it TESS}}
\newcommand\teff{{\log T_{\rm {eff}}}}
\newcommand\lum{{\log L/L_\odot}}

\usepackage{amsmath,graphicx,multirow,mathtools}

\defcitealias{neugent12_ysg}{N12}
\defcitealias{yang21}{Y21}
 
\received{21 July, 2023}
\accepted{21 October, 2023}
\submitjournal{ApJ}

\shorttitle{Cool Supergiants with \Gaia~XP Spectra}
\shortauthors{Dorn-Wallenstein et al.}

\begin{document}

\title{Physical Properties of 5,000 Cool LMC Supergiants with \Gaia XP Spectra: A Detailed Portrait of the Upper HR Diagram Hints at Missing Supernova Progenitors}

\correspondingauthor{Trevor Z. Dorn-Wallenstein}
\email{tdorn-wallenstein@carnegiescience.edu}

\author[0000-0003-3601-3180]{Trevor Z. Dorn-Wallenstein}\thanks{Carnegie Fellow}
\affiliation{The Observatories of the Carnegie Institution for Science \\ 813 Santa Barbara Street \\ Pasadena, CA 91101, USA}

\author[0000-0002-5787-138X]{Kathryn F. Neugent}
\thanks{NASA Hubble Fellow}
\affiliation{Center for Astrophysics, Harvard \& Smithsonian, 60 Garden St., Cambridge, MA 02138, USA}


\author[0000-0003-2184-1581]{Emily M. Levesque}
\affiliation{University of Washington Astronomy Department \\
Physics and Astronomy Building, 3910 15th Ave NE  \\
Seattle, WA 98105, USA}


\begin{abstract}

Characterizing the physical properties of cool supergiants allows us to probe the final stages of a massive star's evolution before it undergoes core collapse. Despite their importance, the fundamental properties for these stars --- $\teff$ and $\lum$ --- are only known for a limited number of objects. The third data release of the \Gaia~mission contains precise photometry and low-resolution spectroscopy of hundreds of cool supergiants in the LMC with well-constrained properties. Using these data, we train a simple and easily-interpretable machine learning model to regress effective temperatures and luminosities with high accuracy and precision comparable to the training data. We then apply our model to 5000 cool supergiants, many of which have no previously-published $T_{\rm eff}$ or $L$ estimates. The resulting Hertzprung-Russell diagram is well-populated, allowing us to study the distribution of cool supergiants in great detail. Examining the luminosity functions of our sample, we find a notable flattening in the luminosity function of yellow supergiants above $\lum=5$, and a corresponding steepening of the red supergiant luminosity function. We place this finding in context with previous results, and present its implications for the infamous red supergiant problem.  

\end{abstract}

\section{Introduction}\label{sec:intro}

Measurements of the fundamental properties of stars --- their effective temperatures, luminosities, masses, radii, surface gravities, chemical composition, etc. --- are critical observational benchmarks. By comparing these properties with predictions from stellar evolutionary models, we can place tight constraints on a star's evolutionary pathway from zero age main sequence to core collapse. Stars with initial masses $M \gtrsim 10 M_\odot$\footnote{We note that this threshold is approximate, and highly-dependent upon uncertain stellar physics that dictate the structure of pre-supernova stellar cores \citep[e.g.][]{sukhbold18,tsang22}.} are rare, by dint of both their short lifetimes and the initial mass function, and are poorly-represented in large spectroscopic surveys, making the assembly of a statistical sample of massive stars a daunting task. This is especially the case for evolved massive stars, particularly cool supergiants in the final evolutionary phases before core collapse.

As a result, the evolution and final fates of massive stars remain unclear. Stars descended from Zero-Age Main Sequence (ZAMS) progenitors of $10-30 M_\odot$ are especially interesting, as recent observational and theoretical evidence has revealed our lack of understanding of whether or not they explode after core collapse, what evolutionary stage they explode in, and what the resulting supernova (SN) explosion looks like \citep[e.g.][]{smith14,smartt15,sukhbold18,neustadt21}. Notably, significant debate has arisen surrounding the discrepancy between observed red supergiant (RSG) populations (which can be found up to a maximum luminosity around $\lum\sim5.5$; see e.g., \citealt[][]{levesque05,mcdonald22,massey22}) and the observed or inferred RSG progenitors of Type II-P supernovae (SNe), which appear to have a maximum luminosity around $\lum\sim5.0$ \citep{smartt09}. While this issue was dubbed the {\it red supergiant problem}, it extends to core-collapse supernovae of all subtypes \citep{smartt15}, is detectable in both direct progenitor imaging and indirect methods \citep[e.g.][]{prentice22} and the statistical significance of the discrepancy has only grown (e.g. \citealt{kochanek20,rodriguez22}, but see also \citealt{davies20}). 

The key processes that impact the red supergiant problem --- the interior physics that governs the lifetimes of the yellow supergiant (YSG) and RSG phases \citep[e.g.][]{ekstrom12}, and mass loss both from winds and binary interactions \citep[e.g.][]{eldridge17} --- are encoded in the distribution of stars in the upper Hertzprung-Russell (HR) Diagram. Given a sufficiently large sample of stars, such effects can be directly observed, and there is a rich history of work surveying Local Group galaxies for cool supergiants and comparing their luminosity distributions with model predictions. For example, in studies of M31 and the Small Magellanic Cloud (SMC) respectively, \citet{drout09} and \citet{neugent10} found that the ``old'' Geneva models \citep[e.g.][]{meynet03} overpredicted the relative numbers of luminous YSGs. Subsequently, \citet{neugent12_ysg} and \citet{drout12} used samples in the Large Magellanic Cloud (LMC) and M33 respectively along the with ``new'' Geneva models \citep{ekstrom12}, and found excellent agreement. This agreement was driven by an updated mass loss prescription in the models that included a supra-Eddington mass loss rate, which become important for stars above $\sim$20 $M_\odot$.\footnote{This RSG mass loss prescription is also in agreement with recent work by \citet{humphreys20}, who find a sharp increase in RSG mass loss rates above $\lum=5$. However, \citet{beasor20,beasor21} have simultaneously argued for a {\it reduction} in theoretically-employed RSG mass loss rates. The work continues.} However, all of these studies used limited samples of only a few hundred YSGs at most, and the resulting comparison between the models and data was done in coarse bins of initial mass/luminosity. 

More recent work \citep[e.g.][]{neugent20b,neugent20,mcdonald22,massey22} has focused exclusively on RSGs, which are more abundant, allowing for a detailed comparison between the observed and predicted RSG luminosity functions. In particular, \citet{massey22} recently demonstrated that the impact of supra-Eddington winds --- a decreased RSG lifetime above $\lum\approx5$ caused by the loss of the envelope, leading to a reduction in the percentage of high-luminosity RSGs relative to the overall population  --- is readily observable. As a consequence of this envelope loss, stripped RSGs are predicted to evolve bluewards in the HR diagram before undergoing core-collapse as yellow or blue supergiants (BSGs). Confirming this scenario using an updated statistical sample of YSGs in context with its corresponding RSG population is therefore crucial.

With its unparalleled astrometric and photometric precision, the \Gaia~mission \citep{gaia16} has already observed evolved massive stars in sufficient quantity to perform studies like this, allowing for tremendous headway to be made in studying massive star properties {\it en masse} by leveraging well-understood machine learning techniques \citep{dornwallenstein21,maravelias22}. In particular, the recent third data release (DR3; \citealt{gaia23}) includes flux-calibrated $R\sim50$ optical spectroscopy for all sources brighter than $G<17.65$ \citep{deangeli22}. In this work, we demonstrate that these data are sufficiently information-rich that we can use them to train a regressor to predict effective temperatures $T_{\rm eff}$ and luminosities $L$ for supergiants in the Large Magellanic Cloud (LMC) of all spectral types later than A0. This sample allows us to study, with incredible precision, the distribution of cool supergiants in the HR diagram.

In \S\ref{sec:regressing}, we describe our training data, the range of regression methods we experimented with, and the performance of these regressors. In \S\ref{sec:results} we apply our regressor to a sample of $\sim$5000 YSGs and RSGs, $\sim3500$ of which have no previously published temperatures or luminosities, and show that our regressor performs significantly better than temperature estimates from broadband photometry. We then examine the distribution of stars in the upper HR diagram in detail, finding strong evidence that the luminosity function of YSGs flattens above $\lum=5$, while the RSG luminosity function steepens. We discuss the implications of this result in \S\ref{sec:discussion}, and present both an effective temperature scale and bolometric corrections derived from our observations, before concluding in \S\ref{sec:conclusion}.

\section{Regressing \texorpdfstring{$\teff$ \& $\lum$}{}}\label{sec:regressing}

\subsection{Training Sample}\label{subsec:training}

The fundamental problem in regression is to predict some target variable(s) --- in this case, $\teff$ and $\lum$ --- given a set of input observations, or features. We must first identify a sample of cool supergiants whose luminosities and temperatures are already known to train and test this model. \citet{neugent12_ysg} (\citetalias{neugent12_ysg} hereafter) present a large sample of YSGs and RSGs in the LMC with temperatures ($\teff$) and luminosities ($\lum$) estimated using updated formulae derived from Kurucz \citep{kurucz92} and MARCS \citep{gustafsson08} models and near-infrared photometry. These measurements have a typical precision of 0.015 dex and 0.10 dex in $\teff$ and $\lum$ respectively. While other, more complete catalogs of Magellanic Cloud supergiants exist \citep[e.g.][]{dorda16,dorda18,yang19,neugent20b,yang21}, these samples either do not include temperature and luminosity estimates, or focus solely on RSGs. Additionally, in the case of the RSGs from \citet{neugent20b}, the temperatures are discretized to 50 K increments. Because we wish to construct a homogeneous training sample, we choose to only use the \citetalias{neugent12_ysg} catalog, and not augment it with more recent work. 

We first cross-match the stars from \citetalias{neugent12_ysg} with \Gaia~DR3 using the \Gaia~archive,\footnote{\url{https://gea.esac.esa.int/archive/}, DOI: \citet{gaiarchive2022}} and select all objects with the {\tt has\_xp\_continuous} flag set to True. Because \citetalias{neugent12_ysg} identified their targets with data from the Two Micron All-Sky Survey (2MASS, \citealt{cutri03}), all stars in their sample have 2MASS identification numbers, and so we were able to perform this cross-match using the existing 2MASS-\Gaia~DR3 cross-match on the \Gaia~archive instead of performing a separate spatial cross-match. This results in 818 objects.

A known issue with this sample is the accidental inclusion of a number of stars that have since been identified as OB supergiants; because of the near-infrared excess in these stars caused by free-free emission in their winds, the derived temperatures from $J-K_S$ photometry appear sensible for AFG supergiants. Using SIMBAD,\footnote{\url{https://simbad.u-strasbg.fr/simbad/}} we identify the 112 objects with ``O'' or ``B'' in their spectral types and remove them.\footnote{We note that this cut also rejects a number of binary stars with spectral types like ``M4Ia + B7III,'' as OB stars are the most common companions to evolved supergiants \citep{neugent20b}. While the resulting sample is by-no-means free of binary stars, it is less contaminated by objects with composite spectra that might impact the performance of the regressor down the line.} 

\begin{figure}[t!]
\centering
\includegraphics[width=0.5\textwidth]{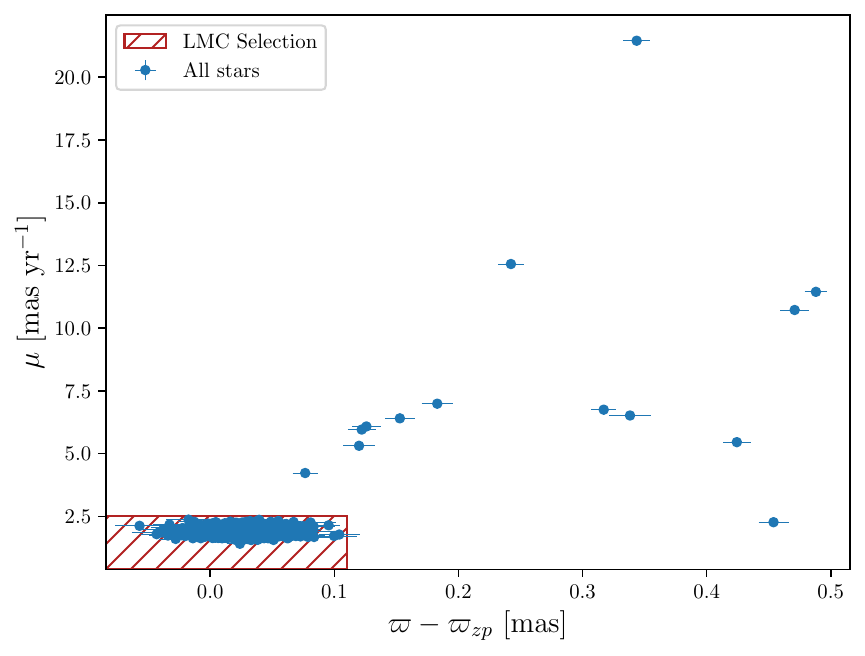}
\caption{Zero-point corrected parallax versus total proper motion for 658 stars in our sample. The LMC is the obvious clump of stars in the lower left corner. We show the box used to select these stars in red.}\label{fig:plx_pm}
\end{figure}

Another known issue with this sample is contamination by foreground objects whose radial velocities are similar to the radial velocities of LMC stars. Thankfully, \Gaia~allows us to easily identify and remove such objects. We first remove stars in our sample without a listed parallax, $\varpi$, and with a renormalized unit weight error $\mathrm{RUWE}>1.4$ \citep[which ensures a good astrometric solution; see][]{gaiacollab18}; we also require stars to have a 5- or 6-parameter astrometric solution ({\tt astrometric\_params\_solved}$=$31 or 95, respectively). This allows us to use the {\sc gaiadr3\_zeropoint} python package \citep{lindegren21} to compute the color- and spatially-dependent \Gaia~parallax zero-point, $\varpi_{\rm zpt}$ for each star. After rejecting OB stars and objects that do not satisfy our astrometric cuts, we are left with 658 objects. We show $\varpi-\varpi_{\rm zpt}$ and the total proper motion $\mu$ for the stars in our sample in Figure \ref{fig:plx_pm}. LMC stars are clearly visible as a clump in the lower-left of the Figure. We select these stars using $\varpi-\varpi_{\rm zpt} \leq 0.11$ mas\footnote{A simple inversion of 0.11 mas yields a distance of $\sim$10 kpc. However, extensive work has shown that naive inverse parallaxes are only trustworthy for the most nearby stars \citep[e.g.][]{bailerjones15,bailerjones18,bailerjones21,olivares20}.} and $\mu\leq2.5$ mas yr$^{-1}$. Applying this selection (shown as a red box in Figure \ref{fig:plx_pm}) results in 641 likely LMC supergiants\footnote{We also remove 5 objects with anomalously low luminosities ($\lum\leq3$).} with low-resolution \Gaia~spectra.

In order to validate our regression methods, we use the {\tt train\_test\_split} function in {\sc scikit-learn} to randomly select 70\% of this sample to use as a training set which will be used to fit the models, leaving the remaining 30\% as a test set that we can use the evaluate their performance. Because the sample is dominated by RSGs due to the longer lifetime of that phase relative to the YSG phase,\footnote{This is an additional reason not to include more RSGs from \citet{neugent20b}: RSGs are already overrepresented in the training data.} we bin the sample into 0.05 dex wide bins in $\teff$, label all stars in each bin with the $\teff$ at the center of the bin, and use those bin labels to stratify the train/test split and ensure that each bin is well-represented in both the training and test sets. We note that this binning is only used to stratify the train/test split, and that only the actual observed $\teff$ values are used in the regressor.

\begin{figure}[t!]
\centering
\includegraphics[width=0.5\textwidth]{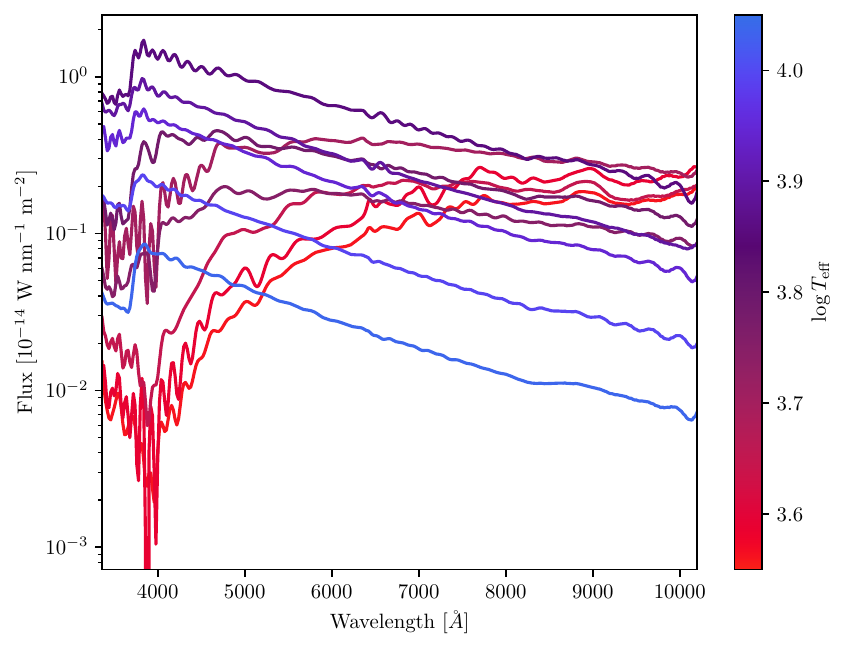}
\caption{Sampled \Gaia~spectra of the brightest star in each 0.05 dex wide bin in $\teff$ between 3.55 and 4.05. The spectra are color-coded by their effective temperature as indicated by the colorbar.}\label{fig:ex_spec}
\end{figure}

\subsubsection{Low-resolution Spectra from \Gaia~DR3}\label{subsubsec:gaiadata}

Spectra from \Gaia~DR3 are described by a set of 110 coefficients (55 each from the BP/RP instruments, which we denote $b_i$/$r_i$ for the $i^{\rm th}$ coefficient from the BP/RP spectrum respectively), which are generated by decomposing the observed spectra into basis functions,\footnote{In this case, Hermite polynomials, centered on the wavelength-center of each instrument, and multiplied by a Gaussian to simulate the throughput of the instrument.} then rotating the basis functions into a space that maximizes the amount of variance in the observed spectra described by each subsequent component. It is important to note that, because of how the basis spectra are constructed, there is no one-to-one mapping from a single spectral coefficient and a star's effective temperature, or the presence of a specific absorption or emission line. Instead, this process, similar to principal component analysis, allows the data to be represented by significantly fewer numbers than the pixels in the original spectra. The user can then easily reconstruct the data without significant loss of information, even when using a relatively small number of coefficients. Readers interested in more details about the \Gaia~XP spectral representation are referred to \citet{deangeli22}. 

Figure \ref{fig:ex_spec} shows the reconstructed spectra of the brightest star selected from 0.05 dex-wide bins in $\teff$. The reconstruction is performed using the coefficients downloaded from the \Gaia~archive and the {\sc GaiaXPy} package.\footnote{\url{https://gaia-dpci.github.io/GaiaXPy-website/}} Examining these spectra, it is clear that hotter stars do indeed have bluer spectra, as one might hope. An experienced spectroscopist may even be able to identify a star's spectral type, and sophisticated analyses might be employed to test the presence of specific absorption or emission lines \citep[e.g.][]{weiler22}. However, the spectral resolution is not sufficient to measure the diagnostics one might use to measure an effective temperature or surface gravity for a given star. Thankfully, the coefficients themselves are still information-rich \citep{witten22}; for example, \citet{deangeli22} demonstrated that only the first few coefficients can be used to reliably classify stars by spectral type. The coefficients are also linearly independent --- a subset of coefficients for a star cannot be used to predict the other coefficients --- making them perfect quantities to use when performing regression. 

\subsection{Feature Selection}\label{subsec:featureselection}

While the spectral coefficients are information-rich, they also encode the star's apparent brightness, in the sense that brighter stars have larger spectral coefficients. Imagining each star's spectral coefficients for a given instrument as a vector in a 55-dimensional space, we wish to separate the magnitude of that vector (the apparent brightness) from its orientation (the actual spectral shape). Therefore, we normalize the coefficients for each star by the Euclidean norm of the coefficients in each instrument ($|b|$/$|r|$, respectively); from here, we change our notation slightly so that $b_i$/$r_i$ refer to the {\it normalized} coefficients. 

In addition to the spectral coefficients, we also compute each star's absolute magnitude in the \Gaia~bandpass, $M_G$, assuming a distance modulus for the LMC of 18.52 \citep{kovacs00}, as well as the $G_{BP}-G_{RP}$ color. We correct neither quantity for extinction; instead, we use the dust map of the LMC from \citet{chen22} and the {\sc dustmaps} package \citep{green18} to compute $E(B-V)$ values for each star based on its position to use as a feature. As corrections to $M_G$ and $G_{BP}-G_{RP}$ are usually linear in $E(B-V)$, the regressor should be able to account for the effects of extinction and reddening. In \S\ref{subsubsec:featureimportance}, we explore the performance of the regressor with and without $E(B-V)$ and test this assumption. 

This results in a total of 115 possible features for us to use:
\begin{itemize}
    \item 55 normalized BP spectral coefficients, $b_i$,
    \item 55 normalized RP spectral coefficients, $r_i$,
    \item The Euclidean norm of the coefficients for each instrument, $|b|$ and $|r|$, and
    \item The three photometric features, $M_G$, $G_{BP}-G_{RP}$, and $E(B-V)$.
\end{itemize}
In practice, because $|b|$ and $|r|$ span a large dynamic range, we use the base-10 logarithm of these values as features instead. 

Of these 115 features, which ones can we anticipate being useful predictors of $\teff$ or $\lum$? As a rough pass, we can identify only those features that show some correlation with the target variables. We compute the correlation matrix of the data, selecting all features whose correlations with either $\teff$ or $\lum$ are greater than $\pm0.2$.\footnote{Unsurprisingly, negative correlations are present in the data --- e.g., $M_G$ is negatively correlated with $\lum$, $G_{BP}-G_{RP}$ is negatively correlated with $\teff$.} Note that this correlation threshold is an arbitrary choice intended to produce a large set of features that we can further winnow down to those that best predict our target variables (see \S\ref{subsubsec:featureimportance}).  We then also include $E(B-V)$. This results in a total of 84 features along with the two target variables. We list these features in Table \ref{tab:features}. We then scale each feature to have 0 mean and unit variance by fitting a {\tt StandardScaler} object from {\sc scikit-learn} on the training set before transforming both the training and test sets. This ensures that features with a high variance do not overly influence the regression.

\begin{deluxetable*}{l|c}
\tabletypesize{\scriptsize}
\tablecaption{The 84 features selected for our regressor.\label{tab:features}}
\tablehead{\colhead{Category} & \colhead{Features}}
\startdata
BP Spectral Coefficients & $\log |b|$, 0-8, 12, 15-24, 26-43, 46, 48, 49, 51-53 \\ 
RP Spectral Coefficients &  $\log |r|$, 0-3, 5-11, 13-16, 18-28, 30, 31, 34, 39-41, 46, 53, 54  \\ 
Photometry & $M_G$, $G_{BP}-G_{RP}$, $E(B-V)$
\enddata
\end{deluxetable*}

A plethora of methods for regression can be found in the literature, many of which are implemented in the frequently-used python package, {\sc scikit-learn}. Appendix \ref{app:regression} describes our testing of these methods in detail, including our development of various metrics tailored to our training set, and our strategies to perform regression with two output variables. In brief, of the 11 regressors we experimented with, optimal performance was achieved by the Support Vector Machine (SVM) regressor (SVR; \citealt{cortes95,smola04}) and the Bayesian Ridge Regressor (RR, \S3.3 in \citealt{bishop06}). While each regressor performed slightly better or worse across the metrics we measured, ultimately, we chose the Bayesian RR, an extension of least-squares regression that attempts to simultaneously minimize the sum of square residuals (i.e., lease-squares regression) {\it and} the sum of the square of the model coefficients. This constraint reduces the variance, or the sensitivity of the model to different realizations of the training data.

\subsubsection{Feature Importance}\label{subsubsec:featureimportance}

Given the number of features at our disposal, it is reasonable to now turn our attention to whether we even need this information in order to regress $\lum$ and $\teff$. In particular, we will now explore the necessity of including $E(B-V)$, $M_G$ and $G_{BP}-G_{RP}$, or even the spectral coefficients themselves in the regressor. Using the Bayesian RR regressor, we fit the scaled training set and predict $\lum$ and $\teff$ for the test set using four different subsets of features:
\begin{enumerate}
    \item All features
    \item All features except $E(B-V)$ (i.e., no reddening)
    \item All features except $E(B-V)$, $M_G$, and $G_{BP}-G_{RP}$ (i.e., no photometry)
    \item Only $E(B-V)$, $M_G$, and $G_{BP}-G_{RP}$ (i.e., no spectroscopy).
\end{enumerate}

Each row in Table \ref{tab:without_features_sum} shows the results of this procedure for each subset of features, summarized using seven numerical diagnostics. While we describe the motivation and derivation of these diagnostics in detail in Appendix \ref{app:regression}, they are all based on residuals between the predicted and true values of $\teff$, $\lum$, or both quantities simultaneously. $d_{M,95}^2$ captures the distribution of distances between the true and predicted locations of stars in the HR diagram with respect to the typical observational uncertainties: better predictions yield smaller $d_{M,95}^2$ values. We also compute the mean and standard deviation of differences between actual and predicted temperature ($\langle\Delta\teff\rangle$ and $\sigma_{\Delta\teff}$), and similar metrics for $\lum$ ($\langle\Delta\lum\rangle$ and $\sigma_{\Delta\lum}$). Finally, we compute the percentage of predicted values that fall within the error ellipse in the HR diagram generated using the corresponding true values, and both 1 and 2 times the observational uncertainties (\%$\leq1\sigma$ and \%$\leq2\sigma$).

The Bayesian RR trained with no spectroscopy performs the worst by far across all seven metrics. Furthermore, while the predicted $\lum$ values are normally distributed about the true values (albeit with more scatter than expected by the observational uncertainty), the residuals in $\teff$ are actually structured (i.e., they depend on $\teff$). We computed the correlation of the residuals with the values of the features in the test set, and find that the residuals are strongly correlated with low order spectral coefficients (e.g., the correlation between residuals and $r_0$ is 0.81, and the correlation with $b_2$ is -0.63), and weakly correlated with higher order coefficients (e.g., seven coefficients with a subscript greater than or equal to ten have correlations with the residuals around $\pm$0.4-0.5). These results are unsurprising; the relationship between colors and effective temperature is known to be nonlinear, and the Bayesian RR is a linear model. 

The remaining sets of features perform well. Training the regressor without reddening as a feature yields a modest decrease in performance compared to the fiducial model trained on all of the available features. Similarly, the regressor trained exclusively on the spectral coefficients (which also don't include reddening) yields a modest {\it increase} in performance. While this would seem to indicate that the regressor does not need photometry (including $E(B-V)$, though we discuss this result further in \S\ref{sec:discussion}), we note that this is only true for our sample where the most-reddened star has $E(B-V)=0.26$. This result may not be (and likely isn't) applicable to more extincted sources. However, testing this statement further would require artificially reddening the observed spectra, converting back into spectral coefficients, using these coefficients and the trained regressor to predict $\teff$ and $\lum$, and examining the accuracy of the results as a function of the added reddening. Unfortunately, {\sc GaiaXPy} does not yet have the capability of converting observed or modelled spectra into basis coefficients, so such a test is not currently feasible. Finally, the differences in performance across these three examples are relatively small. The only firm conclusion that we draw is that the spectral coefficients absolutely are necessary.

\begin{deluxetable*}{l|ccccccc}
\tabletypesize{\scriptsize}
\tablecaption{Summary of Bayesian Ridge Regression performance on $\teff$ and $\lum$ simultaneously, using only specific subsets of features.\label{tab:without_features_sum}}
\tablehead{\colhead{Features} & \colhead{$d_{M,95}^2$} & \colhead{$\langle\Delta\teff\rangle$} & \colhead{$\sigma_{\Delta\teff}$} & \colhead{$\langle\Delta\lum\rangle$} & \colhead{$\sigma_{\Delta\lum}$} & \colhead{\%$\leq1\sigma$} & \colhead{\%$\leq2\sigma$}}
\startdata
No Spectroscopy & 33.234 & 0.0024 & 0.0397 & 0.0053 & 0.0721 & 37.82 & 64.77 \\ 
No $E(B-V)$ & 7.619 & 0.0020 & 0.0190 & 0.0051 & 0.0508 & 86.01 & 93.78 \\ 
All Features & 7.487 & 0.0020 & 0.0189 & 0.0050 & 0.0503 & 85.49 & 93.78 \\ 
No Photometry & 7.387 & 0.0020 & 0.0192 & 0.0051 & 0.0512 & 84.97 & 93.78 \\ 
\enddata
\end{deluxetable*}

While this experiment conclusively demonstrates the utility of the \Gaia~spectral coefficients, we can go further and determine exactly what contribution to the model each feature is making. In principle, the Bayesian RR minimizes the sum of squared coefficients in the underlying linear model, which ideally ensures that many of the coefficients remain small (i.e., only a small number of features contribute to the prediction). We can directly assess this by performing a so-called``greedy search'' to determine which features contribute the most to the classifier performance. For each feature in the scaled training set, we train the regressor on just this feature using a five-fold cross-validation procedure\footnote{This is a commonly-applied method that involves randomly dividing the data into $n$ equally-sized parts. Each part (or ``fold'') is used as a test-set for the regressor or classifier trained on the remaining $n-1$ folds, and the performance is averaged across folds.}, recording $d_{M,95}^2$, and the percentage of residuals falling within both one and two $\sigma$ as in Tables \ref{tab:tl_sum} and \ref{tab:without_features_sum} for each fold. We select the feature that yields the lowest average $d_{M,95}^2$ across the folds, ensuring the contribution of each feature to the performance of the regressor is stable across subsets of the data. We then train the regressor on all combinations of this feature and the remaining features, selecting which combination again yields the lowest average $d_{M,95}^2$. We repeat this process, iteratively adding successive features into the regressor until all features are used. 

\begin{figure*}[t!]
\plotone{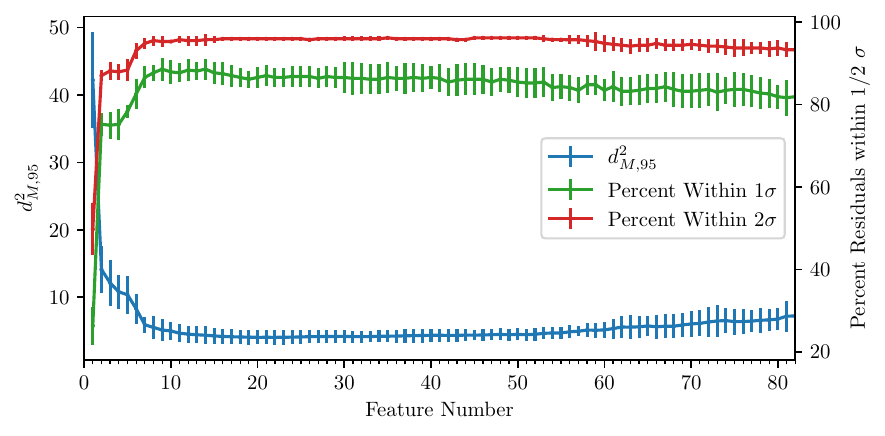}
\caption{Results from a greedy search over features averaged over five folds of the training set for the Bayesian Ridge Regressor trained on an interatively grown subset of features. The blue line and errors show the mean and standard deviation of $d_{M,95}^2$ across five folds of the training set. Green/red show the percentage of residuals within one and two $\sigma$ of their expected values.}\label{fig:feature_importance}
\end{figure*}

Figure \ref{fig:feature_importance} shows the results of our greedy search over features. The blue line shows the mean $d_{M,95}^2$ value as a function of successively added features, with the error bars showing the standard deviation of $d_{M,95}^2$ values across folds. We also show the mean and standard deviation of the percentage of residuals within one (green) and two (red) sigma. Interestingly, the values of all three quantities plateau at $\sim$5, 88\%, and 95\% respectively after with only ten features: $b_{1}$, $\log|r|$, $b_{19}$, $r_{27}$, $r_{1}$, $b_{0}$, $b_{3}$, $b_{21}$, $b_{20}$, and $r_{28}$. A further $\sim$10 features are required before $d_{M,95}^2$ reaches a minimum around 4. Notably, the final twenty features actually cause an {\it increase} in $d_{M,95}^2$, along with modest decreases in the other two metrics; after a certain point, additional features make the regressor perform worse.

\begin{deluxetable}{l|l}
\tabletypesize{\scriptsize}
\tablecaption{The first 20 features found via the greedy search method, which we use to derive our final sample of luminosities and temperatures.\label{tab:feature importance}}
\tablehead{\colhead{Category} & \colhead{Features}}
\startdata
Blue Spectral Coefficients & 0, 1, 3, 7, 18, 19, 20,   \\ 
 & 21, 34, 35, 36, 48, $\log |b|$  \\ 
 \hline
Red Spectral Coefficients & 0, 1, 3, 16, 27, 28, $\log |r|$ 
\enddata
\end{deluxetable}

Finally, we can assess which of these features are the most important for the performance on the until-now withheld test set --- i.e., the most important for the generalizability of our regressor on unseen data --- using a ``permutation importance'' strategy. We first train the Bayesian Ridge Regressor on the entire scaled training set using only the first twenty features recovered by the greedy search. We then use the regressor to predict temperatures and luminosities for the scaled test set, and compute the negative of the $d_{M,95}^2$ score, which we define as $s$; the negative is to ensure that higher values of $s$ correspond to better-performing models. We also define $N_r$, the number of iterations of the following procedure used to assess the importance of each feature, which we set to 200. Then, for each feature $j$, on each iteration $r$, we randomly shuffle the corresponding column of the scaled test set, use the corrupted data to predict the temperatures and luminosities, and again compute the negative $d_{M,95}^2$ score, which we define as $s_{j,r}$. The importance of each feature $j$ can then be defined as
\begin{equation}
    i_j = s - \frac{1}{N_r}\sum_{r=1}^{N_r}{s_{j,r}}
\end{equation}
i.e., the average change in the $d_{M,95}^2$ score when a given feature has been shuffled. If the feature has no importance, the regressor won't ``notice'' the corrupted data, each $s_{j,r}$ will be approximately equal to $s$, and $i_j\approx0$. We note that this entire procedure is implemented within {\sc scikit-learn}'s {\tt permutation\_importance} object.

We find that there are six features for which the mean importance $i_j$ minus two times the standard deviation of the importances across all repeats is greater than zero: $r_{1}$  ($i=124.792 \pm 14.159$), $b_{0}$  ($64.116 \pm 5.324$), $b_{1}$  ($53.798 \pm 4.873$), $\log|r|$ ($45.991 \pm 5.502$), $\log|b|$ ($3.271 \pm 0.949$), and $b_{34}$ ($0.245 \pm 0.092$). We also repeated this permutation importance procedure, using the percentage of the residuals within $1\sigma$ as the scorer, and found nine statistically significantly important features:  $r_{1}$  ($0.636 \pm 0.027$), $\log|r|$ ($0.568 \pm 0.031$), $b_{1}$  ($0.315 \pm 0.022$), $b_{0}$  ($0.309 \pm 0.024$), $\log|b|$ ($0.222 \pm 0.024$), $b_{3}$  ($0.146 \pm 0.022$), $r_{0}$  ($0.102 \pm 0.020$), $r_{3}$  ($0.075 \pm 0.019$), and $b_{21}$ ($0.034 \pm 0.014$). Note that the listed importances are now defined differently because of the different choice of scorer. Finally, using the percentage of residuals within $2\sigma$ as the scorer, we again find nine features: $r_{1}$  ($0.585 \pm 0.032$), $\log|r|$ ($0.440 \pm 0.032$), $b_{1}$  ($0.314 \pm 0.023$), $b_{0}$  ($0.254 \pm 0.019$), $\log|b|$ ($0.077 \pm 0.016$), $b_{3}$  ($0.053 \pm 0.014$), $r_{0}$  ($0.034 \pm 0.010$), $b_{19}$ ($0.014 \pm 0.005$), and $b_{20}$ ($0.009 \pm 0.004$). The three lists of features are mostly overlapping, and any one of these lists could represent a ``bare bones'' list of features necessary for the model to be generalizable. Readers looking to adopt their methodology for their own scientific goals should carefully select a scoring metric that is best suited to their work and needs.

Given the the results of the three different feature importance experiments described above, we continue through the rest of the paper to use the Bayesian Ridge Regressor, with only the first 20 features identified using the greedy search process, which yields a mean $d_{M,95}^2$ of $4.05\pm1.06$, and $86.6\pm2.5/96.0\pm0.5$\% of residuals within 1/2 sigma (respectively). We list these features in Table \ref{tab:feature importance}, and pause here to note two important details:
\begin{enumerate}
    \item The spectral coefficients include both low-order features (e.g., $b_0$, $b_1$, $r_0$, and $r_1$ are all listed), and much higher-order features (e.g., $b_{48}$, $r_46$), once again confirming that the \Gaia~spectra are information dense even at the small wavelength scales probed by the higher-order coefficients.
    \item None of the best twenty coefficients use the \Gaia~photometry --- $E(B-V)$ is barely excluded as the 21$^{\rm st}$ feature, while $G_{BP}-G_{RP}$ and $M_G$ are the 45$^{\rm th}$ and 46$^{\rm th}$ features respectively. Our methodology is robust without extinctions, at least for the relatively low levels of extinction in our sample, and does not require broadband photometry.
\end{enumerate}

\section{Temperatures and Luminosities for 5000 LMC Cool Supergiants}\label{sec:results}

We now apply our regressor to an order-of-magnitude larger sample of cool supergiants. This sample is derived from \citet{yang21} (\citetalias{yang21} hereafter), who used {\it Spitzer} and \Gaia~photometry to identify candidate RSGs, YSGs, and BSGs in the LMC. While some of the RSGs have temperatures and luminosities derived in \citet{yang18}, that work only adopted the \citetalias{neugent12_ysg} prescription, and did not include YSGs, so the $\teff$ and $\lum$ measurements we present here are the first published estimates for many of these objects.

We begin by first assembling a clean set of stars with features from the \citetalias{yang21} catalog. We cross-matched \citetalias{yang21} with 2MASS in order to perform a similar cross-match to \Gaia~using 2MASS identifiers (yielding 7670 stars), before identifying likely LMC stars and removing OB stars as above.\footnote{We note that many stars do not have a spectral type listed on SIMBAD, resulting in quite a few likely OB stars whose temperatures fall well-outside the bounds of our training set, which only extends up to $\sim$10 kK. We address this issue below.} This results in a total of 5914 LMC cool supergiants from \citetalias{yang21} that have \Gaia~spectra and pass our astrometric cuts. We then assemble features as described above (i.e., $M_G$, $G_{BP}-G_{RP}$, $E(B-V)$, the normalized spectral coefficients $b_i$ and $r_i$, and the log-norms $\log|b|$ and $\log|r|$). Finally, because there is some overlap between the samples from \citetalias{neugent12_ysg} and \citetalias{yang21}, we remove all stars from \citetalias{yang21} with a \Gaia~DR3 source number matching a source in the \citetalias{neugent12_ysg} sample. This leaves us with 5505 stars with no previous $\lum$ or $\teff$ measurements. Next, we drop all features but the 20 most important listed in Table \ref{tab:feature importance}, scale these features as described above, and use the trained Bayesian RR to predict $\teff$ and $\lum$ for these stars. We highlight that the feature scaler and the Bayesian RR remain trained only on the training set, and that the photometric features $M_G$, $G_{BP}-G_{RP}$, and $E(B-V)$ are {\it not} included in the final list of features.

We now wish to trim the stars the from \citetalias{yang21} that may fall outside of the bounds of our training set. This includes likely OB supergiants without a spectral type listed on SIMBAD. In a low-dimensional feature space, algorithms such as sigma-clipping can be employed to identify and remove outliers. However, identifying outliers in the 20-dimensional feature space we are using here is a non-trivial problem. This is especially the case when the data (even after scaling) are not normally distributed, and may have underlying covariances. While various approaches exist that we experimented with (e.g., the OGK algorithm; \citealt{maronna02}),
we instead choose to identify outliers in the target variables, $\teff$ and $\lum$. This allows us to visually identify stars for which our regressor performs poorly.

We first flag all stars from \citetalias{yang21} that have predicted $\teff$ below 3.452 or above 4.037 (i.e., the coolest and warmest stars in the training set), which we denote as out-of-boundary, OOB stars. We then bin the training sample into 20 equally-sized bins in $\teff$, and identify the lowest luminosity object in each bin, allowing us to form a minimum-luminosity boundary in the HR diagram; this boundary occurs at $\lum\approx4.5$, and dips to slightly lower luminosities for the RSGs. We separately flag stars that aren't OOB, but whose predicted $\lum$ falls below the minimum-luminosity boundary that we derive (which we denote as low-$L$ stars). Finally, we note that none of our predicted luminosities for the \citetalias{yang21} stars lie above the highest luminosity stars from \citetalias{neugent12_ysg}. 


\begin{figure}[t!]
\plotone{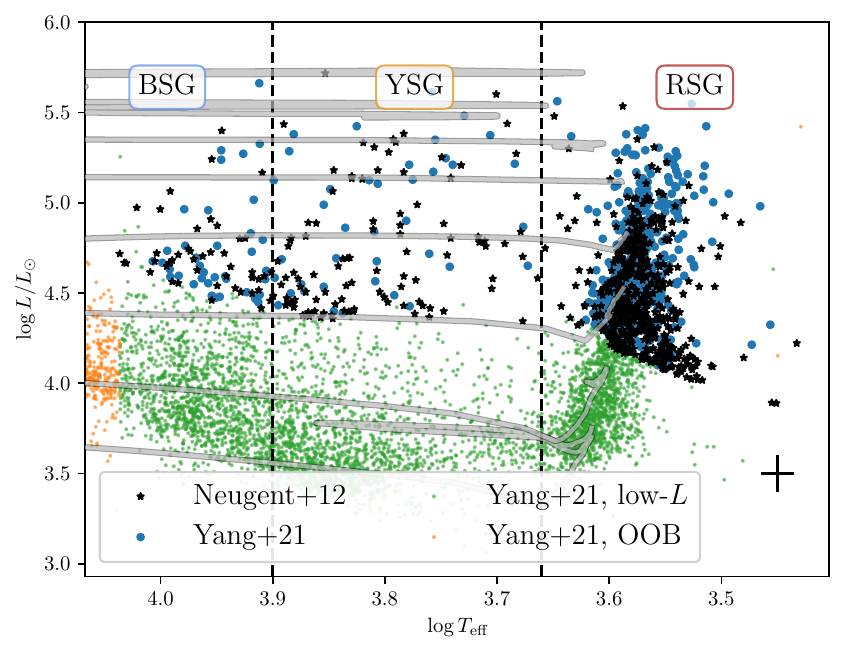}
\caption{Hertzprung-Russell diagram for the combined \citetalias{neugent12_ysg} (black points) and \citetalias{yang21} catalog (all others). Orange points fall outside of the temperature range of the training sample (OOB stars). Blue/green points are the stars whose luminosities are brighter/fainter than the minimum-luminosity training star in a narrow $\teff$ bin. The black cross in the lower right indicates the typical uncertainties in the training set. We use vertical black lines to divide the HR diagram into RSGs, YSGs, and BSGs based on effective temperature. Grey lines show non-rotating, $Z=0.006$ evolutionary tracks for 7, 9, 12, 15, 20, 25, 32, and 40 $M_\odot$ stars from \citet{eggenberger21} for comparison.}\label{fig:hrd}
\end{figure}

Figure \ref{fig:hrd} shows the resulting HR diagram. When comparing with predictions from evolutionary models, it is useful to roughly distinguish between RSGs, YSGs, and BSGs. Therefore, we use the boundaries from \citet{dornwallenstein18,dornwallenstein20} (which are based on evolutionary tracks from \citealt{ekstrom12} and \citealt{eldridge17}\footnote{Note that in general, effective temperature scales are metallicity-dependent \citep[e.g.][]{levesque05}.}) to define RSGs (KM supergiants) as all stars with $\teff<3.66$, YSGs (FG supergiants) as stars with $3.66\leq\teff<3.9$, and BSGs (here mostly A supergiants) as stars with $\teff\geq3.9$. We show the dividing lines between assigned classes as vertical dashed black lines, and denote each class with a suggestively-colored text box. Non-rotating evolutionary tracks at LMC metallicity ($Z=0.006$) from \citet{eggenberger21} are shown in grey for stars with 7, 9, 12, 15, 20, 25, 32, and 40 $M_\odot$ in initial mass.

\begin{figure}[t!]
\plotone{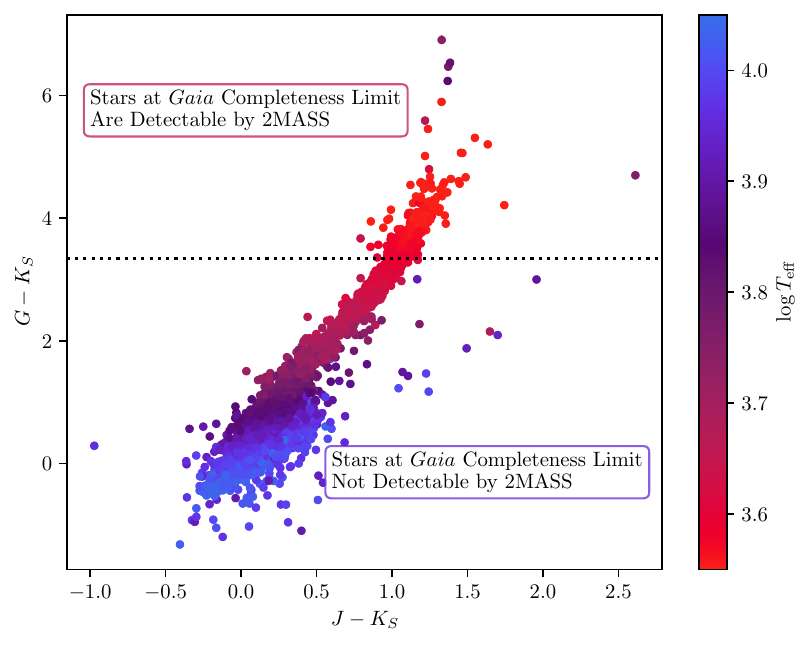}
\caption{$G-K_S$ versus $J-K_S$ color-color diagram for the non-OOB \citetalias{yang21} stars, colored by our newly-derived $\teff$ values. The horizontal black line shows the $G-K_S$ color of a star that is at the completeness limit for both \Gaia~low resolution spectroscopy and 2MASS $K_S$. A star at the \Gaia~completeness limit but bluer than this line will not be detected by 2MASS; i.e., they are just bright enough in the optical for \Gaia~spectroscopy, but not enough in the infrared to derive temperatures via the $J-K_S$ method. Many massive stars in the Local Group satisfy this criterion.}\label{fig:gaia_2mass_completeness}
\end{figure}

All stars from both the training and test set are indicated as black star-shaped points. The black error bar in the lower right shows the typical uncertainties in $\lum$ and $\teff$ reported by \citetalias{neugent12_ysg}; given the results of our testing, we believe these uncertainties to be applicable to the newly-predicted values as well. Blue points indicate the predicted values for stars from \citetalias{yang21} that fall within the bounds of the training set. OOB stars are shown as small orange points, and low-$L$ stars are shown as small green points. Examining the low-$L$ stars, we see that their predicted locations in the HR diagram fall roughly where we would expect; the sequence of RSGs seen in both the training data and the blue points extends into the low-$L$ regime, and the distribution of YSGs appears to be consistent between the two populations (a statement that we will examine in detail below). This is unsurprising, as the spectrum of a star (and thus its coefficients) is most-dependent upon its effective temperature, with surface gravity/luminosity mostly impacting the strengths of gravity-sensitive lines. All told, not counting the OOB stars, our derived sample contains 5,721 unique stars with $\teff$ and $\lum$ measurements.

\begin{figure*}[t!]
\plottwo{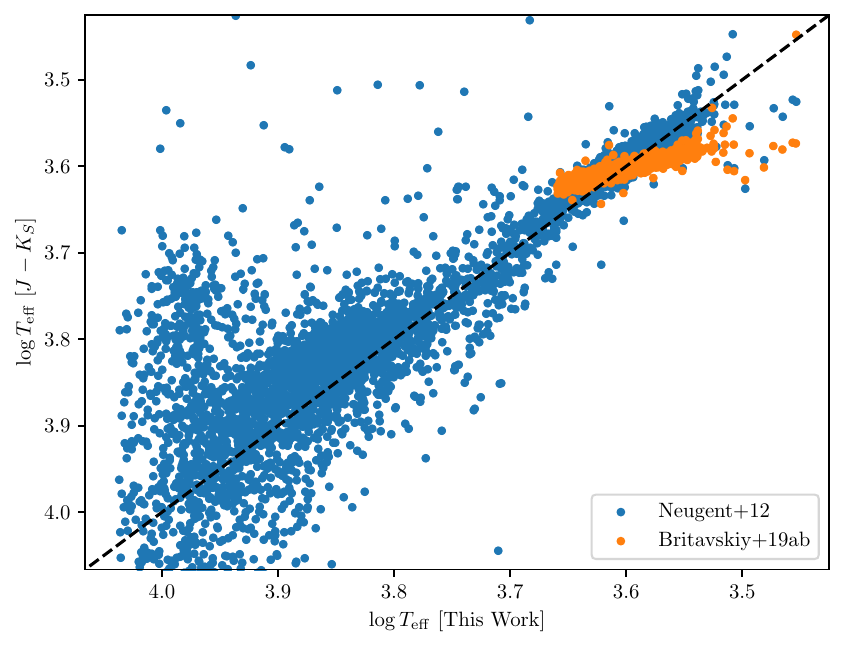}{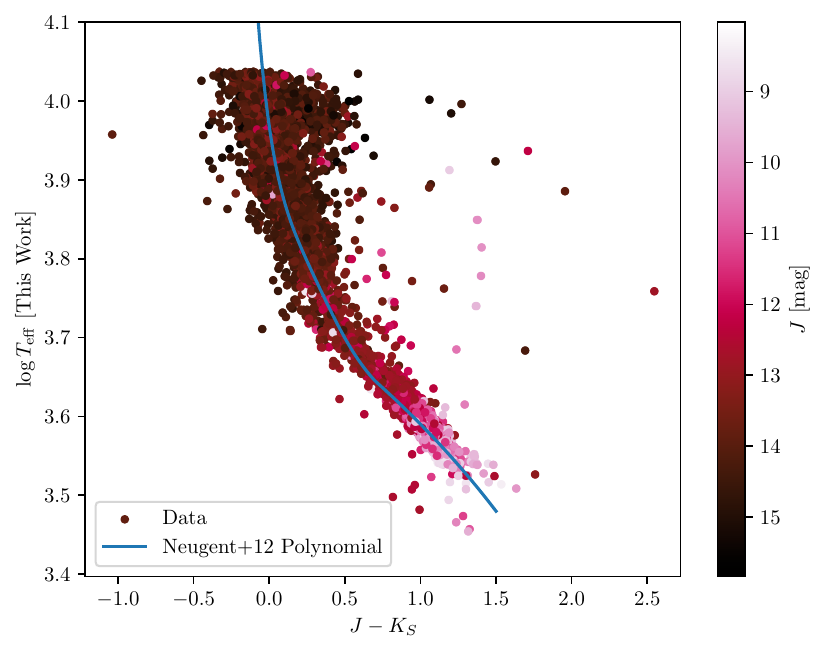}
\caption{({\it Left}) Comparison between $\teff$ values for \citetalias{yang21} stars derived in this work and $\teff$ derived using $J-K_S$ photometry using the relations from \citetalias{neugent12_ysg} (blue points) and \citet{britavskiy19b} (orange points, RSGs only). The one-to-one line is shown in black. The $J-K_S$ temperatures show significantly more scatter around $\teff\approx3.9-4.0$. ({\it Right}) $\teff$ values derived in this work as a function of $J-K_S$. Points are colored by 2MASS $J$-band magnitude. The sixth-order $J-K_S$-$\teff$ relation from \citetalias{neugent12_ysg} is shown in blue. The scatter in the left panel at warmer temperatures is naturally explained between the steepness of the relation at higher effective temperatures and the overall faintness of these warmer stars in the near-infrared, resulting in noisier $J-K_S$ measurements.}\label{fig:tjk_comp}
\end{figure*}

\begin{figure}[ht!]
\plotone{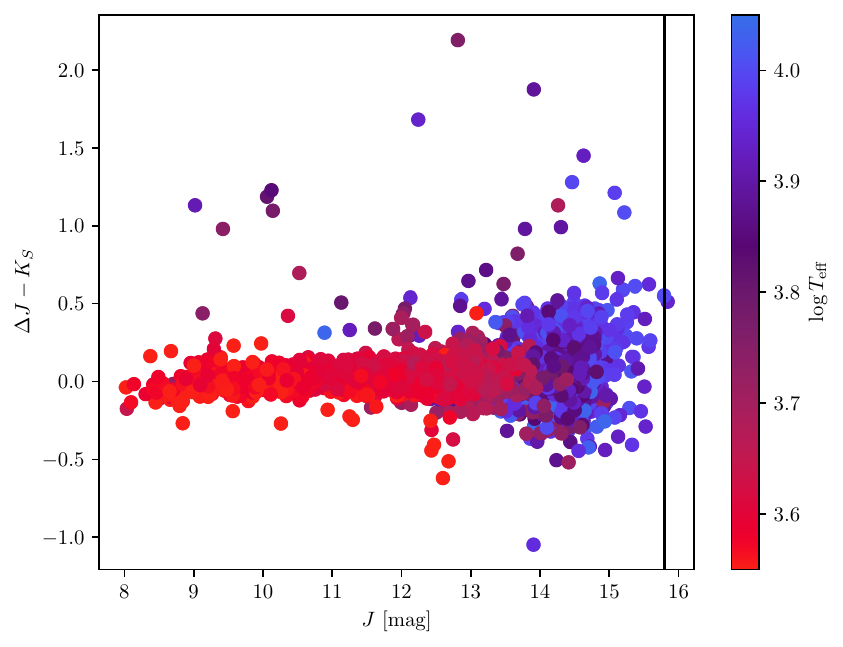}
\caption{The difference between the observed and corrected $J-K_S$ color and the value of $J-K_S$ implied using our machine-learned $\teff$ values and inverting the \citetalias{neugent12_ysg} polynomial, $\Delta J-K_S$, as a function of the observed $J$-band magnitude, colored by our derived $\teff$ values.}\label{fig:delta_jk}
\end{figure}

As we mentioned in \S\ref{subsec:training}, in the intervening years, \citet{neugent20b} published an updated catalog of $\sim$4000 LMC RSGs that includes 369 stars from our training set, and 1867 stars from our final derived sample. Because \citet{neugent20b} discretized their RSG temperatures to a 50K grid (unlike in \citetalias{neugent12_ysg}), we did not use these updated temperatures for the stars in our training data out of a desire to have homogeneous data quality and precision across the portion of the HR diagram we were interested in, as well as avoiding overrepresentation of RSGs in the training data. However, examining our temperature and luminosity predictions for the stars in both \citetalias{yang21} and \citet{neugent20b} but not in \citet{neugent12_ysg}, our derived values are in good agreement with the values reported by \citet{neugent20b}, including the low-$L$ stars. Moreover, our sample includes 3624 stars that are in neither \citetalias{neugent12_ysg} nor \citet{neugent20b}, of which 497 are RSGs. The remaining 3127 stars in our sample are YSGs and BSGs with $\teff>3.66$, compared to 281 such stars from \citetalias{neugent12_ysg}. This represents an order of magnitude increase in the number of LMC AFG supergiants with known positions in the HR diagram compared to previous work.\footnote{Of course, the intent of \citetalias{neugent12_ysg} was not to derive a complete sample of YSGs in the LMC. Regardless, studies of massive stars have long been hindered by poor sample sizes.}

\subsection{Do We Need Machine Learning Anyway?}

While this result is impressive, whenever machine learning is invoked, it is critical to determine whether it is necessary. The temperature and luminosity measurements from \citetalias{neugent12_ysg} were directly determined from 2MASS $J$ and $K_S$ band photometry, whereas the measurements we report here rely on a high-dimensional feature space derived from spectroscopy. Could we have made these measurements with 2MASS photometry alone and written a significantly shorter paper?\footnote{The astute reader will notice that this did not happen.} 

One issue that arises is that the \Gaia~spectroscopy and 2MASS photometry have different completeness limits: the 2MASS magnitude limit in the $K_S$ band is 14.3 \citep{cutri03}, while \Gaia~reports spectroscopy for all stars brighter than $G=17.65$ \citep{deangeli22}. Figure \ref{fig:gaia_2mass_completeness} shows a two-color diagram of $J-K_S$ (i.e., the color used by \citetalias{neugent12_ysg} to predict effective temperature.) vs. $G-K_S$.\footnote{Both \citetalias{neugent12_ysg} and \citetalias{yang21} provide 2MASS identifiers for their samples, allowing us to access NIR photometry for our combined sample.} Points are the observed values for all non-OOB stars from \citetalias{yang21}, colored by our derived effective temperature as shown by the colorbar at right. The horizontal black dashed line show where $G-K_S=3.35$, i.e., the color of a star at the completeness limits for both \Gaia~spectroscopy and 2MASS. Any star that is at the \Gaia~completeness limit but is bluer than $G-K_S=3.35$ --- effectively, all but the coolest RSGs --- will not be detected by 2MASS. We note that this is only for sources with low reddening like the ones in our sample.

Of course, massive stars in the Magellanic Clouds are bright in both the near-infrared and optical, and such completeness concerns do not apply. However, a number of evolved massive stars in M31 which have already been observed by \Gaia~do in fact fall below the 2MASS completeness limit, to say nothing of the crowding and highly-spatially variable extinction \citep{dalcanton15} that can further hinder infrared observations (e.g. \citealt{sick14,massey21b}). A simple query for all \Gaia~DR3 sources with low-resolution spectroscopy and parallax less than 0.1 mas that were included in the \Gaia~Andromeda Photometric Survey (GAPS; \citealt{evans22}) yields 9,949 objects, 471 of which are within one degree of the center of Andromeda, the overwhelming majority of which (445 stars) are brighter than $G=17.65$. The method that we present is therefore a viable way of estimating temperatures and luminosities for the brightest supergiants in the Local Group without spectroscopy, provided a suitable training set can be constructed (e.g., \citealt{dalcanton12}). Finally, as we allude to above, the 2MASS spatial resolution is a factor of a few larger than the \Gaia~resolution \citep{cutri03,gaia16}, making \Gaia~the more reliable tool to use in crowded regions like the Galactic plane where many massive stars can be found.

Even for the massive stars in uncrowded fields within our Galaxy or its satellites which are likely to be brighter than the 2MASS completeness limits, how do our measurements compare with those derived from 2MASS photometry? To answer this question, we follow \citetalias{neugent12_ysg}, and convert observed 2MASS colors for the \citetalias{yang21} to the \citet{bessell88} system, which we then deredden assuming $E(J-K) = 0.535 \times E(B-V)$ \citep{schlegel98}. We then derive effective temperatures using the piecewise function presented by \citetalias{neugent12_ysg}. For the RSGs specifically, \citet{britavskiy19a,britavskiy19b} also present an alternative function to convert between $J-K_S$ and $T_{\rm eff}$. We plot both temperature estimates against the temperatures derived in this work (after removing OOB stars) in the left panel of Figure \ref{fig:tjk_comp}. Points in blue/orange show the comparison with $J-K_S$ temperatures derived following \citetalias{neugent12_ysg}/\citeauthor{britavskiy19b}, respectively; for the latter points, we only plot temperatures for RSGs (i.e., stars with \Gaia-derived log-temperatures below 3.66). A one-to-one line is shown as a dashed black line. 

The \citetalias{neugent12_ysg} prescription follows the one-to-one line closely, especially for the cool stars, which is unsurprising given that the effective temperatures for our training set are {\it from} \citetalias{neugent12_ysg}. However, we see significant scatter for warmer stars. The reasons for this scatter are two-fold: first, the function from \citetalias{neugent12_ysg} is a sixth-order polynomial in $J-K_S$ that becomes incredibly steep bluer than $J-K_S\lesssim0.5$. This is seen in the right panel of Figure \ref{fig:tjk_comp}, which shows our effective temperature values as a function of $J-K_S$. The sixth-order polynomial relation from \citetalias{neugent12_ysg} is shown in blue.\footnote{We pause here to point out a secondary issue, which is that the polynomial has some quite large coefficients (e.g., the fourth-order coefficient is 22.098), which is likely an overfit to the data.} The points are colored by 2MASS $J$ magnitude, which demonstrates the second reason for the scatter observed in the warmer stars in the left panel: these stars are fainter overall in the near-infrared (some even approaching the 2MASS $J$-band completeness limit of 15.8). These two effects combine such that warmer stars have noisier 2MASS photometry, and that noise is significantly more impactful in the steeper part of the relation from \citetalias{neugent12_ysg}. 

Another way of visualizing this is presented in Figure \ref{fig:delta_jk}. We numerically invert the relation from \citetalias{neugent12_ysg}, and compute the difference between the observed color and the color implied by our newly-derived $\teff$ values, $\Delta J-K_S$. We plot this value against the observed $J$-band magnitude, and color the points by our inferred $\teff$. While the cooler, brighter stars have $\Delta J-K_S\approx0$, the warmer, fainter stars have an observed $J-K_S$ that is discrepant from what it ``should'' be given our derived temperatures.

Ultimately, these results demonstrate a key benefit of using a large number of features and the machinery developed above: we mitigate the effect of measurement uncertainty in a single color ($J-K_S$) on the derived temperatures for the bluer stars in our sample. However, it is critical to recall that the temperatures of stars in our training set are still derived from $J-K_S$. Therefore, while our derived temperatures for any individual star are less sensitive to measurement uncertainty than they would be by using $J-K_S$, any bias or error in the underlying relation from \citetalias{neugent12_ysg} is still ``baked in'' when fitting the Bayesian RR to the training data.

\begin{figure*}[t!]
\plotone{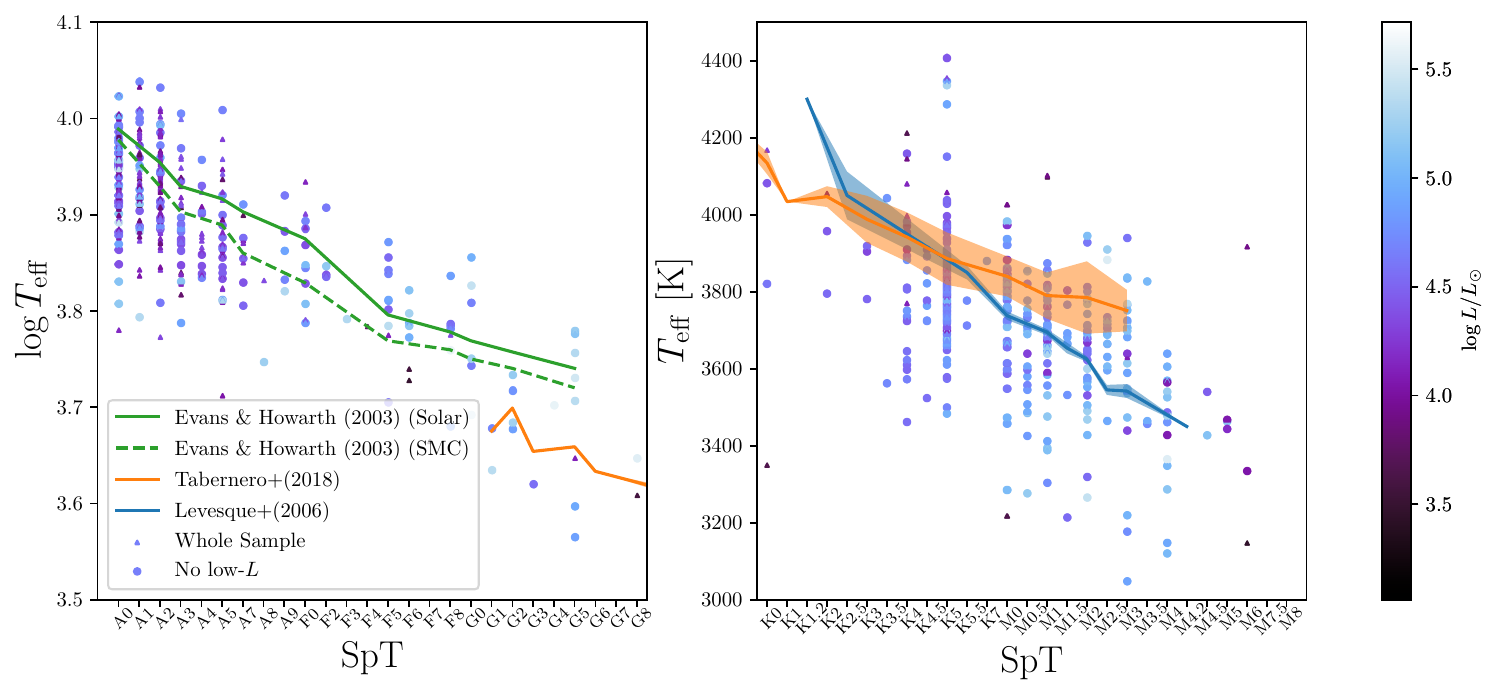}
\caption{Effective temperature scale for the LMC cool supergiants in our sample; temperatures for stars later than K0 are plotted in linear scaling in the right panel. Overplotted are effective temperature scales from the literature: the temperature scales for solar (solid line) and SMC (dashed line) AFG supergiants from \citet{evans03} are shown in green, the scale for LMC GKM supergiants from \citet{tabernero18} is shown in orange, and the LMC RSG effective temperature scale from \citet{levesque06} is shown in blue.}\label{fig:teff_scale}
\end{figure*}

\begin{figure*}[ht!]
\plottwo{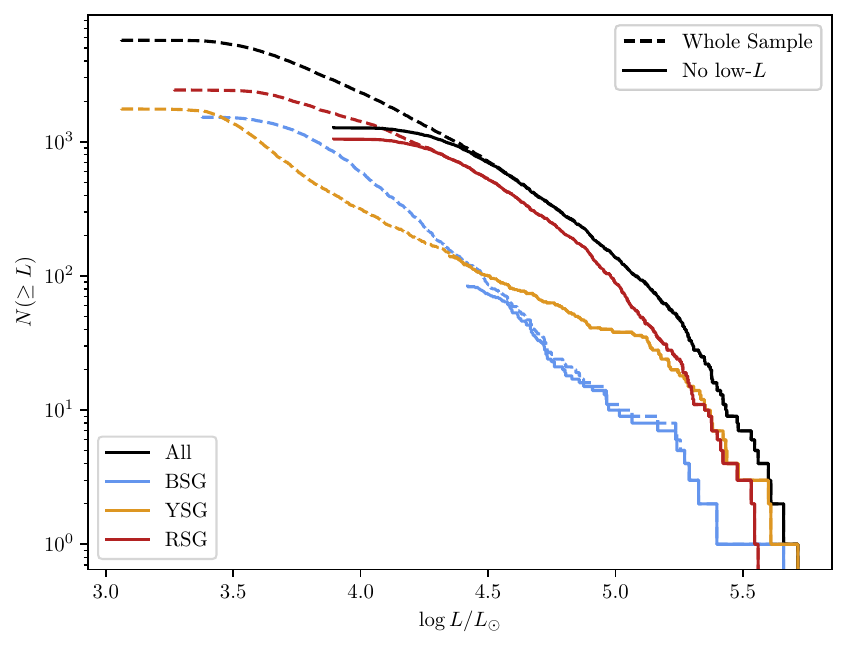}{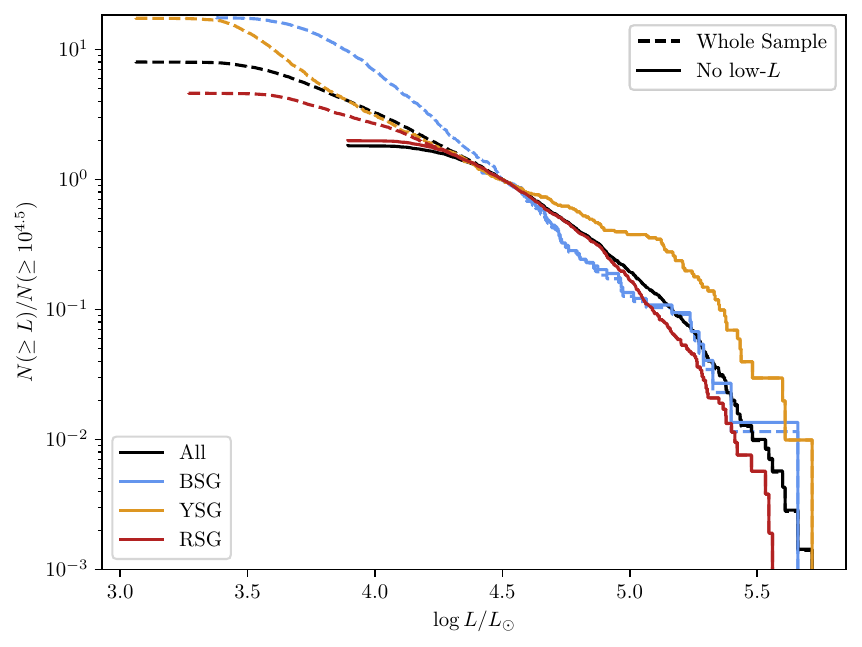}
\caption{({\it Left}) Luminosity functions for all non-OOB stars in our sample, expressed as the number of stars with luminosity greater than or equal to $L$, $N(\geq L)$. Dashed/solid lines include/exclude the low-$L$ stars, respectively. Black lines are for stars at all temperatures, while the luminosity functions for RSGs, YSGs, and BSGs are correspondingly colored. ({\it Right}) Same, but normalized to unity at $\lum=4.5$. In both panels, the slopes of the luminosity functions vary, with the YSG luminosity function becoming notably flatter above $\lum\approx5$.}\label{fig:lum_func}
\end{figure*}

\begin{figure}[ht!]
\plotone{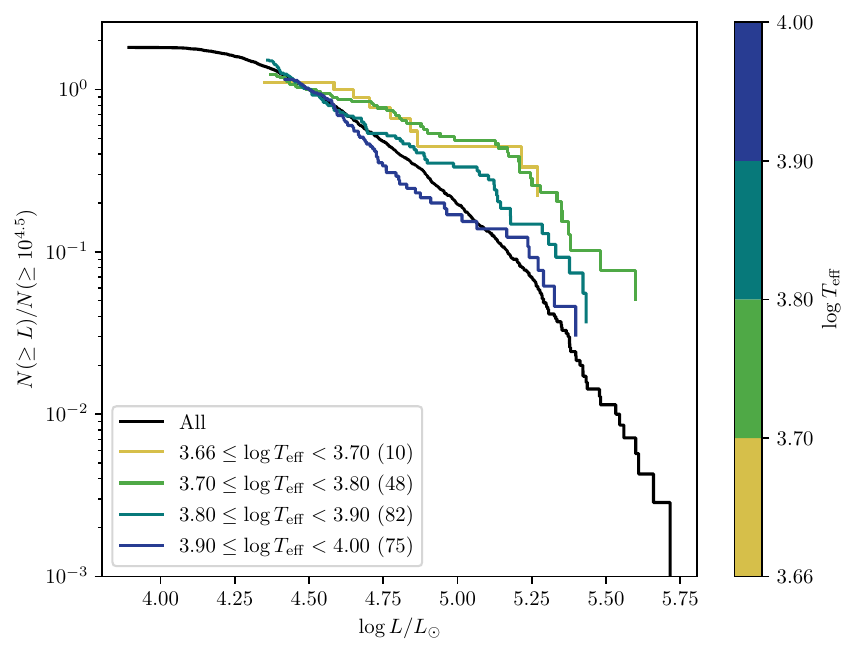}
\caption{A breakdown of the luminosity function of YSGs and BSGs, demonstrating the change in slope as a function of $\teff$ (normalized at $\lum=4.5$). This sample excludes OOB and low-$L$ stars. The black line is for all stars for references, while the colored lines are for the stars in temperature bins indicated in the legend (with the number of stars in each bin in parentheses), with the temperature of the bin increasing from yellow to blue}\label{fig:lum_func_tbins}
\end{figure}

\subsection{Comparison with Spectroscopically-Derived Effective Temperature Scales}

Both the method that we present here, as well as the sheer number of stars with newly-reported $\teff$ and $\lum$ measurements, are potentially powerful probes of massive star evolution. However, we wish to know how reliable our reported values are. As a first test, we can compare our effective temperature estimates with the spectral types of the stars in our sample. Using literature spectral types from SIMBAD where available, we bin stars by spectral type; for objects with a variable or uncertain spectral type, we simply take the first listed type.\footnote{We note that, in general, the quality of spectral types available for a significant number of massive stars is sometimes quite poor \citep[e.g.][]{dornwallenstein21}, which is a dominant source of uncertainty here.} Figure \ref{fig:teff_scale} shows the effective temperatures of stars in our sample as a function of their spectral types. Stars from the whole sample are plotted as small triangles, while circles show only stars that aren't identified as low-$L$ objects. Each point is color-coded by its $\lum$. The left panel shows $\teff$ for the AFG stars in our sample, while the right panel shows linear $T_{\rm eff}$ for the KM stars. 

Unsurprisingly given the difference in precision between our temperatures and those derived by fitting model atmospheres to flux-calibrated spectra, the scatter within each spectral type is large. However, to within the errors, the effective temperature scale is monotonically decreasing: as we would hope, later spectral types have lower effective temperatures. While this is hardly surprising, it is an excellent sanity check on our work. We can also compare with previous studies of the effective temperature scales of cool supergiants in the LMC: the blue line in the right panel shows the RSG effective temperature scale from \citet{levesque06}, while the orange line in both panels shows the scale from \citet{tabernero18}; both scales are derived using spectroscopic methods. For both temperature scales, we also show the listed uncertainties for spectral types where those authors' samples contained more than one star. These uncertainties are too small to be seen in log-scale (the left panel). Additionally, while not for the LMC, we also show the AFG supergiant effective temperature scales for solar/SMC AFG supergiants from \citet{evans03} in solid/dashed green lines in the left panel.\footnote{Unfortunately, \citet{evans10} did not quote uncertainties for their temperature scales.} Note that the effective temperature scale from \citeauthor{levesque06} occasionally contains temperatures for multiple spectral types (e.g., 4300 K for K1-K1.5 I); for plotting purposes, we assign these values to an intermediate type (in this example, K1.25, which is not a real spectral type). Our effective temperatures broadly agree with the effective temperature scales from all three works. For spectral types later than $\sim$K5, our derived temperatures are somewhat cooler than the \citeauthor{tabernero18} scale, while agreeing better with the \citeauthor{levesque06} scale. This is likely due to the different choices of atmosphere models between the two works; \citeauthor{levesque06} used MARCS models, while \citeauthor{tabernero18} used ATLAS-APOGEE (KURUCZ) models \citep{meszaros12}. For the earlier supergiants, our results agree well with the \citeauthor{evans06} temperature scale, which predicts overall warmer temperatures for the G supergiants compared to \citeauthor{tabernero18}; in this case, both works used ATLAS atmosphere models, though \citeauthor{evans06} used the older ATLAS9 models \citep{kurucz92}.

Of course, the relative merits of these different sets of atmosphere models and the radiative transfer codes used to produce synthetic spectra are not the focus of this work. Rather, it is reassuring that our temperatures --- which used a training set with photometrically-derived temperatures --- compare favorably to effective temperature scales derived by spectroscopy. We note that the scatter of our derived temperatures in each spectral type bin is $\sim$2-3 times larger than the quoted uncertainty of 0.015 dex. We speculate that this could be due to the fact that we are plotting our temperatures against literature spectral types, many of which date back to objective prism surveys of the LMC from the 1960s and 70s \citep[e.g.][]{feast60,sanduleak70}. This is in contrast to the spectroscopic works against which we are comparing, which use spectral types derived simultaneously with the effective temperatures. Some cool supergiants are known to be spectroscopically variable on year-to-decade timescales by as much as a full spectral type \citep{massey07c,levesque07b,wasatonic15}. This is likely artificially inflating the scatter seen in Figure \ref{fig:teff_scale}.

Finally, we note the presence in our sample of some quite late M stars, 7 of which are M5 or later. This is in contrast with the effective temperature scale of e.g. \citet{levesque06}, which goes as late as M4.5 in the LMC (corresponding to 3450 K in their scale). For most of these late M stars, the derived luminosities are generally low enough that they are almost certainly AGBs. AGBs reach cooler effective temperatures than RSGs, and can contaminate populations of RSGs below $\lum=4.5$, see e.g. \citet{boyer11,massey21}. However, two objects have quite high derived luminosities: HV 2602 ($\lum=5.15$, SpT M5Ia; \citealt{dorda18}), and HV 11984 ($\lum=5.12$, SpT M8; \citealt{westerlund60}). Both stars are from \citetalias{yang21}, and as such only have temperatures derived in the optical from \Gaia~spectra. Of these two, HV 2602's spectral type is only half of a sub-type later than published LMC RSG effective temperature scales, is a bona fide supergiant with luminosity class Ia, and has a derived temperature is appropriately low for a star of its spectral type. However, HV 11984 is a known Mira variable, making it unlikely that our regressor performed well on this object.

\subsection{The Hayashi Limit}

As a final check on our effective temperatures, we can perform an easy test on their physical validity. Almost any linear model has infinite domain and range, the exception being a model that predicts a constant value. However, the effective temperatures of stars {\it do} have a limited range, from a few $10^5$ K \citep[WO stars, e.g.][]{aadland22} at the hot end, down to the Hayashi limit at the cool end \citep[i.e., the coolest temperature a star can reach while being in hydrostatic equillibrium;][]{hayashi61}. While the limit is dependent on metallicity, theoretical \citep{eggenberger21} and observational \citep{levesque06} work place it around $\teff\approx3.55$ ($\sim$3450 K) in the LMC. Having imposed no physics-based prior on our derived temperatures, are the coolest stars in our sample consistent with the Hayashi limit? While observational noise scatters some of the RSGs in Figure \ref{fig:hrd} well-below the Hayashi limit, it is clear the the bulk of the RSG branch occurs roughly where it should, especially in comparison to the \citet{eggenberger21} evolutionary tracks. To be more quantitative, we borrow from the methodologies used to detect the tip of the red giant branch (TRGB), and use an edge detection procedure similar to what is described in \S3.2 of \citet{durbin20}. 

We first compute a histogram of $\teff$ between 3.5 and 3.65, with a bin size of 0.003 dex (i.e., five times smaller than the typical 0.015 dex uncertainty). We then smooth this histogram using a Savitsky-Golay filter \citep{savitzky64} implemented in {\sc SciPy} \citep{scipy:2001,scipy:2020}, with a window size of 15, and polynomial order 3. To identify the edge of the temperature distibution, we convolve the resulting smoothed histogram with a Sobel kernel ({\tt scipy.ndimage.sobel}). We then take the effective temperature at the maximum of the resulting edge response function is the detected Hayashi limit, which we find to be $\teff=3.56$ (3650 K): exactly where it should be. While we caution against adopting this value over more precise values obtained from spectroscopic studies focused exclusively on RSGs, it is nevertheless reassuring to recover known physics from our derived HR diagram. 



\section{Discussion}\label{sec:discussion}

\subsection{The Distribution of AFGKM Supergiants in the HR Diagram}

Having derived robust temperatures and luminosities for $\sim$5,000 stars, we can now turn to studying their distribution in the HR diagram in detail. In particular, we now turn out attention to the luminosity function, i.e., $N(\geq L)$, the number of stars in the sample with luminosity greater than or equal to $L$. Massive stars in this part of the HR diagram evolve at approximately constant luminosity until they become RSGs, at which point they evolve vertically in the HR diagram. As a result, the slope of the luminosity function encodes the initial mass function (IMF), recent star formation history,\footnote{The massive stars in our sample are $<50$ Myr old \citep{eggenberger21}. On this timescale, the bulk properties of massive star populations are unaffected by variations in SFH of a few tens of percent \citep{massey21}, at least outside of regions of active star formation such as 30 Dor. Therefore, for the following discussion, we ignore the impact of star formation history.} the relationship between initial mass and the post-main sequence luminosity, and the mass-dependence of the lifetimes of various post-main sequence evolutionary stages. Assuming that there are no physical effects that might lead to drastically different evolutionary behaviors at a given luminosity (and neglecting recent star formation history) these quantities take simple functional forms like power laws, the slopes of which inform the slope of the observed luminosity function. Of course, this assumption is incorrect, which makes the luminosity function a fantastic probe of massive star evolution.

We plot the empirical luminosity function in the left panel of Figure \ref{fig:lum_func}. The solid/dashed line shows the luminosity function excluding/including the low-$L$ stars. Black lines show the luminosity function for all stars, whereas red/yellow/blue lines show the luminosity function for RSGs/YSGs/BSGs respectively as delineated by the boundaries shown in Figure \ref{fig:hrd}. From this comparison alone, we can see that the slopes of the luminosity functions for the three subpopulations differ markedly for luminosities above $\lum=5$. In particular, the RSG luminosity function steepens relative to the whole population while the YSG luminosity function becomes flatter. The change in slope becomes even easier to see in the right-hand panel of Figure \ref{fig:lum_func}, where we have normalized all luminosity functions to unity at $\lum=4.5$. We again see that the RSG luminosity function steepens while the YSG luminosity function flattens above $\lum=5$. 

Interestingly, the BSG luminosity function roughly follows the behavior of the overall luminosity function, except in the luminosity range $4.5\leq\lum\leq5$, we see a brief steepening followed by a flattening, along with an additional flattening in the YSG luminosity function.\footnote{Note that, while we can expect our sample to be reasonably complete for the YSGs and RSGs, our sample deliberately excludes OB stars, and so, when talking about BSGs, we are really referring only to A supergiants. To make this explicit, we enforce an upper temperature cutoff at $\teff=4$ for this discussion.} To examine this behavior further, we further subdivide the sample into four bins of $\teff$: one from $3.66\leq\teff<3.7$, and then three 0.1 dex wide bins from $\teff=3.7$ to 4.0. In each bin, we discard the low-$L$ and OOB objects, and compute the luminosity function, which we show (normalized at $\lum=4.5$) in Figure \ref{fig:lum_func_tbins}, again compared to the luminosity function of the entire temperature range. For context, we note the number of stars in each temperature bin in the legend. We can see that as we move from cooler (yellow line) to warmer (blue line) stars, the flattening of the YSG luminosity function above $\lum=5$ becomes weaker and weaker. Meanwhile, the decrease in the BSG luminosity function around $\lum=4.5$ is not present in the next-coolest temperature bin. 

\begin{figure*}[ht!]
\plotone{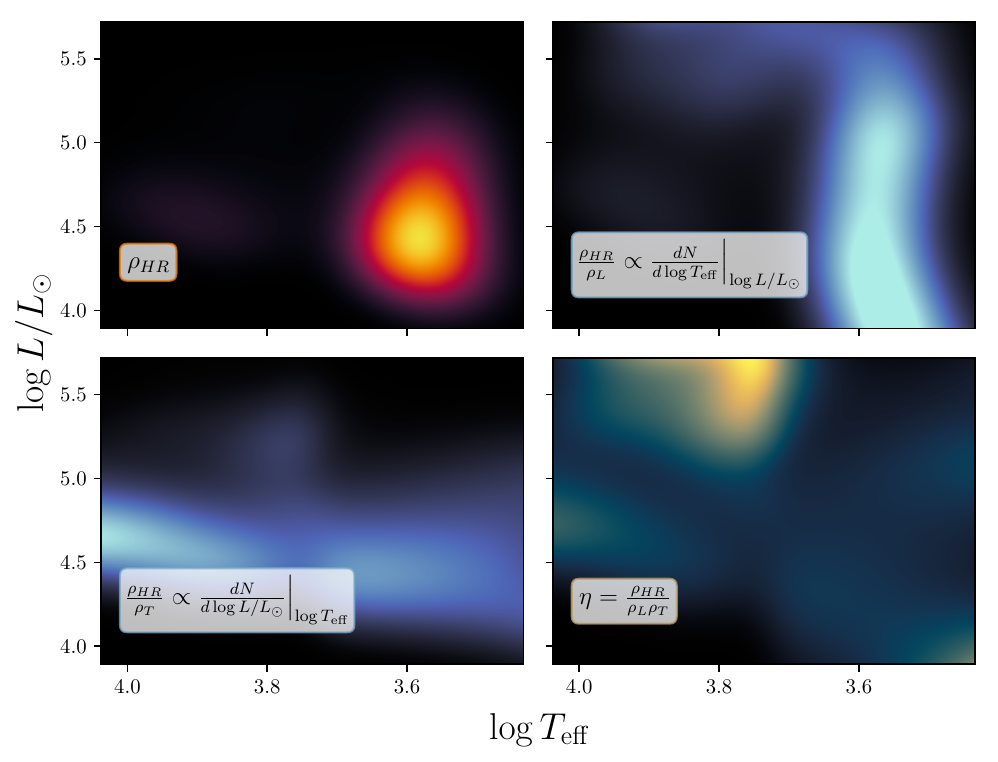}
\caption{({\it Top left}) KDE of the HR diagram, $\rho_{HR}$. Low-luminosity RSGs are the most common stars in the sample. ({\it Bottom left}) $\rho_{HR}$ normalized by the KDE of $\teff$, $\rho_T$; equivalent to the differential luminosity function as a function of temperature. Lower-luminosity stars are more common, but the luminosity function flattens at $\teff\approx3.8$ ({\it (Top right}) $\rho_{HR}$ normalized by the KDE of $\lum$, $\rho_L$; brighter regions are temperatures with longer lifetimes. The RSG branch can clearly be seen, but the YSG and BSG lifetimes relative to the RSG phase increase at high luminosity. ({\it Bottom right}): $\rho_{HR}$ normalized by $\rho_L\rho_T$. Brighter regions show overdensities in the HR diagram, and dark regions show underdensities.}\label{fig:kde_breakdown}
\end{figure*}

\begin{figure*}[ht!]
\plotone{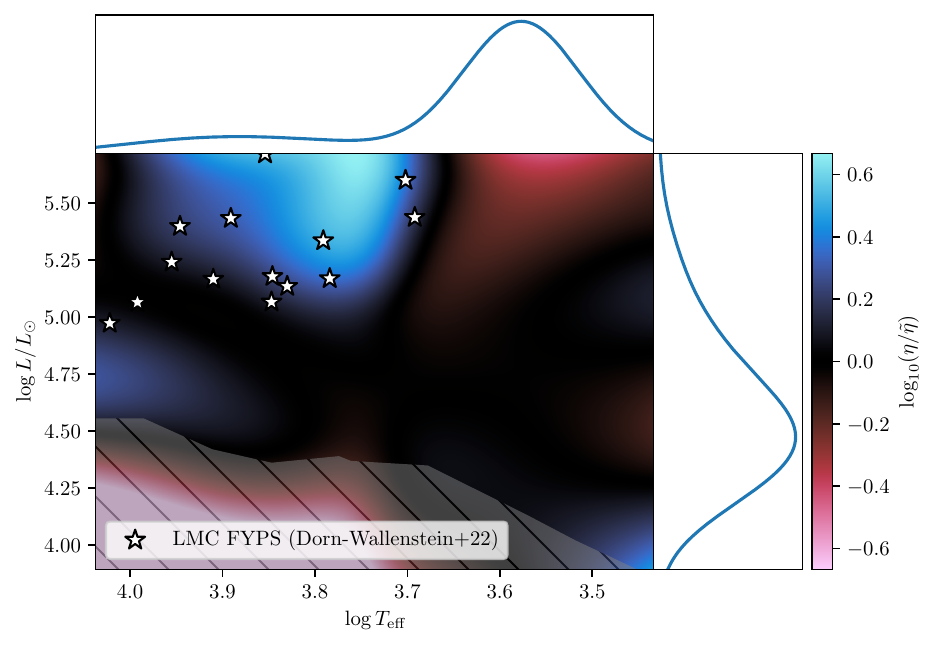}
\caption{Similar to the bottom right panel of Figure \ref{fig:kde_breakdown}, after subtracting and dividing by the median value. The one-dimensional KDEs of $\teff$ and $\lum$ are shown above/to the right of the plot respectively. Regions with colors close to black contain roughly as many stars as one would expect by simply drawing values from each one-dimensional KDE, while regions in red/blue have fewer/more stars respectively. We postulate that the region in blue contains post-RSGs; the white points are the positions in the HR diagram of the recently-discovered fast yellow pulsating supergiants (FYPS), which are putative post-RSGs \citep{dornwallenstein22}. The cross-hatched region shows the minimum-luminosity boundary of the sample used to compute the KDE.}\label{fig:hr_kde}
\end{figure*}

The observed variation in the luminosity function slope as a function of temperature encodes the variation in the relative lifetimes of each phase as a function of mass. As Figure \ref{fig:lum_func_tbins} illustrates, this variation occurs over relatively narrow bins of effective temperature. We now wish to examine this behavior in greater detail. We first use a {\tt StandardScaler} to scale the $\teff$ and $\lum$ measurements for our sample after removing the low-$L$ and OOB objects. We then compute a kernel density estimate (KDE) of the scaled values. The KDE approximates the density of stars in the HR diagram, which we denote $\rho_{HR}$, by replacing each point with a two-dimensional Gaussian, before summing these Gaussians together. We do this using the {\tt KernelDensity} estimator from {\sc scikit-learn}. Because the bandwidth of the Gaussian kernel is a free parameter and can't be set independently for each axis of the HR diagram, the scaling ensures that this bandwidth captures the variance of both $\teff$ and $\lum$. We set the bandwidth to 0.5 (i.e., half the standard deviation of the scaled variables), and compute the KDE on a grid of 1000x1000 points that span the maximum/minimum scaled values, which we then transform back into the observed space of $\teff$ and $\lum$. We plot the resulting estimate of $\rho_{HR}$ in the top left corner of Figure \ref{fig:kde_breakdown}. The figure is dominated by the lowest-luminosity RSGs in the sample. This is because this region of the HR diagram contains only the Hertzprung gap and the RSG branch --- not the main sequence --- making these low-luminosity RSGs the most common objects.  

To find regions that deviate from this overall pattern --- analogous to variations in the slopes of the luminosity function --- we also compute the one-dimensional KDE of both $\teff$ and $\lum$, which we denote $\rho_T$ and $\rho_L$, respectively. The bottom left panel of Figure \ref{fig:kde_breakdown} shows the result of dividing each row in the grid of $\rho_{HR}$ values by $\rho_T$. Each column of the resulting grid of values is thus proportional to an estimate of the differential luminosity function, $dN/d\lum$ evaluated at constant $\teff$. While broadly this panel of the figure is dominated by the lowest-luminosity stars (as we expect from Figure \ref{fig:lum_func}), we can now star to see a flattening of the luminosity function above $\lum=5$ around $\teff=3.8$. 

Similarly, we can also compute the ratio of $\rho_{HR}$ to $\rho_L$ in each column of the grid, which we plot in the upper right panel of Figure \ref{fig:kde_breakdown}. Now each row in the resulting image is an estimate of the differential distribution of temperatures, $dN/d\teff$, evaluated at constant luminosity. This value is reflective of the relative lifetimes as a function of temperature. As such, the RSG branch clearly stands out as the longest-lived phase in this part of the HR diagram.\footnote{Again, note that our sample does not extend into the temperature regime containing OB stars.} However, at high-luminosity, the RSG branch fades, while the area containing YSGs and BSGs brightens.

Clearly, something is happening at high luminosities to produce deviations from the smooth behavior that we would expect if the processes governing the distribution of stars in the HR diagram followed simple power laws in initial mass. To visualize local deviations in the numbers of stars relative to both the average distributions of luminosity and temperature, we compute $\rho_{HR}/(\rho_L\rho_T)$,\footnote{Note that the ratios are computed column- and row-wise as indicated above.} a quantity that we denote $\eta$, which we plot in the bottom right panel of Figure \ref{fig:kde_breakdown}. As we might expect, the median value of $\eta$ is approximately unity to within 1\%: i.e., at most locations in the HR diagram, $\rho_{HR}$ looks similar to if one were to draw the observed sample from $\rho_T$ and $\rho_L$ independently. However, there are clear regions that deviate significantly from the median: regions where the structure of stars is more complicated than what the individual one-dimensional KDEs capture.  

To visualize these deviations, we plot $\log_{10}{\eta/\tilde{\eta}}$ ($\tilde{\eta}$ is the median value) in the main panel of Figure \ref{fig:hr_kde}. Regions in red/blue show strong negative/positive deviations from the ``expected'' number of stars. The cross-hatched region shows the area underneath the minimum-luminosity boundary used to identify low-$L$ stars, i.e., where the KDE is extrapolated beyond the boundaries of the sample. The top- and right-hand panels show $\rho_T$ and $\rho_L$ respectively. We now clearly see the information contained within Figures \ref{fig:lum_func} and \ref{fig:lum_func_tbins}: the steepening of the RSG luminosity function is seen as the red region in the top right corner, while the flattening of the YSG luminosity function manifests as the wide blue region above $\lum=5$. Interestingly, the YSG behavior appears to be structured; the luminosity at which the blue region appears increases with temperature, and around $\teff\approx3.95$, an additional band of blue appears at $\lum=4.75$: exactly where the luminosity function flattens in Figure \ref{fig:lum_func_tbins}. 

\subsection{The YSG Plateau Above \texorpdfstring{$\lum=5$}{} In Context With the Red Supergiant Problem}

The red supergiant problem is the statistically-significant non-detection of supernova progenitors with luminosities above $\lum=5$ \citep[][]{smartt09,smartt15,kochanek20,rodriguez22}. While a number of solutions\footnote{undoubtedly too many} exist in the literature, one solution that agrees well with observations of both massive star populations and individual objects \citep{neugent20,humphreys20} is that luminous RSGs lose enough of their envelopes through (perhaps eruptive) mass loss or interactions with a binary companion to evolve blueward in the HR diagram as post-red supergiant objects before undergoing core collapse \citep[e.g.][]{ekstrom12,neugent20}. 

We would expect this scenario to then be encoded in the distribution of stars in the HR diagram. Indeed, there is a rich history of work examining the relative numbers of evolved supergiants at different luminosities. In the last 15 years, \citet{drout09}, \citet{neugent10}, \citet{neugent12}, \citet{drout12}, \citet{neugent20}, \citet{mcdonald22}, and \citet{massey22} have all shown that stellar evolution models with prescriptions for RSG mass loss that are simple power-laws in luminosity overpredict the number of high luminosity RSGs. If, as in the most recent set of Geneva evolutionary models \citep{ekstrom12,georgy13,groh19,eggenberger21}, there is a sudden increase in the RSG mass loss rates above $\lum\approx5$, the lifetimes of these luminous RSGs decrease, resulting in populations that are in agreement with what we observe in the Local Group. These RSGs become (partially-) stripped, and evolve back across the HR diagram, resulting in an apparent increase in the lifetime of YSGs. Here, we have extended this work using an extremely large sample of stars with reliable $\teff$ and $\lum$ measurements (especially for warmer stars), and illustrate the potential onset of this effect at $\lum=5$. 

We might therefore expect many of the stars that occupy the strong overdensity of YSGs shown in blue in Figure \ref{fig:hr_kde} to be post-RSG objects. While the most luminous post-RSGs can be readily identified via the presence of circumstellar material \citep[CSM, e.g.][]{jones93,humphreys97,humphreys02,shenoy16}, high-amplitude photometric variability \citep[e.g.][]{nieuwenhuijzen95,stothers01} or spectroscopic signatures \citep[e.g.][]{humphreys13,gordon16,kourniotis22}, it is only recently that a method of identifying lower-luminosity post-RSGs has been proposed: \citet{dornwallenstein20b,dornwallenstein22} have identified a new class of pulsating variable star, fast yellow pulsating supergiants (FYPS). They present evidence that FYPS pulsate due to having lost much of their envelope during a previous RSG phase, an evolutionary status that is in agreement with the high luminosities of FYPS (all above $\lum=5$). We show the HR diagram positions of the candidate FYPS in the LMC identified by \citet{dornwallenstein22} as white stars in Figure \ref{fig:hr_kde}. With the exception of two objects, FYPS reside completely within the overdensity of YSGs that we report above. All of them are found above a luminosity of $\lum=5$. Additionally, of the 39 LMC YSGs above $\lum=5$ studied by \citet{dornwallenstein22}, 13 ($\sim$33\%) are FYPS. This fraction is consistent with the findings from \citet{gordon16}, who argued that 30-40\% of luminous YSGs should be in a post-RSG evolutionary state. Combined, these pieces of evidence paint a clear picture of the evolution of massive stars above $\lum=5$: either via winds or binary interactions, these objects' RSG phase is interrupted (resulting in the underdensity of RSGs in the HR diagram), and they evolve bluewards in the HR diagram (resulting in the overdensity of YSGs), where they can be observed pulsating as FYPS. This scenario is independent of whether stars above $\lum=5$ explode at all; recent work has revealed a complex landscape of explodability \citep{sukhbold16,sukhbold18,patton20,tsang22}, and the lack of high-luminosity SN progenitors extends to all SN subtypes \citep{smartt15}. The red supergiant problem is by-no-means solved.

We note that the picture we paint hinges upon mass lost during the RSG phase, which is notably dusty \citep{massey05,walmswell12}; at least some post-RSGs are likely to be embedded in a dusty CSM as well \citep{humphreys23}. This circumstellar dust contributes an additional reddening component in excess of the $E(B-V)$ values derived using the dust maps from \citet{chen22}. Therefore, it is possible that some stars in our sample are actually quite reddened, while our final regressor does not include reddening as a feature. As we mention in \S\ref{subsubsec:featureimportance}, determining the impact of unaccounted-for reddening on our regressor isn't currently possible given the capabilities of {\sc GaiaXPy}. However, for 353 stars in our sample, we were able to access both literature spectral types and archive photometry using SIMBAD, which we used to derive $E(B-V)$ using the intrinsic colors from \citet{fitzerald70} and \citet{ducati01} available online,\footnote{\url{https://www.stsci.edu/~inr/intrins.html}} and found these $E(B-V)$ values to be 0.17 mag higher than the dustmap-derived values with a comparitively high dispersion of 0.46 mag. This indicates that there are indeed stars in our sample that are notably more reddened than the stars around them, whether because of line-of-sight effects, or excess circumstellar reddening. Ultimately, as we demonstrate in \S\ref{sec:regressing} and \ref{sec:results}, our temperatures {\it in aggregate} are accurate and trustworthy. Furthermore, when we inspected the residuals between the preditions of our regressor and the values of the temperatures and luminosities in the test set, we found no correlation between the residuals and the $E(B-V)$ values (both from the dustmap, and computed from the intrinsic and observed colors), indicating that our regressor performs just as well on the more reddened stars in our sample, despite not including $E(B-V)$ as a feature. Including mid-infrared photometry to identify individual stars that may have been impacted by a dusty CSM could in theory yield marginal improvements to our results, which we will explore in future work. In the meantime, readers looking to use our catalog of luminosities and temperatures are encouraged to determine whether potentially dusty stars may impact their analysis.

Finally, we note that binary interactions can also increase the YSG lifetime of a star on its first crossing of the HR diagram. Instead of strong mass transfer stripping the RSG primary and causing it to experience a second YSG phase, moderate amounts of mass transfer can increase the YSG lifetime of the mass gainer by up to an order of magnitude depending on when the interaction occurs in the secondary's evolution. For a constantly-star forming population, this can result in a $\sim$22\% increase in the number of YSGs (J. Eldridge, private communication).

\section{Summary \& Conclusion}\label{sec:conclusion}

Using a well-vetted training sample of 641 cool supergiants, we develop a model to predict effective temperatures and luminosities for $\sim$5,000 cool supergiants in the Large Magellanic Cloud (Figure \ref{fig:hrd}); this is the first such measurement for many of these objects. Our measurements are comparable with --- and for warmer AF supergiants, overall more reliable than --- values derived from 2MASS photometry. The resulting sample allows us to study the distribution of stars in the upper HR diagram in detail, yielding the following results:
\begin{itemize}
    \item We find a transition point in the luminosity functions of RSGs and YSGs above $\lum=5$: the RSG luminosity function steepens in comparison with the overall luminosity function of the sample, while the YSG luminosity function flattens (Figures \ref{fig:lum_func} \& \ref{fig:lum_func_tbins}). This luminosity boundary is incredibly interesting, as it corresponds with the brightest directly-detected progenitors of core collapse SNe. 
    \item Indeed, by visualizing over- and under-densities in the HR diagram (Figures \ref{fig:kde_breakdown} \& \ref{fig:hr_kde}), we again find a plateau of YSGs and a dropoff of RSGs above $\lum=5$. This is a high-fidelity illustration that is in agreement with almost 15 years of previous literature on the relative numbers of evolved supergiants at different luminosities in the Local Group. In particular, the location of the YSG plateau matches the positions in the HR diagram of the FYPS reported by \citet{dornwallenstein22}. This paints a clean picture of the end states of massive stellar evolution, wherein stars below $\lum=5$ end their lives as RSGs, while more luminous objects experience enhanced mass loss through some unknown process (decreasing the RSG lifetime and steepening the RSG luminosity function), and become YSGs (flattening the YSG luminosity function) that may pulsate as FYPS. However, there are many caveats to this particular explanation including binary evolution, RSG mass-loss rates and episodic events, unaccounted-for contamination in the \tess~lightcurves of FYPS \citep{pedersen23}, and more.
    \item Regardless of the evolutionary scenario, this large sample of objects significantly increases our ability to study evolved massive stars as ensembles of objects. 
\end{itemize}

\newpage

\acknowledgments

The authors acknowledge that the work presented was in part conducted on occupied land originally and still inhabited and cared for by the first peoples of Seattle --- the Duwamish People --- Los Angeles --- the Tongva, Kizh, and Chumash peoples --- and Colorado --- the Cheyenne, Arapaho, and Ute people. We honor with gratitude the land itself, and the original inhabitants of these places. We are grateful to have the opportunity to live and work on their ancestral lands.

T.D.W. thanks Jan Eldridge and Emma Beasor for a productive discussion about yellow supergiants, as well as Phil Massey for his comments on the manuscript.

Support for K.F.N. was provided by NASA through the NASA Hubble Fellowship grant \#HST-HF2-51516 awarded by the Space Telescope Science Institute, which is operated by the Association of Universities for Research in Astronomy, Inc., for NASA, under contract NAS5-26555. 

This research has made use of the SIMBAD database, operated at CDS, Strasbourg, France. This research has made use of the VizieR catalogue access tool, CDS, Strasbourg, France (DOI: 10.26093/cds/vizier). The original description of the VizieR service was published in A\&AS 143, 23. 

This work has made use of data from the European Space Agency (ESA) mission {\it Gaia} (\url{https://www.cosmos.esa.int/gaia}), processed by the {\it Gaia} Data Processing and Analysis Consortium (DPAC, \url{https://www.cosmos.esa.int/web/gaia/dpac/consortium}). Funding for the DPAC has been provided by national institutions, in particular the institutions participating in the {\it Gaia} Multilateral Agreement. This work has made use of the Python package {\sc GaiaXPy}, developed and maintained by members of the Gaia Data Processing and Analysis Consortium (DPAC) and in  particular, Coordination Unit 5 (CU5), and the Data Processing Centre located at the Institute of Astronomy, Cambridge, UK (DPCI).

This work made use of the following software and facilities:

\vspace{5mm}

\facility{\Gaia \citep{gaia16,gaiarchive2022}}

\software{
Astropy v4.3.1 \citep{astropy13,astropy18,astropy22},
Astroquery v0.4.6 \citep{astroquery},
Cmasher v1.6.3 \citep{vandervelden20},
Dustmaps v1.0.10 \citep{green18},
Gaiadr3\_zeropoint v0.0.1 \citep{lindegren21},
GaiaXPy v1.0.0 \doi{10.5281/zenodo.6563093},
Matplotlib v3.5.2 \citep{Hunter:2007}, 
NumPy v1.21.5 \citep{numpy:2011,harris20}, 
Pandas v1.3.5 \citep{pandas:2010}, 
Python 3.7.8, 
Scikit-learn v1.0.2 \citep{scikit-learn11},
Scipy v1.7.3 \citep{scipy:2001,scipy:2020}
}

\newpage

\bibliography{bib}

\begin{thebibliography}{}
\expandafter\ifx\csname natexlab\endcsname\relax\def\natexlab#1{#1}\fi
\providecommand{\url}[1]{\href{#1}{#1}}
\providecommand{\dodoi}[1]{doi:~\href{http://doi.org/#1}{\nolinkurl{#1}}}
\providecommand{\doeprint}[1]{\href{http://ascl.net/#1}{\nolinkurl{http://ascl.net/#1}}}
\providecommand{\doarXiv}[1]{\href{https://arxiv.org/abs/#1}{\nolinkurl{https://arxiv.org/abs/#1}}}

\bibitem[{gai(2022)}]{gaiarchive2022}
 2022, Gaia {DR}3,  European Space Agency, \dodoi{10.5270/esa-qa4lep3}.
\newblock \url{https://doi.org/10.5270/esa-qa4lep3}

\bibitem[{{Aadland} {et~al.}(2022){Aadland}, {Massey}, {Hillier}, {Morrell},
  {Neugent}, \& {Eldridge}}]{aadland22}
{Aadland}, E., {Massey}, P., {Hillier}, D.~J., {et~al.} 2022, \apj, 931, 157,
  \dodoi{10.3847/1538-4357/ac66e7}

\bibitem[{{Astropy Collaboration} {et~al.}(2013){Astropy Collaboration},
  {Robitaille}, {Tollerud}, {Greenfield}, {Droettboom}, {Bray}, {Aldcroft},
  {Davis}, {Ginsburg}, {Price-Whelan}, {Kerzendorf}, {Conley}, {Crighton},
  {Barbary}, {Muna}, {Ferguson}, {Grollier}, {Parikh}, {Nair}, {Unther},
  {Deil}, {Woillez}, {Conseil}, {Kramer}, {Turner}, {Singer}, {Fox}, {Weaver},
  {Zabalza}, {Edwards}, {Azalee Bostroem}, {Burke}, {Casey}, {Crawford},
  {Dencheva}, {Ely}, {Jenness}, {Labrie}, {Lim}, {Pierfederici}, {Pontzen},
  {Ptak}, {Refsdal}, {Servillat}, \& {Streicher}}]{astropy13}
{Astropy Collaboration}, {Robitaille}, T.~P., {Tollerud}, E.~J., {et~al.} 2013,
  \aap, 558, A33, \dodoi{10.1051/0004-6361/201322068}

\bibitem[{{Astropy Collaboration} {et~al.}(2022){Astropy Collaboration},
  {Price-Whelan}, {Lim}, {Earl}, {Starkman}, {Bradley}, {Shupe}, {Patil},
  {Corrales}, {Brasseur}, {N{\"o}the}, {Donath}, {Tollerud}, {Morris},
  {Ginsburg}, {Vaher}, {Weaver}, {Tocknell}, {Jamieson}, {van Kerkwijk},
  {Robitaille}, {Merry}, {Bachetti}, {G{\"u}nther}, {Aldcroft},
  {Alvarado-Montes}, {Archibald}, {B{\'o}di}, {Bapat}, {Barentsen},
  {Baz{\'a}n}, {Biswas}, {Boquien}, {Burke}, {Cara}, {Cara}, {Conroy},
  {Conseil}, {Craig}, {Cross}, {Cruz}, {D'Eugenio}, {Dencheva}, {Devillepoix},
  {Dietrich}, {Eigenbrot}, {Erben}, {Ferreira}, {Foreman-Mackey}, {Fox},
  {Freij}, {Garg}, {Geda}, {Glattly}, {Gondhalekar}, {Gordon}, {Grant},
  {Greenfield}, {Groener}, {Guest}, {Gurovich}, {Handberg}, {Hart},
  {Hatfield-Dodds}, {Homeier}, {Hosseinzadeh}, {Jenness}, {Jones}, {Joseph},
  {Kalmbach}, {Karamehmetoglu}, {Ka{\l}uszy{\'n}ski}, {Kelley}, {Kern},
  {Kerzendorf}, {Koch}, {Kulumani}, {Lee}, {Ly}, {Ma}, {MacBride}, {Maljaars},
  {Muna}, {Murphy}, {Norman}, {O'Steen}, {Oman}, {Pacifici}, {Pascual},
  {Pascual-Granado}, {Patil}, {Perren}, {Pickering}, {Rastogi}, {Roulston},
  {Ryan}, {Rykoff}, {Sabater}, {Sakurikar}, {Salgado}, {Sanghi}, {Saunders},
  {Savchenko}, {Schwardt}, {Seifert-Eckert}, {Shih}, {Jain}, {Shukla}, {Sick},
  {Simpson}, {Singanamalla}, {Singer}, {Singhal}, {Sinha}, {Sip{\H{o}}cz},
  {Spitler}, {Stansby}, {Streicher}, {{\v{S}}umak}, {Swinbank}, {Taranu},
  {Tewary}, {Tremblay}, {de Val-Borro}, {Van Kooten}, {Vasovi{\'c}}, {Verma},
  {de Miranda Cardoso}, {Williams}, {Wilson}, {Winkel}, {Wood-Vasey}, {Xue},
  {Yoachim}, {Zhang}, {Zonca}, \& {Astropy Project Contributors}}]{astropy22}
{Astropy Collaboration}, {Price-Whelan}, A.~M., {Lim}, P.~L., {et~al.} 2022,
  \apj, 935, 167, \dodoi{10.3847/1538-4357/ac7c74}

\bibitem[{{Bailer-Jones}(2015)}]{bailerjones15}
{Bailer-Jones}, C. A.~L. 2015, \pasp, 127, 994, \dodoi{10.1086/683116}

\bibitem[{{Bailer-Jones} {et~al.}(2021){Bailer-Jones}, {Rybizki}, {Fouesneau},
  {Demleitner}, \& {Andrae}}]{bailerjones21}
{Bailer-Jones}, C.~A.~L., {Rybizki}, J., {Fouesneau}, M., {Demleitner}, M., \&
  {Andrae}, R. 2021, \aj, 161, 147, \dodoi{10.3847/1538-3881/abd806}

\bibitem[{{Bailer-Jones} {et~al.}(2018){Bailer-Jones}, {Rybizki}, {Fouesneau},
  {Mantelet}, \& {Andrae}}]{bailerjones18}
{Bailer-Jones}, C.~A.~L., {Rybizki}, J., {Fouesneau}, M., {Mantelet}, G., \&
  {Andrae}, R. 2018, \aj, 156, 58, \dodoi{10.3847/1538-3881/aacb21}

\bibitem[{{Beasor} {et~al.}(2021){Beasor}, {Davies}, \& {Smith}}]{beasor21}
{Beasor}, E.~R., {Davies}, B., \& {Smith}, N. 2021, \apj, 922, 55,
  \dodoi{10.3847/1538-4357/ac2574}

\bibitem[{{Beasor} {et~al.}(2020){Beasor}, {Davies}, {Smith}, {van Loon},
  {Gehrz}, \& {Figer}}]{beasor20}
{Beasor}, E.~R., {Davies}, B., {Smith}, N., {et~al.} 2020, \mnras, 492, 5994,
  \dodoi{10.1093/mnras/staa255}

\bibitem[{{Bessell} \& {Brett}(1988)}]{bessell88}
{Bessell}, M.~S., \& {Brett}, J.~M. 1988, \pasp, 100, 1134,
  \dodoi{10.1086/132281}

\bibitem[{Bishop({2006})}]{bishop06}
Bishop, C.~M. {2006}, Pattern recognition and machine learning (New York :
  Springer, {2006} {\copyright}2006).
\newblock \url{https://search.library.wisc.edu/catalog/9910032530902121}

\bibitem[{{Boyer} {et~al.}(2011){Boyer}, {Srinivasan}, {van Loon}, {McDonald},
  {Meixner}, {Zaritsky}, {Gordon}, {Kemper}, {Babler}, {Block}, {Bracker},
  {Engelbracht}, {Hora}, {Indebetouw}, {Meade}, {Misselt}, {Robitaille},
  {Sewi{\l}o}, {Shiao}, \& {Whitney}}]{boyer11}
{Boyer}, M.~L., {Srinivasan}, S., {van Loon}, J.~T., {et~al.} 2011, \aj, 142,
  103, \dodoi{10.1088/0004-6256/142/4/103}

\bibitem[{Breiman(2001)}]{breiman01}
Breiman, L. 2001, Mach. Learn., 45, 5–32, \dodoi{10.1023/A:1010933404324}

\bibitem[{{Britavskiy} {et~al.}(2019{\natexlab{a}}){Britavskiy}, {Lennon},
  {Patrick}, {Evans}, {Herrero}, {Langer}, {van Loon}, {Clark}, {Schneider},
  {Almeida}, {Sana}, {de Koter}, \& {Taylor}}]{britavskiy19a}
{Britavskiy}, N., {Lennon}, D.~J., {Patrick}, L.~R., {et~al.}
  2019{\natexlab{a}}, \aap, 624, A128, \dodoi{10.1051/0004-6361/201834564}

\bibitem[{{Britavskiy} {et~al.}(2019{\natexlab{b}}){Britavskiy}, {Bonanos},
  {Herrero}, {Cervi{\~n}o}, {Garc{\'\i}a-{\'A}lvarez}, {Boyer}, {Masseron},
  {Mehner}, \& {McQuinn}}]{britavskiy19b}
{Britavskiy}, N.~E., {Bonanos}, A.~Z., {Herrero}, A., {et~al.}
  2019{\natexlab{b}}, \aap, 631, A95, \dodoi{10.1051/0004-6361/201935212}

\bibitem[{{Chen} {et~al.}(2022){Chen}, {Guo}, {Gao}, {Yang}, {Liu}, \&
  {Jiang}}]{chen22}
{Chen}, B.~Q., {Guo}, H.~L., {Gao}, J., {et~al.} 2022, \mnras, 511, 1317,
  \dodoi{10.1093/mnras/stac072}

\bibitem[{Cortes \& Vapnik(1995)}]{cortes95}
Cortes, C., \& Vapnik, V. 1995, Machine Learning, 20, 273,
  \dodoi{10.1007/BF00994018}

\bibitem[{{Cutri} {et~al.}(2003){Cutri}, {Skrutskie}, {van Dyk}, {Beichman},
  {Carpenter}, {Chester}, {Cambresy}, {Evans}, {Fowler}, {Gizis}, {Howard},
  {Huchra}, {Jarrett}, {Kopan}, {Kirkpatrick}, {Light}, {Marsh}, {McCallon},
  {Schneider}, {Stiening}, {Sykes}, {Weinberg}, {Wheaton}, {Wheelock}, \&
  {Zacarias}}]{cutri03}
{Cutri}, R.~M., {Skrutskie}, M.~F., {van Dyk}, S., {et~al.} 2003, VizieR Online
  Data Catalog, 2246

\bibitem[{{Dalcanton} {et~al.}(2012){Dalcanton}, {Williams}, {Lang}, {Lauer},
  {Kalirai}, {Seth}, {Dolphin}, {Rosenfield}, {Weisz}, {Bell}, {Bianchi},
  {Boyer}, {Caldwell}, {Dong}, {Dorman}, {Gilbert}, {Girardi}, {Gogarten},
  {Gordon}, {Guhathakurta}, {Hodge}, {Holtzman}, {Johnson}, {Larsen}, {Lewis},
  {Melbourne}, {Olsen}, {Rix}, {Rosema}, {Saha}, {Sarajedini}, {Skillman}, \&
  {Stanek}}]{dalcanton12}
{Dalcanton}, J.~J., {Williams}, B.~F., {Lang}, D., {et~al.} 2012, \apjs, 200,
  18, \dodoi{10.1088/0067-0049/200/2/18}

\bibitem[{{Dalcanton} {et~al.}(2015){Dalcanton}, {Fouesneau}, {Hogg}, {Lang},
  {Leroy}, {Gordon}, {Sandstrom}, {Weisz}, {Williams}, {Bell}, {Dong},
  {Gilbert}, {Gouliermis}, {Guhathakurta}, {Lauer}, {Schruba}, {Seth}, \&
  {Skillman}}]{dalcanton15}
{Dalcanton}, J.~J., {Fouesneau}, M., {Hogg}, D.~W., {et~al.} 2015, \apj, 814,
  3, \dodoi{10.1088/0004-637X/814/1/3}

\bibitem[{{Davies} \& {Beasor}(2020)}]{davies20}
{Davies}, B., \& {Beasor}, E.~R. 2020, \mnras, \dodoi{10.1093/mnras/staa174}

\bibitem[{{De Angeli} {et~al.}(2022){De Angeli}, {Weiler}, {Montegriffo},
  {Evans}, {Riello}, {Andrae}, {Carrasco}, {Busso}, {Burgess}, {Cacciari},
  {Davidson}, {Harrison}, {Hodgkin}, {Jordi}, {Osborne}, {Pancino},
  {Altavilla}, {Barstow}, {Bailer-Jones}, {Bellazzini}, {Brown}, {Castellani},
  {Cowell}, {Delchambre}, {De Luise}, {Diener}, {Fabricius}, {Fouesneau},
  {Fremat}, {Gilmore}, {Giuffrida}, {Hambly}, {Hidalgo}, {Holland},
  {Kostrzewa-Rutkowska}, {van Leeuwen}, {Lobel}, {Marinoni}, {Miller},
  {Pagani}, {Palaversa}, {Piersimoni}, {Pulone}, {Ragaini}, {Rainer},
  {Richards}, {Rixon}, {Ruz-Mieres}, {Sanna}, {Sarro}, {Rowell}, {Sordo},
  {Walton}, \& {Yoldas}}]{deangeli22}
{De Angeli}, F., {Weiler}, M., {Montegriffo}, P., {et~al.} 2022, arXiv
  e-prints, arXiv:2206.06143.
\newblock \doarXiv{2206.06143}

\bibitem[{{Dorda} {et~al.}(2018){Dorda}, {Negueruela},
  {Gonz{\'a}lez-Fern{\'a}ndez}, \& {Marco}}]{dorda18}
{Dorda}, R., {Negueruela}, I., {Gonz{\'a}lez-Fern{\'a}ndez}, C., \& {Marco}, A.
  2018, \aap, 618, A137, \dodoi{10.1051/0004-6361/201833219}

\bibitem[{{Dorda} {et~al.}(2016){Dorda}, {Negueruela},
  {Gonz{\'a}lez-Fern{\'a}ndez}, \& {Tabernero}}]{dorda16}
{Dorda}, R., {Negueruela}, I., {Gonz{\'a}lez-Fern{\'a}ndez}, C., \&
  {Tabernero}, H.~M. 2016, \aap, 592, A16, \dodoi{10.1051/0004-6361/201528024}

\bibitem[{{Dorn-Wallenstein} {et~al.}(2021){Dorn-Wallenstein}, {Davenport},
  {Huppenkothen}, \& {Levesque}}]{dornwallenstein21}
{Dorn-Wallenstein}, T.~Z., {Davenport}, J. R.~A., {Huppenkothen}, D., \&
  {Levesque}, E.~M. 2021, \apj, 913, 32, \dodoi{10.3847/1538-4357/abf1f2}

\bibitem[{{Dorn-Wallenstein} \& {Levesque}(2018)}]{dornwallenstein18}
{Dorn-Wallenstein}, T.~Z., \& {Levesque}, E.~M. 2018, \apj, 867, 125,
  \dodoi{10.3847/1538-4357/aae5d6}

\bibitem[{{Dorn-Wallenstein} \& {Levesque}(2020)}]{dornwallenstein20}
---. 2020, \apj, 896, 164, \dodoi{10.3847/1538-4357/ab8d28}

\bibitem[{{Dorn-Wallenstein} {et~al.}(2022){Dorn-Wallenstein}, {Levesque},
  {Davenport}, {Neugent}, {Morris}, \& {Bostroem}}]{dornwallenstein22}
{Dorn-Wallenstein}, T.~Z., {Levesque}, E.~M., {Davenport}, J. R.~A., {et~al.}
  2022, arXiv e-prints, arXiv:2206.11917.
\newblock \doarXiv{2206.11917}

\bibitem[{{Dorn-Wallenstein} {et~al.}(2020){Dorn-Wallenstein}, {Levesque},
  {Neugent}, {Davenport}, {Morris}, \& {Gootkin}}]{dornwallenstein20b}
{Dorn-Wallenstein}, T.~Z., {Levesque}, E.~M., {Neugent}, K.~F., {et~al.} 2020,
  \apj, 902, 24, \dodoi{10.3847/1538-4357/abb318}

\bibitem[{{Drout} {et~al.}(2012){Drout}, {Massey}, \& {Meynet}}]{drout12}
{Drout}, M.~R., {Massey}, P., \& {Meynet}, G. 2012, \apj, 750, 97,
  \dodoi{10.1088/0004-637X/750/2/97}

\bibitem[{{Drout} {et~al.}(2009){Drout}, {Massey}, {Meynet}, {Tokarz}, \&
  {Caldwell}}]{drout09}
{Drout}, M.~R., {Massey}, P., {Meynet}, G., {Tokarz}, S., \& {Caldwell}, N.
  2009, \apj, 703, 441, \dodoi{10.1088/0004-637X/703/1/441}

\bibitem[{{Ducati} {et~al.}(2001){Ducati}, {Bevilacqua}, {Rembold}, \&
  {Ribeiro}}]{ducati01}
{Ducati}, J.~R., {Bevilacqua}, C.~M., {Rembold}, S.~B., \& {Ribeiro}, D. 2001,
  \apj, 558, 309, \dodoi{10.1086/322439}

\bibitem[{{Durbin} {et~al.}(2020){Durbin}, {Beaton}, {Dalcanton}, {Williams},
  \& {Boyer}}]{durbin20}
{Durbin}, M.~J., {Beaton}, R.~L., {Dalcanton}, J.~J., {Williams}, B.~F., \&
  {Boyer}, M.~L. 2020, \apj, 898, 57, \dodoi{10.3847/1538-4357/ab9cbb}

\bibitem[{{Eggenberger} {et~al.}(2021){Eggenberger}, {Ekstr{\"o}m}, {Georgy},
  {Martinet}, {Pezzotti}, {Nandal}, {Meynet}, {Buldgen}, {Salmon},
  {Haemmerl{\'e}}, {Maeder}, {Hirschi}, {Yusof}, {Groh}, {Farrell}, {Murphy},
  \& {Choplin}}]{eggenberger21}
{Eggenberger}, P., {Ekstr{\"o}m}, S., {Georgy}, C., {et~al.} 2021, \aap, 652,
  A137, \dodoi{10.1051/0004-6361/202141222}

\bibitem[{{Ekstr{\"o}m} {et~al.}(2012){Ekstr{\"o}m}, {Georgy}, {Eggenberger},
  {Meynet}, {Mowlavi}, {Wyttenbach}, {Granada}, {Decressin}, {Hirschi},
  {Frischknecht}, {Charbonnel}, \& {Maeder}}]{ekstrom12}
{Ekstr{\"o}m}, S., {Georgy}, C., {Eggenberger}, P., {et~al.} 2012, \aap, 537,
  A146, \dodoi{10.1051/0004-6361/201117751}

\bibitem[{{Eldridge} {et~al.}(2017){Eldridge}, {Stanway}, {Xiao}, {McClelland},
  {Taylor}, {Ng}, {Greis}, \& {Bray}}]{eldridge17}
{Eldridge}, J.~J., {Stanway}, E.~R., {Xiao}, L., {et~al.} 2017, \pasa, 34,
  e058, \dodoi{10.1017/pasa.2017.51}

\bibitem[{{Evans} \& {Howarth}(2003)}]{evans03}
{Evans}, C.~J., \& {Howarth}, I.~D. 2003, \mnras, 345, 1223,
  \dodoi{10.1046/j.1365-2966.2003.07038.x}

\bibitem[{{Evans} {et~al.}(2006){Evans}, {Lennon}, {Smartt}, \&
  {Trundle}}]{evans06}
{Evans}, C.~J., {Lennon}, D.~J., {Smartt}, S.~J., \& {Trundle}, C. 2006, \aap,
  456, 623, \dodoi{10.1051/0004-6361:20064988}

\bibitem[{{Evans} {et~al.}(2022){Evans}, {Eyer}, {Busso}, {Riello}, {De
  Angeli}, {Burgess}, {Audard}, {Clementini}, {Garofalo}, {Holl}, {Jevardat de
  Fombelle}, {Lanzafame}, {Lecoeur-Taibi}, {Mowlavi}, {Nienartowicz},
  {Palaversa}, \& {Rimoldini}}]{evans22}
{Evans}, D.~W., {Eyer}, L., {Busso}, G., {et~al.} 2022, arXiv e-prints,
  arXiv:2206.05591.
\newblock \doarXiv{2206.05591}

\bibitem[{{Evans} {et~al.}(2010){Evans}, {Primini}, {Glotfelty}, {Anderson},
  {Bonaventura}, {Chen}, {Davis}, {Doe}, {Evans}, {Fabbiano}, {Galle}, {Gibbs},
  {Grier}, {Hain}, {Hall}, {Harbo}, {(Helen He}, {Houck}, {Karovska},
  {Kashyap}, {Lauer}, {McCollough}, {McDowell}, {Miller}, {Mitschang},
  {Morgan}, {Mossman}, {Nichols}, {Nowak}, {Plummer}, {Refsdal}, {Rots},
  {Siemiginowska}, {Sundheim}, {Tibbetts}, {Van Stone}, {Winkelman}, \&
  {Zografou}}]{evans10}
{Evans}, I.~N., {Primini}, F.~A., {Glotfelty}, K.~J., {et~al.} 2010, \apjs,
  189, 37, \dodoi{10.1088/0067-0049/189/1/37}

\bibitem[{{Feast} {et~al.}(1960){Feast}, {Thackeray}, \& {Wesselink}}]{feast60}
{Feast}, M.~W., {Thackeray}, A.~D., \& {Wesselink}, A.~J. 1960, \mnras, 121,
  337, \dodoi{10.1093/mnras/121.4.337}

\bibitem[{{Fitzgerald}(1970)}]{fitzerald70}
{Fitzgerald}, M.~P. 1970, \aap, 4, 234

\bibitem[{{Gaia Collaboration} {et~al.}(2016){Gaia Collaboration}, {Prusti},
  {de Bruijne}, {Brown}, {Vallenari}, {Babusiaux}, {Bailer-Jones}, {Bastian},
  {Biermann}, {Evans}, {Eyer}, {Jansen}, {Jordi}, {Klioner}, {Lammers},
  {Lindegren}, {Luri}, {Mignard}, {Milligan}, {Panem}, {Poinsignon},
  {Pourbaix}, {Randich}, {Sarri}, {Sartoretti}, {Siddiqui}, {Soubiran},
  {Valette}, {van Leeuwen}, {Walton}, {Aerts}, {Arenou}, {Cropper}, {Drimmel},
  {H{\o}g}, {Katz}, {Lattanzi}, {O'Mullane}, {Grebel}, {Holland}, {Huc},
  {Passot}, {Bramante}, {Cacciari}, {Casta{\~n}eda}, {Chaoul}, {Cheek}, {De
  Angeli}, {Fabricius}, {Guerra}, {Hern{\'a}ndez}, {Jean-Antoine-Piccolo},
  {Masana}, {Messineo}, {Mowlavi}, {Nienartowicz}, {Ord{\'o}{\~n}ez-Blanco},
  {Panuzzo}, {Portell}, {Richards}, {Riello}, {Seabroke}, {Tanga},
  {Th{\'e}venin}, {Torra}, {Els}, {Gracia-Abril}, {Comoretto},
  {Garcia-Reinaldos}, {Lock}, {Mercier}, {Altmann}, {Andrae}, {Astraatmadja},
  {Bellas-Velidis}, {Benson}, {Berthier}, {Blomme}, {Busso}, {Carry},
  {Cellino}, {Clementini}, {Cowell}, {Creevey}, {Cuypers}, {Davidson}, {De
  Ridder}, {de Torres}, {Delchambre}, {Dell'Oro}, {Ducourant}, {Fr{\'e}mat},
  {Garc{\'\i}a-Torres}, {Gosset}, {Halbwachs}, {Hambly}, {Harrison}, {Hauser},
  {Hestroffer}, {Hodgkin}, {Huckle}, {Hutton}, {Jasniewicz}, {Jordan},
  {Kontizas}, {Korn}, {Lanzafame}, {Manteiga}, {Moitinho}, {Muinonen},
  {Osinde}, {Pancino}, {Pauwels}, {Petit}, {Recio-Blanco}, {Robin}, {Sarro},
  {Siopis}, {Smith}, {Smith}, {Sozzetti}, {Thuillot}, {van Reeven}, {Viala},
  {Abbas}, {Abreu Aramburu}, {Accart}, {Aguado}, {Allan}, {Allasia},
  {Altavilla}, {{\'A}lvarez}, {Alves}, {Anderson}, {Andrei}, {Anglada Varela},
  {Antiche}, {Antoja}, {Ant{\'o}n}, {Arcay}, {Atzei}, {Ayache}, {Bach},
  {Baker}, {Balaguer-N{\'u}{\~n}ez}, {Barache}, {Barata}, {Barbier}, {Barblan},
  {Baroni}, {Barrado y Navascu{\'e}s}, {Barros}, {Barstow}, {Becciani},
  {Bellazzini}, {Bellei}, {Bello Garc{\'\i}a}, {Belokurov}, {Bendjoya},
  {Berihuete}, {Bianchi}, {Bienaym{\'e}}, {Billebaud}, {Blagorodnova},
  {Blanco-Cuaresma}, {Boch}, {Bombrun}, {Borrachero}, {Bouquillon}, {Bourda},
  {Bouy}, {Bragaglia}, {Breddels}, {Brouillet}, {Br{\"u}semeister},
  {Bucciarelli}, {Budnik}, {Burgess}, {Burgon}, {Burlacu}, {Busonero}, {Buzzi},
  {Caffau}, {Cambras}, {Campbell}, {Cancelliere}, {Cantat-Gaudin}, {Carlucci},
  {Carrasco}, {Castellani}, {Charlot}, {Charnas}, {Charvet}, {Chassat},
  {Chiavassa}, {Clotet}, {Cocozza}, {Collins}, {Collins}, {Costigan}, {Crifo},
  {Cross}, {Crosta}, {Crowley}, {Dafonte}, {Damerdji}, {Dapergolas}, {David},
  {David}, {De Cat}, {de Felice}, {de Laverny}, {De Luise}, {De March}, {de
  Martino}, {de Souza}, {Debosscher}, {del Pozo}, {Delbo}, {Delgado},
  {Delgado}, {di Marco}, {Di Matteo}, {Diakite}, {Distefano}, {Dolding}, {Dos
  Anjos}, {Drazinos}, {Dur{\'a}n}, {Dzigan}, {Ecale}, {Edvardsson}, {Enke},
  {Erdmann}, {Escolar}, {Espina}, {Evans}, {Eynard Bontemps}, {Fabre},
  {Fabrizio}, {Faigler}, {Falc{\~a}o}, {Farr{\`a}s Casas}, {Faye}, {Federici},
  {Fedorets}, {Fern{\'a}ndez-Hern{\'a}ndez}, {Fernique}, {Fienga}, {Figueras},
  {Filippi}, {Findeisen}, {Fonti}, {Fouesneau}, {Fraile}, {Fraser}, {Fuchs},
  {Furnell}, {Gai}, {Galleti}, {Galluccio}, {Garabato}, {Garc{\'\i}a-Sedano},
  {Gar{\'e}}, {Garofalo}, {Garralda}, {Gavras}, {Gerssen}, {Geyer}, {Gilmore},
  {Girona}, {Giuffrida}, {Gomes}, {Gonz{\'a}lez-Marcos},
  {Gonz{\'a}lez-N{\'u}{\~n}ez}, {Gonz{\'a}lez-Vidal}, {Granvik}, {Guerrier},
  {Guillout}, {Guiraud}, {G{\'u}rpide}, {Guti{\'e}rrez-S{\'a}nchez}, {Guy},
  {Haigron}, {Hatzidimitriou}, {Haywood}, {Heiter}, {Helmi}, {Hobbs},
  {Hofmann}, {Holl}, {Holland}, {Hunt}, {Hypki}, {Icardi}, {Irwin}, {Jevardat
  de Fombelle}, {Jofr{\'e}}, {Jonker}, {Jorissen}, {Julbe}, {Karampelas},
  {Kochoska}, {Kohley}, {Kolenberg}, {Kontizas}, {Koposov}, {Kordopatis},
  {Koubsky}, {Kowalczyk}, {Krone-Martins}, {Kudryashova}, {Kull}, {Bachchan},
  {Lacoste-Seris}, {Lanza}, {Lavigne}, {Le Poncin-Lafitte}, {Lebreton},
  {Lebzelter}, {Leccia}, {Leclerc}, {Lecoeur-Taibi}, {Lemaitre}, {Lenhardt},
  {Leroux}, {Liao}, {Licata}, {Lindstr{\o}m}, {Lister}, {Livanou}, {Lobel},
  {L{\"o}ffler}, {L{\'o}pez}, {Lopez-Lozano}, {Lorenz}, {Loureiro},
  {MacDonald}, {Magalh{\~a}es Fernandes}, {Managau}, {Mann}, {Mantelet},
  {Marchal}, {Marchant}, {Marconi}, {Marie}, {Marinoni}, {Marrese},
  {Marschalk{\'o}}, {Marshall}, {Mart{\'\i}n-Fleitas}, {Martino}, {Mary},
  {Matijevi{\v{c}}}, {Mazeh}, {McMillan}, {Messina}, {Mestre}, {Michalik},
  {Millar}, {Miranda}, {Molina}, {Molinaro}, {Molinaro}, {Moln{\'a}r},
  {Moniez}, {Montegriffo}, {Monteiro}, {Mor}, {Mora}, {Morbidelli}, {Morel},
  {Morgenthaler}, {Morley}, {Morris}, {Mulone}, {Muraveva}, {Musella},
  {Narbonne}, {Nelemans}, {Nicastro}, {Noval}, {Ord{\'e}novic},
  {Ordieres-Mer{\'e}}, {Osborne}, {Pagani}, {Pagano}, {Pailler}, {Palacin},
  {Palaversa}, {Parsons}, {Paulsen}, {Pecoraro}, {Pedrosa}, {Pentik{\"a}inen},
  {Pereira}, {Pichon}, {Piersimoni}, {Pineau}, {Plachy}, {Plum}, {Poujoulet},
  {Pr{\v{s}}a}, {Pulone}, {Ragaini}, {Rago}, {Rambaux}, {Ramos-Lerate},
  {Ranalli}, {Rauw}, {Read}, {Regibo}, {Renk}, {Reyl{\'e}}, {Ribeiro},
  {Rimoldini}, {Ripepi}, {Riva}, {Rixon}, {Roelens}, {Romero-G{\'o}mez},
  {Rowell}, {Royer}, {Rudolph}, {Ruiz-Dern}, {Sadowski}, {Sagrist{\`a}
  Sell{\'e}s}, {Sahlmann}, {Salgado}, {Salguero}, {Sarasso}, {Savietto},
  {Schnorhk}, {Schultheis}, {Sciacca}, {Segol}, {Segovia}, {Segransan},
  {Serpell}, {Shih}, {Smareglia}, {Smart}, {Smith}, {Solano}, {Solitro},
  {Sordo}, {Soria Nieto}, {Souchay}, {Spagna}, {Spoto}, {Stampa}, {Steele},
  {Steidelm{\"u}ller}, {Stephenson}, {Stoev}, {Suess}, {S{\"u}veges}, {Surdej},
  {Szabados}, {Szegedi-Elek}, {Tapiador}, {Taris}, {Tauran}, {Taylor},
  {Teixeira}, {Terrett}, {Tingley}, {Trager}, {Turon}, {Ulla}, {Utrilla},
  {Valentini}, {van Elteren}, {Van Hemelryck}, {van Leeuwen}, {Varadi},
  {Vecchiato}, {Veljanoski}, {Via}, {Vicente}, {Vogt}, {Voss}, {Votruba},
  {Voutsinas}, {Walmsley}, {Weiler}, {Weingrill}, {Werner}, {Wevers},
  {Whitehead}, {Wyrzykowski}, {Yoldas}, {{\v{Z}}erjal}, {Zucker}, {Zurbach},
  {Zwitter}, {Alecu}, {Allen}, {Allende Prieto}, {Amorim},
  {Anglada-Escud{\'e}}, {Arsenijevic}, {Azaz}, {Balm}, {Beck}, {Bernstein},
  {Bigot}, {Bijaoui}, {Blasco}, {Bonfigli}, {Bono}, {Boudreault}, {Bressan},
  {Brown}, {Brunet}, {Bunclark}, {Buonanno}, {Butkevich}, {Carret}, {Carrion},
  {Chemin}, {Ch{\'e}reau}, {Corcione}, {Darmigny}, {de Boer}, {de Teodoro}, {de
  Zeeuw}, {Delle Luche}, {Domingues}, {Dubath}, {Fodor}, {Fr{\'e}zouls},
  {Fries}, {Fustes}, {Fyfe}, {Gallardo}, {Gallegos}, {Gardiol}, {Gebran},
  {Gomboc}, {G{\'o}mez}, {Grux}, {Gueguen}, {Heyrovsky}, {Hoar}, {Iannicola},
  {Isasi Parache}, {Janotto}, {Joliet}, {Jonckheere}, {Keil}, {Kim},
  {Klagyivik}, {Klar}, {Knude}, {Kochukhov}, {Kolka}, {Kos}, {Kutka}, {Lainey},
  {LeBouquin}, {Liu}, {Loreggia}, {Makarov}, {Marseille}, {Martayan},
  {Martinez-Rubi}, {Massart}, {Meynadier}, {Mignot}, {Munari}, {Nguyen},
  {Nordlander}, {Ocvirk}, {O'Flaherty}, {Olias Sanz}, {Ortiz}, {Osorio},
  {Oszkiewicz}, {Ouzounis}, {Palmer}, {Park}, {Pasquato}, {Peltzer}, {Peralta},
  {P{\'e}turaud}, {Pieniluoma}, {Pigozzi}, {Poels}, {Prat}, {Prod'homme},
  {Raison}, {Rebordao}, {Risquez}, {Rocca-Volmerange}, {Rosen}, {Ruiz-Fuertes},
  {Russo}, {Sembay}, {Serraller Vizcaino}, {Short}, {Siebert}, {Silva},
  {Sinachopoulos}, {Slezak}, {Soffel}, {Sosnowska}, {Strai{\v{z}}ys}, {ter
  Linden}, {Terrell}, {Theil}, {Tiede}, {Troisi}, {Tsalmantza}, {Tur},
  {Vaccari}, {Vachier}, {Valles}, {Van Hamme}, {Veltz}, {Virtanen}, {Wallut},
  {Wichmann}, {Wilkinson}, {Ziaeepour}, \& {Zschocke}}]{gaia16}
{Gaia Collaboration}, {Prusti}, T., {de Bruijne}, J.~H.~J., {et~al.} 2016,
  \aap, 595, A1, \dodoi{10.1051/0004-6361/201629272}

\bibitem[{{Gaia Collaboration} {et~al.}(2018){Gaia Collaboration}, {Helmi},
  {van Leeuwen}, {McMillan}, {Massari}, {Antoja}, {Robin}, {Lindegren},
  {Bastian}, {Arenou}, {Babusiaux}, {Biermann}, {Breddels}, {Hobbs}, {Jordi},
  {Pancino}, {Reyl{\'e}}, {Veljanoski}, {Brown}, {Vallenari}, {Prusti}, {de
  Bruijne}, {Bailer-Jones}, {Evans}, {Eyer}, {Jansen}, {Klioner}, {Lammers},
  {Luri}, {Mignard}, {Panem}, {Pourbaix}, {Randich}, {Sartoretti}, {Siddiqui},
  {Soubiran}, {Walton}, {Cropper}, {Drimmel}, {Katz}, {Lattanzi}, {Bakker},
  {Cacciari}, {Casta{\~n}eda}, {Chaoul}, {Cheek}, {De Angeli}, {Fabricius},
  {Guerra}, {Holl}, {Masana}, {Messineo}, {Mowlavi}, {Nienartowicz}, {Panuzzo},
  {Portell}, {Riello}, {Seabroke}, {Tanga}, {Th{\'e}venin}, {Gracia-Abril},
  {Comoretto}, {Garcia-Reinaldos}, {Teyssier}, {Altmann}, {Andrae}, {Audard},
  {Bellas-Velidis}, {Benson}, {Berthier}, {Blomme}, {Burgess}, {Busso},
  {Carry}, {Cellino}, {Clementini}, {Clotet}, {Creevey}, {Davidson}, {De
  Ridder}, {Delchambre}, {Dell'Oro}, {Ducourant},
  {Fern{\'a}ndez-Hern{\'a}ndez}, {Fouesneau}, {Fr{\'e}mat}, {Galluccio},
  {Garc{\'\i}a-Torres}, {Gonz{\'a}lez-N{\'u}{\~n}ez}, {Gonz{\'a}lez-Vidal},
  {Gosset}, {Guy}, {Halbwachs}, {Hambly}, {Harrison}, {Hern{\'a}ndez},
  {Hestroffer}, {Hodgkin}, {Hutton}, {Jasniewicz}, {Jean-Antoine-Piccolo},
  {Jordan}, {Korn}, {Krone-Martins}, {Lanzafame}, {Lebzelter}, {L{\"o}ffler},
  {Manteiga}, {Marrese}, {Mart{\'\i}n-Fleitas}, {Moitinho}, {Mora}, {Muinonen},
  {Osinde}, {Pauwels}, {Petit}, {Recio-Blanco}, {Richards}, {Rimoldini},
  {Sarro}, {Siopis}, {Smith}, {Sozzetti}, {S{\"u}veges}, {Torra}, {van Reeven},
  {Abbas}, {Abreu Aramburu}, {Accart}, {Aerts}, {Altavilla}, {{\'A}lvarez},
  {Alvarez}, {Alves}, {Anderson}, {Andrei}, {Anglada Varela}, {Antiche},
  {Arcay}, {Astraatmadja}, {Bach}, {Baker}, {Balaguer-N{\'u}{\~n}ez}, {Balm},
  {Barache}, {Barata}, {Barbato}, {Barblan}, {Barklem}, {Barrado}, {Barros},
  {Barstow}, {Bartholom{\'e} Mu{\~n}oz}, {Bassilana}, {Becciani}, {Bellazzini},
  {Berihuete}, {Bertone}, {Bianchi}, {Bienaym{\'e}}, {Blanco-Cuaresma}, {Boch},
  {Boeche}, {Bombrun}, {Borrachero}, {Bossini}, {Bouquillon}, {Bourda},
  {Bragaglia}, {Bramante}, {Bressan}, {Brouillet}, {Br{\"u}semeister},
  {Brugaletta}, {Bucciarelli}, {Burlacu}, {Busonero}, {Butkevich}, {Buzzi},
  {Caffau}, {Cancelliere}, {Cannizzaro}, {Cantat-Gaudin}, {Carballo},
  {Carlucci}, {Carrasco}, {Casamiquela}, {Castellani}, {Castro-Ginard},
  {Charlot}, {Chemin}, {Chiavassa}, {Cocozza}, {Costigan}, {Cowell}, {Crifo},
  {Crosta}, {Crowley}, {Cuypers}, {Dafonte}, {Damerdji}, {Dapergolas}, {David},
  {David}, {de Laverny}, {De Luise}, {De March}, {de Martino}, {de Souza}, {de
  Torres}, {Debosscher}, {del Pozo}, {Delbo}, {Delgado}, {Delgado}, {Di
  Matteo}, {Diakite}, {Diener}, {Distefano}, {Dolding}, {Drazinos},
  {Dur{\'a}n}, {Edvardsson}, {Enke}, {Eriksson}, {Esquej}, {Eynard Bontemps},
  {Fabre}, {Fabrizio}, {Faigler}, {Falc{\~a}o}, {Farr{\`a}s Casas}, {Federici},
  {Fedorets}, {Fernique}, {Figueras}, {Filippi}, {Findeisen}, {Fonti},
  {Fraile}, {Fraser}, {Fr{\'e}zouls}, {Gai}, {Galleti}, {Garabato},
  {Garc{\'\i}a-Sedano}, {Garofalo}, {Garralda}, {Gavel}, {Gavras}, {Gerssen},
  {Geyer}, {Giacobbe}, {Gilmore}, {Girona}, {Giuffrida}, {Glass}, {Gomes},
  {Granvik}, {Gueguen}, {Guerrier}, {Guiraud}, {Guti{\'e}rrez-S{\'a}nchez},
  {Hofmann}, {Holland}, {Huckle}, {Hypki}, {Icardi}, {Jan{\ss}en}, {Jevardat de
  Fombelle}, {Jonker}, {Juh{\'a}sz}, {Julbe}, {Karampelas}, {Kewley}, {Klar},
  {Kochoska}, {Kohley}, {Kolenberg}, {Kontizas}, {Kontizas}, {Koposov},
  {Kordopatis}, {Kostrzewa-Rutkowska}, {Koubsky}, {Lambert}, {Lanza}, {Lasne},
  {Lavigne}, {Le Fustec}, {Le Poncin-Lafitte}, {Lebreton}, {Leccia}, {Leclerc},
  {Lecoeur-Taibi}, {Lenhardt}, {Leroux}, {Liao}, {Licata}, {Lindstr{\o}m},
  {Lister}, {Livanou}, {Lobel}, {L{\'o}pez}, {Managau}, {Mann}, {Mantelet},
  {Marchal}, {Marchant}, {Marconi}, {Marinoni}, {Marschalk{\'o}}, {Marshall},
  {Martino}, {Marton}, {Mary}, {Matijevi{\v{c}}}, {Mazeh}, {Messina},
  {Michalik}, {Millar}, {Molina}, {Molinaro}, {Moln{\'a}r}, {Montegriffo},
  {Mor}, {Morbidelli}, {Morel}, {Morris}, {Mulone}, {Muraveva}, {Musella},
  {Nelemans}, {Nicastro}, {Noval}, {O'Mullane}, {Ord{\'e}novic},
  {Ord{\'o}{\~n}ez-Blanco}, {Osborne}, {Pagani}, {Pagano}, {Pailler},
  {Palacin}, {Palaversa}, {Panahi}, {Pawlak}, {Piersimoni}, {Pineau}, {Plachy},
  {Plum}, {Poggio}, {Poujoulet}, {Pr{\v{s}}a}, {Pulone}, {Racero}, {Ragaini},
  {Rambaux}, {Ramos-Lerate}, {Regibo}, {Riclet}, {Ripepi}, {Riva}, {Rivard},
  {Rixon}, {Roegiers}, {Roelens}, {Romero-G{\'o}mez}, {Rowell}, {Royer},
  {Ruiz-Dern}, {Sadowski}, {Sagrist{\`a} Sell{\'e}s}, {Sahlmann}, {Salgado},
  {Salguero}, {Sanna}, {Santana-Ros}, {Sarasso}, {Savietto}, {Schultheis},
  {Sciacca}, {Segol}, {Segovia}, {S{\'e}gransan}, {Shih}, {Siltala}, {Silva},
  {Smart}, {Smith}, {Solano}, {Solitro}, {Sordo}, {Soria Nieto}, {Souchay},
  {Spagna}, {Spoto}, {Stampa}, {Steele}, {Steidelm{\"u}ller}, {Stephenson},
  {Stoev}, {Suess}, {Surdej}, {Szabados}, {Szegedi-Elek}, {Tapiador}, {Taris},
  {Tauran}, {Taylor}, {Teixeira}, {Terrett}, {Teyssand ier}, {Thuillot},
  {Titarenko}, {Torra Clotet}, {Turon}, {Ulla}, {Utrilla}, {Uzzi}, {Vaillant},
  {Valentini}, {Valette}, {van Elteren}, {Van Hemelryck}, {van Leeuwen},
  {Vaschetto}, {Vecchiato}, {Viala}, {Vicente}, {Vogt}, {von Essen}, {Voss},
  {Votruba}, {Voutsinas}, {Walmsley}, {Weiler}, {Wertz}, {Wevems},
  {Wyrzykowski}, {Yoldas}, {{\v{Z}}erjal}, {Ziaeepour}, {Zorec}, {Zschocke},
  {Zucker}, {Zurbach}, \& {Zwitter}}]{gaiacollab18}
{Gaia Collaboration}, {Helmi}, A., {van Leeuwen}, F., {et~al.} 2018, \aap, 616,
  A12, \dodoi{10.1051/0004-6361/201832698}

\bibitem[{{Gaia Collaboration} {et~al.}(2023){Gaia Collaboration}, {Vallenari},
  {Brown}, {Prusti}, {de Bruijne}, {Arenou}, {Babusiaux}, {Biermann},
  {Creevey}, {Ducourant}, {Evans}, {Eyer}, {Guerra}, {Hutton}, {Jordi},
  {Klioner}, {Lammers}, {Lindegren}, {Luri}, {Mignard}, {Panem}, {Pourbaix},
  {Randich}, {Sartoretti}, {Soubiran}, {Tanga}, {Walton}, {Bailer-Jones},
  {Bastian}, {Drimmel}, {Jansen}, {Katz}, {Lattanzi}, {van Leeuwen}, {Bakker},
  {Cacciari}, {Casta{\~n}eda}, {De Angeli}, {Fabricius}, {Fouesneau},
  {Fr{\'e}mat}, {Galluccio}, {Guerrier}, {Heiter}, {Masana}, {Messineo},
  {Mowlavi}, {Nicolas}, {Nienartowicz}, {Pailler}, {Panuzzo}, {Riclet}, {Roux},
  {Seabroke}, {Sordo}, {Th{\'e}venin}, {Gracia-Abril}, {Portell}, {Teyssier},
  {Altmann}, {Andrae}, {Audard}, {Bellas-Velidis}, {Benson}, {Berthier},
  {Blomme}, {Burgess}, {Busonero}, {Busso}, {C{\'a}novas}, {Carry}, {Cellino},
  {Cheek}, {Clementini}, {Damerdji}, {Davidson}, {de Teodoro}, {Nu{\~n}ez
  Campos}, {Delchambre}, {Dell'Oro}, {Esquej}, {Fern{\'a}ndez-Hern{\'a}ndez},
  {Fraile}, {Garabato}, {Garc{\'\i}a-Lario}, {Gosset}, {Haigron}, {Halbwachs},
  {Hambly}, {Harrison}, {Hern{\'a}ndez}, {Hestroffer}, {Hodgkin}, {Holl},
  {Jan{\ss}en}, {Jevardat de Fombelle}, {Jordan}, {Krone-Martins}, {Lanzafame},
  {L{\"o}ffler}, {Marchal}, {Marrese}, {Moitinho}, {Muinonen}, {Osborne},
  {Pancino}, {Pauwels}, {Recio-Blanco}, {Reyl{\'e}}, {Riello}, {Rimoldini},
  {Roegiers}, {Rybizki}, {Sarro}, {Siopis}, {Smith}, {Sozzetti}, {Utrilla},
  {van Leeuwen}, {Abbas}, {{\'A}brah{\'a}m}, {Abreu Aramburu}, {Aerts},
  {Aguado}, {Ajaj}, {Aldea-Montero}, {Altavilla}, {{\'A}lvarez}, {Alves},
  {Anders}, {Anderson}, {Anglada Varela}, {Antoja}, {Baines}, {Baker},
  {Balaguer-N{\'u}{\~n}ez}, {Balbinot}, {Balog}, {Barache}, {Barbato},
  {Barros}, {Barstow}, {Bartolom{\'e}}, {Bassilana}, {Bauchet}, {Becciani},
  {Bellazzini}, {Berihuete}, {Bernet}, {Bertone}, {Bianchi}, {Binnenfeld},
  {Blanco-Cuaresma}, {Blazere}, {Boch}, {Bombrun}, {Bossini}, {Bouquillon},
  {Bragaglia}, {Bramante}, {Breedt}, {Bressan}, {Brouillet}, {Brugaletta},
  {Bucciarelli}, {Burlacu}, {Butkevich}, {Buzzi}, {Caffau}, {Cancelliere},
  {Cantat-Gaudin}, {Carballo}, {Carlucci}, {Carnerero}, {Carrasco},
  {Casamiquela}, {Castellani}, {Castro-Ginard}, {Chaoul}, {Charlot}, {Chemin},
  {Chiaramida}, {Chiavassa}, {Chornay}, {Comoretto}, {Contursi}, {Cooper},
  {Cornez}, {Cowell}, {Crifo}, {Cropper}, {Crosta}, {Crowley}, {Dafonte},
  {Dapergolas}, {David}, {David}, {de Laverny}, {De Luise}, {De March}, {De
  Ridder}, {de Souza}, {de Torres}, {del Peloso}, {del Pozo}, {Delbo},
  {Delgado}, {Delisle}, {Demouchy}, {Dharmawardena}, {Di Matteo}, {Diakite},
  {Diener}, {Distefano}, {Dolding}, {Edvardsson}, {Enke}, {Fabre}, {Fabrizio},
  {Faigler}, {Fedorets}, {Fernique}, {Fienga}, {Figueras}, {Fournier},
  {Fouron}, {Fragkoudi}, {Gai}, {Garcia-Gutierrez}, {Garcia-Reinaldos},
  {Garc{\'\i}a-Torres}, {Garofalo}, {Gavel}, {Gavras}, {Gerlach}, {Geyer},
  {Giacobbe}, {Gilmore}, {Girona}, {Giuffrida}, {Gomel}, {Gomez},
  {Gonz{\'a}lez-N{\'u}{\~n}ez}, {Gonz{\'a}lez-Santamar{\'\i}a},
  {Gonz{\'a}lez-Vidal}, {Granvik}, {Guillout}, {Guiraud},
  {Guti{\'e}rrez-S{\'a}nchez}, {Guy}, {Hatzidimitriou}, {Hauser}, {Haywood},
  {Helmer}, {Helmi}, {Sarmiento}, {Hidalgo}, {Hilger}, {H{\l}adczuk}, {Hobbs},
  {Holland}, {Huckle}, {Jardine}, {Jasniewicz}, {Jean-Antoine Piccolo},
  {Jim{\'e}nez-Arranz}, {Jorissen}, {Juaristi Campillo}, {Julbe}, {Karbevska},
  {Kervella}, {Khanna}, {Kontizas}, {Kordopatis}, {Korn}, {K{\'o}sp{\'a}l},
  {Kostrzewa-Rutkowska}, {Kruszy{\'n}ska}, {Kun}, {Laizeau}, {Lambert},
  {Lanza}, {Lasne}, {Le Campion}, {Lebreton}, {Lebzelter}, {Leccia}, {Leclerc},
  {Lecoeur-Taibi}, {Liao}, {Licata}, {Lindstr{\o}m}, {Lister}, {Livanou},
  {Lobel}, {Lorca}, {Loup}, {Madrero Pardo}, {Magdaleno Romeo}, {Managau},
  {Mann}, {Manteiga}, {Marchant}, {Marconi}, {Marcos}, {Marcos Santos},
  {Mar{\'\i}n Pina}, {Marinoni}, {Marocco}, {Marshall}, {Martin Polo},
  {Mart{\'\i}n-Fleitas}, {Marton}, {Mary}, {Masip}, {Massari},
  {Mastrobuono-Battisti}, {Mazeh}, {McMillan}, {Messina}, {Michalik}, {Millar},
  {Mints}, {Molina}, {Molinaro}, {Moln{\'a}r}, {Monari}, {Mongui{\'o}},
  {Montegriffo}, {Montero}, {Mor}, {Mora}, {Morbidelli}, {Morel}, {Morris},
  {Muraveva}, {Murphy}, {Musella}, {Nagy}, {Noval}, {Oca{\~n}a}, {Ogden},
  {Ordenovic}, {Osinde}, {Pagani}, {Pagano}, {Palaversa}, {Palicio},
  {Pallas-Quintela}, {Panahi}, {Payne-Wardenaar}, {Pe{\~n}alosa Esteller},
  {Penttil{\"a}}, {Pichon}, {Piersimoni}, {Pineau}, {Plachy}, {Plum}, {Poggio},
  {Pr{\v{s}}a}, {Pulone}, {Racero}, {Ragaini}, {Rainer}, {Raiteri}, {Rambaux},
  {Ramos}, {Ramos-Lerate}, {Re Fiorentin}, {Regibo}, {Richards}, {Rios Diaz},
  {Ripepi}, {Riva}, {Rix}, {Rixon}, {Robichon}, {Robin}, {Robin}, {Roelens},
  {Rogues}, {Rohrbasser}, {Romero-G{\'o}mez}, {Rowell}, {Royer}, {Ruz Mieres},
  {Rybicki}, {Sadowski}, {S{\'a}ez N{\'u}{\~n}ez}, {Sagrist{\`a} Sell{\'e}s},
  {Sahlmann}, {Salguero}, {Samaras}, {Sanchez Gimenez}, {Sanna},
  {Santove{\~n}a}, {Sarasso}, {Schultheis}, {Sciacca}, {Segol}, {Segovia},
  {S{\'e}gransan}, {Semeux}, {Shahaf}, {Siddiqui}, {Siebert}, {Siltala},
  {Silvelo}, {Slezak}, {Slezak}, {Smart}, {Snaith}, {Solano}, {Solitro},
  {Souami}, {Souchay}, {Spagna}, {Spina}, {Spoto}, {Steele},
  {Steidelm{\"u}ller}, {Stephenson}, {S{\"u}veges}, {Surdej}, {Szabados},
  {Szegedi-Elek}, {Taris}, {Taylor}, {Teixeira}, {Tolomei}, {Tonello}, {Torra},
  {Torra}, {Torralba Elipe}, {Trabucchi}, {Tsounis}, {Turon}, {Ulla}, {Unger},
  {Vaillant}, {van Dillen}, {van Reeven}, {Vanel}, {Vecchiato}, {Viala},
  {Vicente}, {Voutsinas}, {Weiler}, {Wevers}, {Wyrzykowski}, {Yoldas}, {Yvard},
  {Zhao}, {Zorec}, {Zucker}, \& {Zwitter}}]{gaia23}
{Gaia Collaboration}, {Vallenari}, A., {Brown}, A.~G.~A., {et~al.} 2023, \aap,
  674, A1, \dodoi{10.1051/0004-6361/202243940}

\bibitem[{{Georgy} {et~al.}(2013){Georgy}, {Ekstr{\"o}m}, {Eggenberger},
  {Meynet}, {Haemmerl{\'e}}, {Maeder}, {Granada}, {Groh}, {Hirschi}, {Mowlavi},
  {Yusof}, {Charbonnel}, {Decressin}, \& {Barblan}}]{georgy13}
{Georgy}, C., {Ekstr{\"o}m}, S., {Eggenberger}, P., {et~al.} 2013, \aap, 558,
  A103, \dodoi{10.1051/0004-6361/201322178}

\bibitem[{{Ginsburg} {et~al.}(2019){Ginsburg}, {Sip{\H{o}}cz}, {Brasseur},
  {Cowperthwaite}, {Craig}, {Deil}, {Guillochon}, {Guzman}, {Liedtke}, {Lian
  Lim}, {Lockhart}, {Mommert}, {Morris}, {Norman}, {Parikh}, {Persson},
  {Robitaille}, {Segovia}, {Singer}, {Tollerud}, {de Val-Borro}, {Valtchanov},
  {Woillez}, {Astroquery Collaboration}, \& {a subset of astropy
  Collaboration}}]{astroquery}
{Ginsburg}, A., {Sip{\H{o}}cz}, B.~M., {Brasseur}, C.~E., {et~al.} 2019, \aj,
  157, 98, \dodoi{10.3847/1538-3881/aafc33}

\bibitem[{{Gordon} {et~al.}(2016){Gordon}, {Humphreys}, \& {Jones}}]{gordon16}
{Gordon}, M.~S., {Humphreys}, R.~M., \& {Jones}, T.~J. 2016, \apj, 825, 50,
  \dodoi{10.3847/0004-637X/825/1/50}

\bibitem[{{Green}(2018)}]{green18}
{Green}, G. 2018, The Journal of Open Source Software, 3, 695,
  \dodoi{10.21105/joss.00695}

\bibitem[{{Groh} {et~al.}(2019){Groh}, {Ekstr{\"o}m}, {Georgy}, {Meynet},
  {Choplin}, {Eggenberger}, {Hirschi}, {Maeder}, {Murphy}, {Boian}, \&
  {Farrell}}]{groh19}
{Groh}, J.~H., {Ekstr{\"o}m}, S., {Georgy}, C., {et~al.} 2019, \aap, 627, A24,
  \dodoi{10.1051/0004-6361/201833720}

\bibitem[{{Gustafsson} {et~al.}(2008){Gustafsson}, {Edvardsson}, {Eriksson},
  {J{\o}rgensen}, {Nordlund}, \& {Plez}}]{gustafsson08}
{Gustafsson}, B., {Edvardsson}, B., {Eriksson}, K., {et~al.} 2008, \aap, 486,
  951, \dodoi{10.1051/0004-6361:200809724}

\bibitem[{{Harris} {et~al.}(2020){Harris}, {Millman}, {van der Walt},
  {Gommers}, {Virtanen}, {Cournapeau}, {Wieser}, {Taylor}, {Berg}, {Smith},
  {Kern}, {Picus}, {Hoyer}, {van Kerkwijk}, {Brett}, {Haldane}, {del R{\'\i}o},
  {Wiebe}, {Peterson}, {G{\'e}rard-Marchant}, {Sheppard}, {Reddy}, {Weckesser},
  {Abbasi}, {Gohlke}, \& {Oliphant}}]{harris20}
{Harris}, C.~R., {Millman}, K.~J., {van der Walt}, S.~J., {et~al.} 2020, \nat,
  585, 357, \dodoi{10.1038/s41586-020-2649-2}

\bibitem[{Hastie {et~al.}(2009)Hastie, Tibshirani, \& Friedman}]{hastie09}
Hastie, T., Tibshirani, R., \& Friedman, J. 2009, The Elements of Statistical
  Learning: Data Mining, Inference and Prediction, 2nd edn. (New York :
  Springer {2009} {\copyright}2009).
\newblock \url{http://www-stat.stanford.edu/~tibs/ElemStatLearn/}

\bibitem[{{Hayashi} \& {Hoshi}(1961)}]{hayashi61}
{Hayashi}, C., \& {Hoshi}, R. 1961, \pasj, 13, 442

\bibitem[{{Humphreys} {et~al.}(2013){Humphreys}, {Davidson}, {Grammer},
  {Kneeland}, {Martin}, {Weis}, \& {Burggraf}}]{humphreys13}
{Humphreys}, R.~M., {Davidson}, K., {Grammer}, S., {et~al.} 2013, \apj, 773,
  46, \dodoi{10.1088/0004-637X/773/1/46}

\bibitem[{{Humphreys} {et~al.}(2002){Humphreys}, {Davidson}, \&
  {Smith}}]{humphreys02}
{Humphreys}, R.~M., {Davidson}, K., \& {Smith}, N. 2002, \aj, 124, 1026,
  \dodoi{10.1086/341380}

\bibitem[{{Humphreys} {et~al.}(2020){Humphreys}, {Helmel}, {Jones}, \&
  {Gordon}}]{humphreys20}
{Humphreys}, R.~M., {Helmel}, G., {Jones}, T.~J., \& {Gordon}, M.~S. 2020, \aj,
  160, 145, \dodoi{10.3847/1538-3881/abab15}

\bibitem[{{Humphreys} {et~al.}(2023){Humphreys}, {Jones}, \&
  {Martin}}]{humphreys23}
{Humphreys}, R.~M., {Jones}, T.~J., \& {Martin}, J.~C. 2023, \aj, 166, 50,
  \dodoi{10.3847/1538-3881/acdd6c}

\bibitem[{{Humphreys} {et~al.}(1997){Humphreys}, {Smith}, {Davidson}, {Jones},
  {Gehrz}, {Mason}, {Hayward}, {Houck}, \& {Krautter}}]{humphreys97}
{Humphreys}, R.~M., {Smith}, N., {Davidson}, K., {et~al.} 1997, \aj, 114, 2778,
  \dodoi{10.1086/118686}

\bibitem[{Hunter(2007)}]{Hunter:2007}
Hunter, J.~D. 2007, Computing In Science \& Engineering, 9, 90

\bibitem[{Jones {et~al.}(2001)Jones, Oliphant, Peterson, {et~al.}}]{scipy:2001}
Jones, E., Oliphant, T., Peterson, P., {et~al.} 2001, {SciPy}: Open source
  scientific tools for {Python}.
\newblock \url{http://www.scipy.org/}

\bibitem[{{Jones} {et~al.}(1993){Jones}, {Humphreys}, {Gehrz}, {Lawrence},
  {Zickgraf}, {Moseley}, {Casey}, {Glaccum}, {Koch}, {Pina}, {Jones}, {Venn},
  {Stahl}, \& {Starrfield}}]{jones93}
{Jones}, T.~J., {Humphreys}, R.~M., {Gehrz}, R.~D., {et~al.} 1993, \apj, 411,
  323, \dodoi{10.1086/172832}

\bibitem[{{Kochanek}(2020)}]{kochanek20}
{Kochanek}, C.~S. 2020, \mnras, 493, 4945, \dodoi{10.1093/mnras/staa605}

\bibitem[{{Kourniotis} {et~al.}(2022){Kourniotis}, {Kraus}, {Maryeva}, {Borges
  Fernandes}, \& {Maravelias}}]{kourniotis22}
{Kourniotis}, M., {Kraus}, M., {Maryeva}, O., {Borges Fernandes}, M., \&
  {Maravelias}, G. 2022, \mnras, 511, 4360, \dodoi{10.1093/mnras/stac386}

\bibitem[{{Kov{\'a}cs}(2000)}]{kovacs00}
{Kov{\'a}cs}, G. 2000, \aap, 363, L1.
\newblock \doarXiv{astro-ph/0011056}

\bibitem[{Kumar(1975)}]{kumar75}
Kumar, T.~K. 1975, The Review of Economics and Statistics, 57, 365

\bibitem[{{Kurucz}(1992)}]{kurucz92}
{Kurucz}, R.~L. 1992, in IAU Symposium, Vol. 149, The Stellar Populations of
  Galaxies, ed. B.~{Barbuy} \& A.~{Renzini}, 225

\bibitem[{Legendre(1805)}]{legendre1805}
Legendre, A. 1805, Nouvelles m{\'e}thodes pour la d{\'e}termination des orbites
  des com{\`e}tes, Nineteenth Century Collections Online (NCCO): Science,
  Technology, and Medicine: 1780-1925 (F. Didot).
\newblock \url{https://books.google.com/books?id=FRcOAAAAQAAJ}

\bibitem[{{Levesque} {et~al.}(2007){Levesque}, {Massey}, {Olsen}, \&
  {Plez}}]{levesque07b}
{Levesque}, E.~M., {Massey}, P., {Olsen}, K.~A.~G., \& {Plez}, B. 2007, \apj,
  667, 202, \dodoi{10.1086/520797}

\bibitem[{{Levesque} {et~al.}(2005){Levesque}, {Massey}, {Olsen}, {Plez},
  {Josselin}, {Maeder}, \& {Meynet}}]{levesque05}
{Levesque}, E.~M., {Massey}, P., {Olsen}, K.~A.~G., {et~al.} 2005, \apj, 628,
  973, \dodoi{10.1086/430901}

\bibitem[{{Levesque} {et~al.}(2006){Levesque}, {Massey}, {Olsen}, {Plez},
  {Meynet}, \& {Maeder}}]{levesque06}
---. 2006, \apj, 645, 1102, \dodoi{10.1086/504417}

\bibitem[{{Lindegren} {et~al.}(2021){Lindegren}, {Bastian}, {Biermann},
  {Bombrun}, {de Torres}, {Gerlach}, {Geyer}, {Hern{\'a}ndez}, {Hilger},
  {Hobbs}, {Klioner}, {Lammers}, {McMillan}, {Ramos-Lerate},
  {Steidelm{\"u}ller}, {Stephenson}, \& {van Leeuwen}}]{lindegren21}
{Lindegren}, L., {Bastian}, U., {Biermann}, M., {et~al.} 2021, \aap, 649, A4,
  \dodoi{10.1051/0004-6361/202039653}

\bibitem[{Mahalanobis(1936)}]{mahalanobis36}
Mahalanobis, P.~C. 1936

\bibitem[{{Maravelias} {et~al.}(2022){Maravelias}, {Bonanos}, {Tramper}, {de
  Wit}, {Yang}, \& {Bonfini}}]{maravelias22}
{Maravelias}, G., {Bonanos}, A.~Z., {Tramper}, F., {et~al.} 2022, arXiv
  e-prints, arXiv:2203.08125.
\newblock \doarXiv{2203.08125}

\bibitem[{Maronna \& Zamar(2002)}]{maronna02}
Maronna, R.~A., \& Zamar, R.~H. 2002, Technometrics, 44, 307

\bibitem[{{Massey} {et~al.}(2007){Massey}, {Levesque}, {Olsen}, {Plez}, \&
  {Skiff}}]{massey07c}
{Massey}, P., {Levesque}, E.~M., {Olsen}, K.~A.~G., {Plez}, B., \& {Skiff},
  B.~A. 2007, \apj, 660, 301, \dodoi{10.1086/513182}

\bibitem[{{Massey} {et~al.}(2021{\natexlab{a}}){Massey}, {Neugent},
  {Dorn-Wallenstein}, {Eldridge}, {Stanway}, \& {Levesque}}]{massey21}
{Massey}, P., {Neugent}, K.~F., {Dorn-Wallenstein}, T.~Z., {et~al.}
  2021{\natexlab{a}}, arXiv e-prints, arXiv:2107.08304.
\newblock \doarXiv{2107.08304}

\bibitem[{{Massey} {et~al.}(2022){Massey}, {Neugent}, {Ekstrom}, {Georgy}, \&
  {Meynet}}]{massey22}
{Massey}, P., {Neugent}, K.~F., {Ekstrom}, S., {Georgy}, C., \& {Meynet}, G.
  2022, arXiv e-prints, arXiv:2211.14147.
\newblock \doarXiv{2211.14147}

\bibitem[{{Massey} {et~al.}(2021{\natexlab{b}}){Massey}, {Neugent}, {Levesque},
  {Drout}, \& {Courteau}}]{massey21b}
{Massey}, P., {Neugent}, K.~F., {Levesque}, E.~M., {Drout}, M.~R., \&
  {Courteau}, S. 2021{\natexlab{b}}, \aj, 161, 79,
  \dodoi{10.3847/1538-3881/abd01f}

\bibitem[{{Massey} {et~al.}(2005){Massey}, {Plez}, {Levesque}, {Olsen},
  {Clayton}, \& {Josselin}}]{massey05}
{Massey}, P., {Plez}, B., {Levesque}, E.~M., {et~al.} 2005, \apj, 634, 1286,
  \dodoi{10.1086/497065}

\bibitem[{{McDonald} {et~al.}(2022){McDonald}, {Davies}, \&
  {Beasor}}]{mcdonald22}
{McDonald}, S. L.~E., {Davies}, B., \& {Beasor}, E.~R. 2022, \mnras, 510, 3132,
  \dodoi{10.1093/mnras/stab3453}

\bibitem[{McKinney(2010)}]{pandas:2010}
McKinney, W. 2010, in Proceedings of the 9th Python in Science Conference, ed.
  S.~van~der Walt \& J.~Millman, 51 -- 56

\bibitem[{{M{\'e}sz{\'a}ros} {et~al.}(2012){M{\'e}sz{\'a}ros}, {Allende
  Prieto}, {Edvardsson}, {Castelli}, {Garc{\'\i}a P{\'e}rez}, {Gustafsson},
  {Majewski}, {Plez}, {Schiavon}, {Shetrone}, \& {de Vicente}}]{meszaros12}
{M{\'e}sz{\'a}ros}, S., {Allende Prieto}, C., {Edvardsson}, B., {et~al.} 2012,
  \aj, 144, 120, \dodoi{10.1088/0004-6256/144/4/120}

\bibitem[{{Meynet} \& {Maeder}(2003)}]{meynet03}
{Meynet}, G., \& {Maeder}, A. 2003, \aap, 404, 975,
  \dodoi{10.1051/0004-6361:20030512}

\bibitem[{{Neugent} {et~al.}(2020{\natexlab{a}}){Neugent}, {Levesque},
  {Massey}, {Morrell}, \& {Drout}}]{neugent20b}
{Neugent}, K.~F., {Levesque}, E.~M., {Massey}, P., {Morrell}, N.~I., \&
  {Drout}, M.~R. 2020{\natexlab{a}}, arXiv e-prints, arXiv:2007.15852.
\newblock \doarXiv{2007.15852}

\bibitem[{{Neugent} {et~al.}(2012{\natexlab{a}}){Neugent}, {Massey}, \&
  {Georgy}}]{neugent12}
{Neugent}, K.~F., {Massey}, P., \& {Georgy}, C. 2012{\natexlab{a}}, \apj, 759,
  11, \dodoi{10.1088/0004-637X/759/1/11}

\bibitem[{{Neugent} {et~al.}(2020{\natexlab{b}}){Neugent}, {Massey}, {Georgy},
  {Drout}, {Mommert}, {Levesque}, {Meynet}, \& {Ekstr{\"o}m}}]{neugent20}
{Neugent}, K.~F., {Massey}, P., {Georgy}, C., {et~al.} 2020{\natexlab{b}},
  \apj, 889, 44, \dodoi{10.3847/1538-4357/ab5ba0}

\bibitem[{{Neugent} {et~al.}(2010){Neugent}, {Massey}, {Skiff}, {Drout},
  {Meynet}, \& {Olsen}}]{neugent10}
{Neugent}, K.~F., {Massey}, P., {Skiff}, B., {et~al.} 2010, \apj, 719, 1784,
  \dodoi{10.1088/0004-637X/719/2/1784}

\bibitem[{{Neugent} {et~al.}(2012{\natexlab{b}}){Neugent}, {Massey}, {Skiff},
  \& {Meynet}}]{neugent12_ysg}
{Neugent}, K.~F., {Massey}, P., {Skiff}, B., \& {Meynet}, G.
  2012{\natexlab{b}}, \apj, 749, 177, \dodoi{10.1088/0004-637X/749/2/177}

\bibitem[{{Neustadt} {et~al.}(2021){Neustadt}, {Kochanek}, {Stanek},
  {Basinger}, {Jayasinghe}, {Garling}, {Adams}, \& {Gerke}}]{neustadt21}
{Neustadt}, J.~M.~M., {Kochanek}, C.~S., {Stanek}, K.~Z., {et~al.} 2021,
  \mnras, 508, 516, \dodoi{10.1093/mnras/stab2605}

\bibitem[{{Nieuwenhuijzen} \& {de Jager}(1995)}]{nieuwenhuijzen95}
{Nieuwenhuijzen}, H., \& {de Jager}, C. 1995, \aap, 302, 811

\bibitem[{{Olivares} {et~al.}(2020){Olivares}, {Sarro}, {Bouy}, {Miret-Roig},
  {Casamiquela}, {Galli}, {Berihuete}, \& {Tarricq}}]{olivares20}
{Olivares}, J., {Sarro}, L.~M., {Bouy}, H., {et~al.} 2020, \aap, 644, A7,
  \dodoi{10.1051/0004-6361/202037846}

\bibitem[{{Patton} \& {Sukhbold}(2020)}]{patton20}
{Patton}, R.~A., \& {Sukhbold}, T. 2020, \mnras, 499, 2803,
  \dodoi{10.1093/mnras/staa3029}

\bibitem[{{Pedersen} \& {Bell}(2023)}]{pedersen23}
{Pedersen}, M.~G., \& {Bell}, K.~J. 2023, arXiv e-prints, arXiv:2304.05706,
  \dodoi{10.48550/arXiv.2304.05706}

\bibitem[{Pedregosa {et~al.}(2011)Pedregosa, Varoquaux, Gramfort, Michel,
  Thirion, Grisel, Blondel, Prettenhofer, Weiss, Dubourg, Vanderplas, Passos,
  Cournapeau, Brucher, Perrot, \& Duchesnay}]{scikit-learn11}
Pedregosa, F., Varoquaux, G., Gramfort, A., {et~al.} 2011, Journal of Machine
  Learning Research, 12, 2825

\bibitem[{{Prentice} {et~al.}(2022){Prentice}, {Maguire}, {Siebenaler}, \&
  {Jerkstrand}}]{prentice22}
{Prentice}, S.~J., {Maguire}, K., {Siebenaler}, L., \& {Jerkstrand}, A. 2022,
  \mnras, 514, 5686, \dodoi{10.1093/mnras/stac1657}

\bibitem[{{Rasmussen} \& {Williams}(2006)}]{rasmussen06}
{Rasmussen}, C.~E., \& {Williams}, C. K.~I. 2006, {Gaussian Processes for
  Machine Learning} (MIT Press)

\bibitem[{{Rodr{\'\i}guez}(2022)}]{rodriguez22}
{Rodr{\'\i}guez}, {\'O}. 2022, \mnras, 515, 897, \dodoi{10.1093/mnras/stac1831}

\bibitem[{{Sanduleak}(1970)}]{sanduleak70}
{Sanduleak}, N. 1970, Contributions from the Cerro Tololo Inter-American
  Observatory, 89

\bibitem[{{Savitzky} \& {Golay}(1964)}]{savitzky64}
{Savitzky}, A., \& {Golay}, M.~J.~E. 1964, Analytical Chemistry, 36, 1627,
  \dodoi{10.1021/ac60214a047}

\bibitem[{{Schlegel} {et~al.}(1998){Schlegel}, {Finkbeiner}, \&
  {Davis}}]{schlegel98}
{Schlegel}, D.~J., {Finkbeiner}, D.~P., \& {Davis}, M. 1998, \apj, 500, 525,
  \dodoi{10.1086/305772}

\bibitem[{{Shenoy} {et~al.}(2016){Shenoy}, {Humphreys}, {Jones}, {Marengo},
  {Gehrz}, {Helton}, {Hoffmann}, {Skemer}, \& {Hinz}}]{shenoy16}
{Shenoy}, D., {Humphreys}, R.~M., {Jones}, T.~J., {et~al.} 2016, \aj, 151, 51,
  \dodoi{10.3847/0004-6256/151/3/51}

\bibitem[{{Sick} {et~al.}(2014){Sick}, {Courteau}, {Cuillandre}, {McDonald},
  {de Jong}, \& {Tully}}]{sick14}
{Sick}, J., {Courteau}, S., {Cuillandre}, J.-C., {et~al.} 2014, \aj, 147, 109,
  \dodoi{10.1088/0004-6256/147/5/109}

\bibitem[{{Smartt}(2015)}]{smartt15}
{Smartt}, S.~J. 2015, \pasa, 32, e016, \dodoi{10.1017/pasa.2015.17}

\bibitem[{{Smartt} {et~al.}(2009){Smartt}, {Eldridge}, {Crockett}, \&
  {Maund}}]{smartt09}
{Smartt}, S.~J., {Eldridge}, J.~J., {Crockett}, R.~M., \& {Maund}, J.~R. 2009,
  \mnras, 395, 1409, \dodoi{10.1111/j.1365-2966.2009.14506.x}

\bibitem[{{Smith}(2014)}]{smith14}
{Smith}, N. 2014, \araa, 52, 487, \dodoi{10.1146/annurev-astro-081913-040025}

\bibitem[{Smola \& Sch{\"o}lkopf(2004)}]{smola04}
Smola, A.~J., \& Sch{\"o}lkopf, B. 2004, Statistics and Computing, 14, 199,
  \dodoi{10.1023/B:STCO.0000035301.49549.88}

\bibitem[{{Stothers} \& {Chin}(2001)}]{stothers01}
{Stothers}, R.~B., \& {Chin}, C.-w. 2001, \apj, 560, 934,
  \dodoi{10.1086/322438}

\bibitem[{{Sukhbold} {et~al.}(2016){Sukhbold}, {Ertl}, {Woosley}, {Brown}, \&
  {Janka}}]{sukhbold16}
{Sukhbold}, T., {Ertl}, T., {Woosley}, S.~E., {Brown}, J.~M., \& {Janka}, H.-T.
  2016, \apj, 821, 38, \dodoi{10.3847/0004-637X/821/1/38}

\bibitem[{{Sukhbold} {et~al.}(2018){Sukhbold}, {Woosley}, \&
  {Heger}}]{sukhbold18}
{Sukhbold}, T., {Woosley}, S.~E., \& {Heger}, A. 2018, \apj, 860, 93,
  \dodoi{10.3847/1538-4357/aac2da}

\bibitem[{{Tabernero} {et~al.}(2018){Tabernero}, {Dorda}, {Negueruela}, \&
  {Gonz{\'a}lez-Fern{\'a}ndez}}]{tabernero18}
{Tabernero}, H.~M., {Dorda}, R., {Negueruela}, I., \&
  {Gonz{\'a}lez-Fern{\'a}ndez}, C. 2018, \mnras, 476, 3106,
  \dodoi{10.1093/mnras/sty399}

\bibitem[{{The Astropy Collaboration} {et~al.}(2018){The Astropy
  Collaboration}, {Price-Whelan}, {Sip{\H o}cz}, {G{\"u}nther}, {Lim},
  {Crawford}, {Conseil}, {Shupe}, {Craig}, {Dencheva}, {Ginsburg},
  {VanderPlas}, {Bradley}, {P{\'e}rez-Su{\'a}rez}, {de Val-Borro}, {Aldcroft},
  {Cruz}, {Robitaille}, {Tollerud}, {Ardelean}, {Babej}, {Bachetti}, {Bakanov},
  {Bamford}, {Barentsen}, {Barmby}, {Baumbach}, {Berry}, {Biscani}, {Boquien},
  {Bostroem}, {Bouma}, {Brammer}, {Bray}, {Breytenbach}, {Buddelmeijer},
  {Burke}, {Calderone}, {Cano Rodr{\'{\i}}guez}, {Cara}, {Cardoso},
  {Cheedella}, {Copin}, {Crichton}, {D{\'A}vella}, {Deil}, {Depagne},
  {Dietrich}, {Donath}, {Droettboom}, {Earl}, {Erben}, {Fabbro}, {Ferreira},
  {Finethy}, {Fox}, {Garrison}, {Gibbons}, {Goldstein}, {Gommers}, {Greco},
  {Greenfield}, {Groener}, {Grollier}, {Hagen}, {Hirst}, {Homeier}, {Horton},
  {Hosseinzadeh}, {Hu}, {Hunkeler}, {Ivezi{\'c}}, {Jain}, {Jenness}, {Kanarek},
  {Kendrew}, {Kern}, {Kerzendorf}, {Khvalko}, {King}, {Kirkby}, {Kulkarni},
  {Kumar}, {Lee}, {Lenz}, {Littlefair}, {Ma}, {Macleod}, {Mastropietro},
  {McCully}, {Montagnac}, {Morris}, {Mueller}, {Mumford}, {Muna}, {Murphy},
  {Nelson}, {Nguyen}, {Ninan}, {N{\"o}the}, {Ogaz}, {Oh}, {Parejko}, {Parley},
  {Pascual}, {Patil}, {Patil}, {Plunkett}, {Prochaska}, {Rastogi}, {Reddy
  Janga}, {Sabater}, {Sakurikar}, {Seifert}, {Sherbert}, {Sherwood-Taylor},
  {Shih}, {Sick}, {Silbiger}, {Singanamalla}, {Singer}, {Sladen}, {Sooley},
  {Sornarajah}, {Streicher}, {Teuben}, {Thomas}, {Tremblay}, {Turner},
  {Terr{\'o}n}, {van Kerkwijk}, {de la Vega}, {Watkins}, {Weaver}, {Whitmore},
  {Woillez}, \& {Zabalza}}]{astropy18}
{The Astropy Collaboration}, {Price-Whelan}, A.~M., {Sip{\H o}cz}, B.~M.,
  {et~al.} 2018, ArXiv e-prints.
\newblock \doarXiv{1801.02634}

\bibitem[{{Tsang} {et~al.}(2022){Tsang}, {Vartanyan}, \& {Burrows}}]{tsang22}
{Tsang}, B. T.~H., {Vartanyan}, D., \& {Burrows}, A. 2022, \apjl, 937, L15,
  \dodoi{10.3847/2041-8213/ac8f4b}

\bibitem[{van~der Velden(2020)}]{vandervelden20}
van~der Velden, E. 2020, Journal of Open Source Software, 5, 2004,
  \dodoi{10.21105/joss.02004}

\bibitem[{{Van Der Walt} {et~al.}(2011){Van Der Walt}, {Colbert}, \&
  {Varoquaux}}]{numpy:2011}
{Van Der Walt}, S., {Colbert}, S.~C., \& {Varoquaux}, G. 2011, ArXiv e-prints,
  arXiv:1102.1523.
\newblock \doarXiv{1102.1523}

\bibitem[{Virtanen {et~al.}(2020)Virtanen, Gommers, Oliphant, Haberland, Reddy,
  Cournapeau, Burovski, Peterson, Weckesser, Bright, {van der Walt}, Brett,
  Wilson, Millman, Mayorov, Nelson, Jones, Kern, Larson, Carey, Polat, Feng,
  Moore, {VanderPlas}, Laxalde, Perktold, Cimrman, Henriksen, Quintero, Harris,
  Archibald, Ribeiro, Pedregosa, {van Mulbregt}, \& {SciPy 1.0
  Contributors}}]{scipy:2020}
Virtanen, P., Gommers, R., Oliphant, T.~E., {et~al.} 2020, Nature Methods, 17,
  261, \dodoi{10.1038/s41592-019-0686-2}

\bibitem[{{Walmswell} \& {Eldridge}(2012)}]{walmswell12}
{Walmswell}, J.~J., \& {Eldridge}, J.~J. 2012, \mnras, 419, 2054,
  \dodoi{10.1111/j.1365-2966.2011.19860.x}

\bibitem[{{Wasatonic} {et~al.}(2015){Wasatonic}, {Guinan}, \&
  {Durbin}}]{wasatonic15}
{Wasatonic}, R.~P., {Guinan}, E.~F., \& {Durbin}, A.~J. 2015, \pasp, 127, 1010,
  \dodoi{10.1086/683261}

\bibitem[{{Weiler} {et~al.}(2022){Weiler}, {Carrasco}, {Fabricius}, \&
  {Jordi}}]{weiler22}
{Weiler}, M., {Carrasco}, J.~M., {Fabricius}, C., \& {Jordi}, C. 2022, arXiv
  e-prints, arXiv:2211.06946.
\newblock \doarXiv{2211.06946}

\bibitem[{{Westerlund}(1960)}]{westerlund60}
{Westerlund}, B. 1960, Uppsala Astronomical Observatory Annals, 4, 1

\bibitem[{{Witten} {et~al.}(2022){Witten}, {Aguado}, {Sanders}, {Belokurov},
  {Evans}, {Koposov}, {Allende Prieto}, {De Angeli}, \& {Irwin}}]{witten22}
{Witten}, C. E.~C., {Aguado}, D.~S., {Sanders}, J.~L., {et~al.} 2022, \mnras,
  516, 3254, \dodoi{10.1093/mnras/stac2273}

\bibitem[{{Yang} {et~al.}(2018){Yang}, {Bonanos}, {Jiang}, {Gao}, {Xue},
  {Wang}, {Lam}, {Spetsieri}, {Ren}, \& {Gavras}}]{yang18}
{Yang}, M., {Bonanos}, A.~Z., {Jiang}, B.-W., {et~al.} 2018, \aap, 616, A175,
  \dodoi{10.1051/0004-6361/201832833}

\bibitem[{{Yang} {et~al.}(2019){Yang}, {Bonanos}, {Jiang}, {Gao}, {Gavras},
  {Maravelias}, {Ren}, {Wang}, {Xue}, {Tramper}, {Spetsieri}, \&
  {Pouliasis}}]{yang19}
---. 2019, \aap, 629, A91, \dodoi{10.1051/0004-6361/201935916}

\bibitem[{{Yang} {et~al.}(2021){Yang}, {Bonanos}, {Jiang}, {Gao}, {Gavras},
  {Maravelias}, {Wang}, {Chen}, {Lam}, {Ren}, {Tramper}, \&
  {Spetsieri}}]{yang21}
{Yang}, M., {Bonanos}, A.~Z., {Jiang}, B., {et~al.} 2021, \aap, 646, A141,
  \dodoi{10.1051/0004-6361/202039475}

\bibitem[{{Zhang} {et~al.}(2020){Zhang}, {Ti{\v{n}}o}, {Leonardis}, \&
  {Tang}}]{zhang20}
{Zhang}, Y., {Ti{\v{n}}o}, P., {Leonardis}, A., \& {Tang}, K. 2020, arXiv
  e-prints, arXiv:2012.14261, \dodoi{10.48550/arXiv.2012.14261}

\end{thebibliography}
\bibliographystyle{aasjournal}

\appendix
\restartappendixnumbering

\section{Regressor Testing}\label{app:regression}

Regression is one of the most commonly performed tasks in astronomy. Any astronomer who has fit a function to data, or performed an interpolation, or even reduced spectroscopic data has performed a regression of some sort whether they know it or not. Here, we wish to predict the temperatures and luminosities of objects in our sample, a task for which a variety of methods are more-or-less well-suited. We chose some of the most-commonly applied regression techniques which are available in the frequently used python package, {\sc scikit-learn}. We explored several variations of linear methods: linear regression \citep{legendre1805}; second- and third-order polynomial regressions; Bayesian Ridge Regression (RR; \S{3.3} in \citealt{bishop06}), which offsets the impact of multicolinearity on the model variance \citep{kumar75}; and Support Vector Machine (SVM) regression (SVR; \citealt{cortes95,smola04}), which penalizes the model for points that fall outside of some margin. 

In addition to these linear regressors, we experimented with ``fancier'' methods; we pause here to remind the reader that the best-performing methods in \S\ref{sec:regressing} are all linear. Because none of these more complex methods were chosen for the final regression, we leave a detailed explanation of these methods to other authors. In brief, we experimented with a $k$-nearest neighbors (KNN) regressor; a Gaussian process (GP) regressor \citep{rasmussen06}; a random forest (RF) regressor \citep{breiman01}; and finally, a multi-layer perceptron (MLP; \citealt{hastie09}), a variety of neural network available within {\sc scikit-learn}. In general, we are wary of neural networks; their complexity can make them difficult to troubleshoot and prone to overfitting, and the output can be difficult to interpret \citep{zhang20}. We included the MLP in our experimentation only to demonstrate that even in cases with high dimensional feature spaces with large samples, a neural network is not necessary as simpler methods can perform just has well. We also note that, while none of the authors of this work possess much expertise in neural networks, constructing and training the MLP was a trivial exercise involving a brief internet search, making the MLP the regressor with the highest ratio of complexity to ease of implementation by far. Overall, we caution that astronomers looking to incorporate machine learning into their work should avoid first turning to a neural network.

Because we are simultaneously fitting for $\teff$ and $\lum$, we need to perform {\it multi-output regression}, for which {\sc scikit-learn} offers two strategies. The first, and simpler method ({\tt MultiOutputRegressor}) simultaneously fits an independent instance of the base regressor to each target variable, using identical training data. The second method ({\tt RegressorChain}) fits the base regressor to one target variable, then uses the predicted values as a new feature when fitting the next target variable in the chain, making the order in which the target variables are fitted an additional hyperparameter. Most of the regressors we experimented with already supported multi-output regression using the first method. However, the SVR and Bayesian RR do not, allowing us to experiment with both methods. For these two regressors, we use ``(Multi)''/``(Chain)'' to indicate which variation of multi-output regression we are using.

\subsection{Testing}\label{subsec:testing}

We test a total of nine regressors: linear regression, linear regression with both second- and third-order polynomials, Bayesian RR, SVR, KNN regression, GP regression, RF regression, and the MLP. Among these regressors, the SVR, KNN regressor, GP regressor, RF regressor, and MLP have hyperparameters which can be tuned and about which we have no {\it a priori} information (for example, the margin in the SVR). To select the optimal hyperparameters, we use a grid search cross-validation strategy, splitting the training set further into into 5 folds. At each combination of hyperparameters in the grid, each fold is withheld, the model is fit to the remaining data, and then used to predict the targets for the left-out fold.\footnote{The exception is the RF regressor, where the grid of parameters is too large to search over. Instead, we randomly sample 1000 parameter combinations from the parameter grid, and use three cross-validation folds.} The hyperparameters for each regressor that perform ``best'' across all folds are chosen. 

While each regressor uses its own loss function to optimize the weights and hyperparameters, we need a method of quantifying the performance across regressors that is based on the distance between the observed values and the model predictions. When fitting individual target variables, a common choice when scoring models is the coefficient of determination $R^2$:
\begin{equation}
    R^2 = 1 - \frac{|y - \hat{y}|^2}{|y - \bar{y}|^2}
\end{equation}
where the vector $y$ contains the observed values for each sample, $\hat{y}$ is the predicted values from the model, and $\bar{y}$ is the mean of the true values. The best possible value (i.e., when $y = \hat{y}$) is 1. A model that ignores the features and only predicts the mean target returns $R^2 = 0$.

Greater care needs to be taken when fitting the regressors to both target variables simultaneously. Scorers within {\sc scikit-learn} do allow for multi-output regression, usually by taking a (weighted) average of a score (e.g., $R^2$) for each target variable. Because the observational errors in $\teff$ and $\lum$ are unequal, we wish to use a scorer that computes distances between predicted and true values while taking this into account. We can compute the Mahalanobis distance \citep{mahalanobis36}:
\begin{equation}
    d_M = \sqrt{(\hat{y}_i-\mu)\Sigma^{-1}(\hat{y}_i-\mu)^T}
\end{equation}
where $\mu = y_i$ is a row vector containing the observed values for a given sample, and $\Sigma$ is the covariance matrix of the measurement. Unfortunately, we don't have access to the full covariance matrix, so we assume that $\Sigma$ is a diagonal matrix with elements equal to the squared uncertainties of $\teff$ and $\lum$. In effect, we are determining the likelihood that the predicted value is drawn from a normal distribution with mean $\mu$ and covariance matrix $\Sigma$; lower probabilities correspond to higher $d_M$ values. 

The benefit of using $d_M$ is that $d_M^2$ is distributed like a $\chi^2$ distribution with degrees of freedom equal to the number of dimensions (in this case two: $\lum$ and $\teff$), making it an easy-to-interpret metric of the distance between the data and the model. For a $\chi^2$ distribution with two degrees of freedom, one can expect 95\% of the values to fall within $d_M^2\lesssim6$. Therefore, to score each regressor, we take the distribution of squared Mahalanobis distances between the predicted and true values, and identify the 95$^{\rm th}$-percentile of that distribution, $d_{M,95}^2$. Smaller values of $d_{M,95}^2$ mean that the predictions are closer to the observed values relative to the errors.

\begin{figure*}[ht!]
\plotone{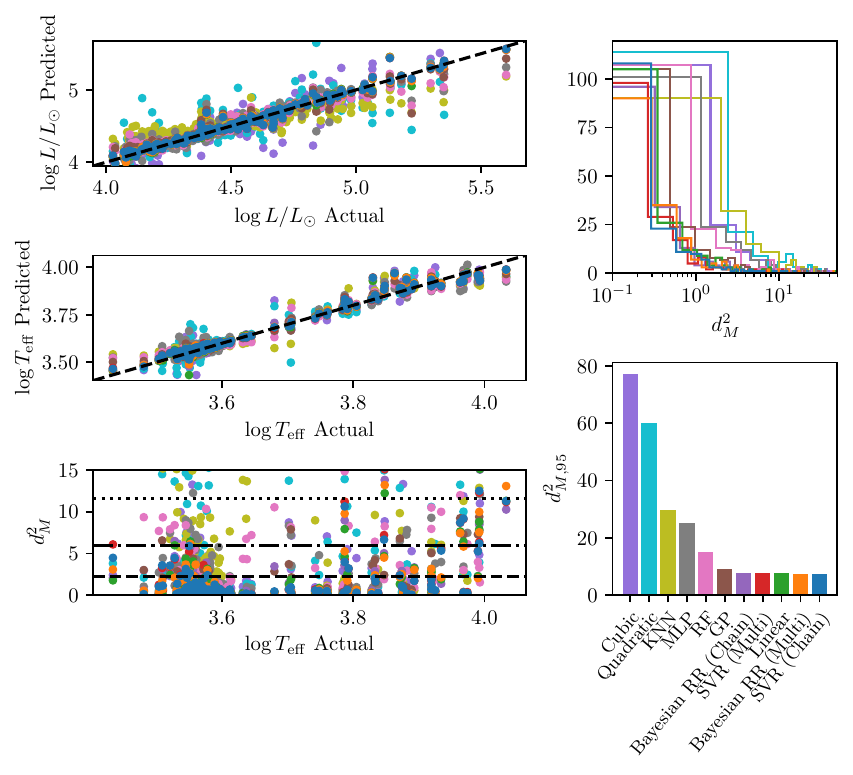}
\caption{A summary of the performance of each regressor we experimented with on both temperature and luminosity simultaneously. The two panels in the top left show the predicted versus actual values of each star the test set for $\lum$ (top left) and $\teff$ (center left). The bottom left panel shows the individual squared Mahalanobis distances $d_M^2$ for each point as a function of $\teff$. Dashed/dash-dotted/dotted lines indicate 64/95/99.7\% confidence levels for a $\chi^2$ distribution with two degrees of freedom, corresponding to 1/2/3$\sigma$ differences between the observed and true values. The top right panel show the histogram of $d_M^2$, and the bottom right panels show the $d_{M,95}^2$ values for each regressor. This panel also serves as a legend to map between the regressors and the colors of the corresponding points in the other panels.}\label{fig:twotarget_summ}
\end{figure*}

\begin{deluxetable*}{l|ccccccc}
\tabletypesize{\scriptsize}
\tablecaption{Summary of regressor performance on $\teff$ and $\lum$ simultaneously.\label{tab:tl_sum}}
\tablehead{\colhead{Regressor} & \colhead{$d_{M,95}^2$} & \colhead{$\langle\Delta\teff\rangle$} & \colhead{$\sigma_{\Delta\teff}$} & \colhead{$\langle\Delta\lum\rangle$} & \colhead{$\sigma_{\Delta\lum}$} & \colhead{\%$\leq1\sigma$} & \colhead{\%$\leq2\sigma$}}
\startdata
Cubic & 77.274 & -0.0016 & 0.0343 & -0.0043 & 0.2956 & 64.77 & 77.72 \\ 
Quadratic & 59.906 & 0.0023 & 0.0511 & 0.0082 & 0.2340 & 56.99 & 72.02 \\ 
KNN & 29.769 & 0.0055 & 0.0304 & 0.0301 & 0.1656 & 48.19 & 69.95 \\ 
MLP & 25.256 & 0.0011 & 0.0286 & 0.0043 & 0.0988 & 64.77 & 82.90 \\ 
RF & 15.190 & 0.0031 & 0.0216 & 0.0090 & 0.0912 & 73.06 & 84.46 \\ 
GP & 9.282 & 0.0021 & 0.0187 & 0.0045 & 0.0728 & 79.27 & 91.19 \\ 
Bayesian RR (Chain) & 7.827 & 0.0019 & 0.0192 & 0.0050 & 0.0503 & 83.42 & 93.26 \\ 
SVR (Multi) & 7.784 & 0.0021 & 0.0190 & 0.0058 & 0.0510 & 85.49 & 93.26 \\ 
Linear & 7.734 & 0.0010 & 0.0199 & 0.0033 & 0.0513 & 86.01 & 93.26 \\ 
Bayesian RR (Multi) & 7.487 & 0.0020 & 0.0189 & 0.0050 & 0.0503 & 85.49 & 93.78 \\ 
SVR (Chain) & 7.463 & 0.0021 & 0.0189 & 0.0058 & 0.0510 & 86.01 & 93.78 \\ 
\enddata
\end{deluxetable*}
\onecolumngrid

After identifying the best hyperparameters for each regressor where necessary, we then fit each regressor to the training set and use the trained regressor to  predict values for the test set. Figure \ref{fig:twotarget_summ} shows a summary of the performance of the regressors. The top/center left panels show the predicted values of $\lum$/$\teff$ respectively versus their observed values. Different colors show the results from different regressors. The one-to-one line is shown in black in both panels. The bottom left panel shows the value of $d_M^2$ for each point in the test set as a function of the observed $\teff$. Recalling that these are distances from the center of a two-dimensional Gaussian, and that $d_M^2$ is $\chi^2$-distributed with two degrees of freedom, we plot the expected 64/95/99.7 percentile thresholds as dashed/dash-dotted/dotted horizontal lines. Note that we only plot $d_M^2$ values below 15; the worst performing regressors resulted in some $d_M^2$ values that are far larger. The top-right panel shows the distribution of $d_M^2$ values for each regressor, and the bottom right panel shows the final $d_{M,95}^2$ scores, sorted from highest (worst performing) to lowest (best performing). The bottom right panel also serves as a legend to interpret the colors assigned to each regressor. 

Tables \ref{tab:tl_sum} shows a numerical summary of Figure \ref{fig:twotarget_summ}, including additional performance metrics. For each regressor, we list $d_{M,95}$, the mean residual between the model and the observed data $\langle\Delta\teff\rangle$/$\langle\Delta\lum\rangle$, the standard deviation of residuals $\sigma_{\Delta\teff}$/$\sigma_{\Delta\lum}$, and the percentage of individual $d_M^2$ values less than the 64\% (\%$\leq1\sigma$) and 95\% (\%$\leq2\sigma$) thresholds. We find that the performance of the regressors peaks around $d_{M,95}^2=7.5$. Recall that the expected 95$^{\rm th}$ value for a $\chi^2$ distribution with two degrees of freedom is around 6, which would indicate that our regressor is predicting values that deviate from the observations by more than what we would expect given the observational uncertainties. However, examining the bottom left panel of Figure \ref{fig:twotarget_summ} in detail, the vast majority of points for the best two regressors in blue --- the SVR (Chain) --- and orange --- the Bayesian RR (Multi) --- fall within the expected 64$^{\rm th}$ or 95$^{\rm th}$ percentile thresholds. There are only twelve points from each of these regressors that are above the 95$^{\rm th}$ percentile line, and for the Bayesian RR, all occur above $\teff\approx3.8$. As these are the objects for which the $J-K_S$ temperatures begin to become increasingly inaccurate (see the discussion around Figures \ref{fig:tjk_comp} and \ref{fig:delta_jk}), this is simply a reflection of the fact that the covariance matrix for these points is likely underestimated.

Continuing to examine Table \ref{tab:tl_sum}, we find that both polynomial regressions, the KNN regressor, and the MLP perform abysmally across all metrics. In the case of everything but the KNN, this is likely the result of overfitting on the training set, causing the regressors to generalize poorly. While the RF and GP regressors begin to perform better, the best performance is achieved by both multi-output options for the SVR and Bayesian RR, as well as the simple linear regressor. As we were performing exploratory work for this paper, these regression methods were consistently the top performers across different preprocessing steps, choices of performance metric, and iterations of the training set. Indeed, given an identical setup, a different split between the training and testing set would likely yield a slightly different ordering of these regressors. Given that:
\begin{enumerate}
    \item We wish to select the simplest possible regressor, and
    \item We have done little to mitigate any multicolinearity in the training data, a problem which Bayesian RR is naturally well-suited for,
\end{enumerate}
we selected the Bayesian RR with the simple multi-output strategy for this paper.

\subsection{Hyperparameter Grids}

As mentioned above, the SVR, KNN regressor, GP regressor, RF regressor, and MLP have a number of tunable hyperparameters that we selected using a grid search (or a random search for the RF regressor). These grids were defined as follows (with details of the meaning of each hyperparameter available in the {\sc scikit-learn} documentation).

For the SVR:
\begin{itemize}
    \item {\tt epsilon}: 0.001, 0.015, 0.1, 0.5
    \item {\tt C}: 10 logarithmically-spaced values between $10^{-2}$ and 10
    \item {\tt kernel}: {\tt linear}, {\tt rbf}, {\tt poly}
    \item {\tt degree}: 2, 3
\end{itemize}

For the KNN regressor:
\begin{itemize}
    \item {\tt n\_neighbors}: all integers in [5,50]
    \item {\tt weights}: {\tt uniform}, {\tt distance}
\end{itemize}

For the GP regressor:
\begin{itemize}
    \item {\tt kernel}: {\tt RBF}, {\tt Matern}, {\tt RBF}+{\tt WhiteKernel}
    \item {\tt noise\_level}$=$0.1, {\tt noise\_level\_bounds}$=$($10^{-10},10^5$)\footnote{Hyperparameters for the {\tt WhiteKernel}}
    \item {\tt normalize\_y}: {\tt True}
    \item {\tt alpha}: 0.1
\end{itemize}

For the RF regressor:
\begin{itemize}
    \item {\tt n\_estimators}: 200, 400, 600, 800, 1000, 1200, 1400, 1600, 1800, 2000
    \item {\tt max\_features}: {\tt auto}, {\tt sqrt}
    \item {\tt max\_depth}: 10, 20, 30, 40, 50, 60, 70, 80, 90, 100, 110, {\tt None}
    \item {\tt min\_samples\_split}: 2, 5, 10
    \item {\tt min\_samples\_leaf}: 1, 2, 4
    \item {\tt bootstrap}: {\tt True}, {\tt False}
\end{itemize}

For the MLP:
\begin{itemize}
    \item {\tt hidden\_layer\_sizes}: (150,100,50), (120,80,40), (100,50,30)
    \item {\tt max\_iter}: 300,500
    \item {\tt activation}: {\tt tanh}, {\tt relu}
    \item {\tt solver}: {\tt sgd}, {\tt adam}
    \item {\tt alpha}: 0.0001, 0.05
    \item {\tt learning\_rate}: {\tt constant}, {\tt adaptive}
\end{itemize}

\section{A Catalog of Temperatures and Luminosities for 6146 AFGMK Supergiants}

Table \ref{tab:catalog} lists the $\teff$ and $\lum$ values for our sample, indexed by the \Gaia~DR3 source ID numbers. In addition, we also indicate the source for each star: N for stars from \citetalias{neugent12_ysg}, Y for stars from \citetalias{yang21} (whose temperatures and luminosities are from this work). We also list the quality of each source's values, which is 2 for the OOB stars, 1 for the low-$L$ stars, and 0 for all others.  

\begin{deluxetable*}{lcccc}
\tabletypesize{\footnotesize}
\tablecaption{Temperatures and Luminosities for 6146 AFGKM Supergiants.\label{tab:catalog}}
\tablehead{\colhead{\Gaia~Source ID} & \colhead{$\teff$} & \colhead{$\lum$} & \colhead{Source\tablenotemark{\footnotesize a}} & \colhead{Quality\tablenotemark{\footnotesize b}}\\
\colhead{} & \colhead{[K]} & \colhead{[$L_\odot$]} & \colhead{} & \colhead{}} 
\startdata
4660104357325381760 & 3.956 & 3.902 & Y & 1 \\ 
4660464928419651328 & 3.580 & 4.864 & Y & 0 \\ 
4656989818831984896 & 3.887 & 3.734 & Y & 1 \\ 
4659350637776084096 & 4.006 & 3.800 & Y & 1 \\ 
4661701535389763328 & 3.645 & 3.805 & Y & 1 \\ 
4657574239298248576 & 3.749 & 4.229 & Y & 1 \\ 
4658301531875937280 & 3.624 & 3.700 & Y & 1 \\ 
4661280555551068800 & 3.587 & 4.779 & Y & 0 \\ 
4655455656504136320 & 3.997 & 4.189 & Y & 1 \\ 
4661226816920305152 & 3.598 & 4.441 & Y & 0 \\ 
\enddata
\tablecomments{A random selection of ten rows are shown to illustrate the content of the table. A complete machine-readable version will be made available online.}
\tablenotetext{a}{N for stars from \citetalias{neugent12_ysg}, Y for stars from \citetalias{yang21}, for which $\teff$ and $\lum$ are derived by our regressor.}
\tablenotetext{b}{0 for stars from \citetalias{neugent12_ysg} or \citetalias{yang21} stars that fall within the temperature and luminosity boundaries of the training set, 1 for low-$L$ stars, and 2 for OOB stars.}
\end{deluxetable*}

\end{document}